\documentclass[
    aps,
    prx,
    letterpaper,
    nobalancelastpage,
    twocolumn,
    superscriptaddress,
    nofootinbib,
    longbibliography
]{revtex4-2}

\usepackage{array}[=2016-10-06]

\usepackage[colorlinks, citecolor=black, linkcolor=black, urlcolor=black]{hyperref}
\usepackage{graphicx}
\usepackage{amsmath}
\usepackage{amssymb}
\usepackage[english]{babel}
\usepackage{color}
\definecolor{revisionred}{rgb}{0,0,0}
\usepackage[version=4]{mhchem}
\usepackage{makecell}
\usepackage{dsfont}
\usepackage{multirow}
\usepackage{booktabs}
\usepackage{stmaryrd}
\usepackage{physics}
\usepackage{bm}
\usepackage{comment}
\usepackage{braket}

\usepackage{soul}

\usepackage{textcomp}

\usepackage{float}
\makeatletter
\let\newfloat\newfloat@ltx
\makeatother

\usepackage{algorithm}
\usepackage{algpseudocode}

\graphicspath{{./Figures/}}

\newcommand{\UIUC}{
    Department of Physics,
    The University of Illinois at Urbana-Champaign,
    Urbana, IL 61801, USA
}

\newcommand{\UC}{
    Department of Physics and James Frank Institute,
    University of Chicago,
    Chicago, IL 60637, USA
}

\newcommand{\PME}{
    Pritzker School of Molecular Engineering,
    University of Chicago, 
    Chicago, IL 60637, USA
}

\newcommand{\ANL}{
    Physics Division, 
    Argonne National Laboratory, 
    Lemont, IL 60439, USA
}

\newcommand{\blue}[1]{\red{#1}}
\DeclareRobustCommand{\red}[1]{\texorpdfstring{{\color{revisionred}\hypersetup{allcolors=revisionred}#1}}{#1}}

\renewcommand{\cite}[1]{\mbox{\citep{#1}}}

\addto\captionsenglish{}
\addto\captionsenglish{\renewcommand\thefigure{\arabic{figure}}}

\newcommand{\bluedel}[1]{\blue{\st{#1}}}
\newsavebox{\aritholdbox}
\newcommand{\bluedelmath}[1]{%
    \begingroup\sbox{\aritholdbox}{\blue{$\displaystyle #1$}}%
    \ifdim\wd\aritholdbox>\linewidth
        \text{\bluedel{\mbox{\resizebox{0.97\linewidth}{!}{\usebox{\aritholdbox}}}}}%
    \else
        \text{\bluedel{\mbox{\usebox{\aritholdbox}}}}%
    \fi\endgroup}

\begin{document}
\makeatletter
\expandafter\let\csname revision@reference@Ataides2025\endcsname\@empty
\expandafter\let\csname revision@reference@Buhler1993\endcsname\@empty
\expandafter\let\csname revision@reference@Chevignard2024\endcsname\@empty
\expandafter\let\csname revision@reference@Ekera2016\endcsname\@empty
\expandafter\let\csname revision@reference@Ekera2020\endcsname\@empty
\expandafter\let\csname revision@reference@Pattison2024\endcsname\@empty
\expandafter\let\csname revision@reference@SahayCultivationData2026\endcsname\@empty
\expandafter\let\csname revision@reference@Schmidt2025\endcsname\@empty
\expandafter\let\csname revision@reference@Sunami2026\endcsname\@empty
\expandafter\let\csname revision@reference@VanMeter2006\endcsname\@empty
\expandafter\let\csname revision@reference@VanMeter2010\endcsname\@empty
\expandafter\let\csname revision@reference@Xiao2022\endcsname\@empty
\expandafter\let\csname revision@reference@Xiao2023\endcsname\@empty
\expandafter\let\csname revision@reference@Yimsiriwattana2004\endcsname\@empty

\let\revision@base@citecolor\@citecolor
\let\revision@base@linkcolor\@linkcolor
\let\revision@base@urlcolor\@urlcolor
\let\revision@bibitem\bibitem
\def\bibitem{\@ifnextchar[{\revision@bibitem@optional}{\revision@bibitem@optional[]}}
\def\revision@bibitem@optional[#1]#2{%
  \ifcsname revision@reference@#2\endcsname
    \color{revisionred}\hypersetup{allcolors=revisionred}%
  \else
    \normalcolor
    \let\@citecolor\revision@base@citecolor
    \let\@linkcolor\revision@base@linkcolor
    \let\@urlcolor\revision@base@urlcolor
  \fi
  \revision@bibitem[#1]{#2}}
\makeatother

\title{Factoring $2048$ bit RSA integers with a half-million-qubit modular atomic processor}
\author{Tian Xue}\email{tianxue2@illinois.edu}
\affiliation{\UIUC}
\author{Jacob P. Covey}\email{jcovey@uchicago.edu}
\affiliation{\UIUC}
\affiliation{\PME}
\affiliation{\UC}
\affiliation{\ANL}

\begin{abstract}
Shor's algorithm is one of the most promising applications of quantum computers. However, since $\sim 10^6$ physical qubits are believed to be required for established approaches, the algorithm will need to be distributed across many modules. In this paper, we provide a distributed compilation of Shor's algorithm \red{(DShor)} on a modular atomic processor. We present an end-to-end compilation and optimization strategy that focuses on the interplay between the inter-module communication and the intra-module clock rate. With a half-million-qubit modular atomic processor with a communication rate of $10^5$ Bell pairs per second and a measurement time of 1 ms in a CPU-inspired architecture, we demonstrate that 2048-bit RSA integers can be factored in \red{approximately $10\%$ more time than a single-module architecture}. Our work presents the first end-to-end analysis and simulation of large-scale integer factorization on modular atomic hardware and it provides a blueprint for the future design of other large-scale modular algorithms.

\end{abstract}
\maketitle

\begingroup
\setlength{\intextsep}{3pt}
\xdef\restoreFirstFigureFloatSpacing{\global\textfloatsep=\the\textfloatsep\relax}
\global\textfloatsep=3pt
\AddToHookNext{shipout/after}{\restoreFirstFigureFloatSpacing}
\begin{figure}[!ht]
    \includegraphics[width=0.72\linewidth]{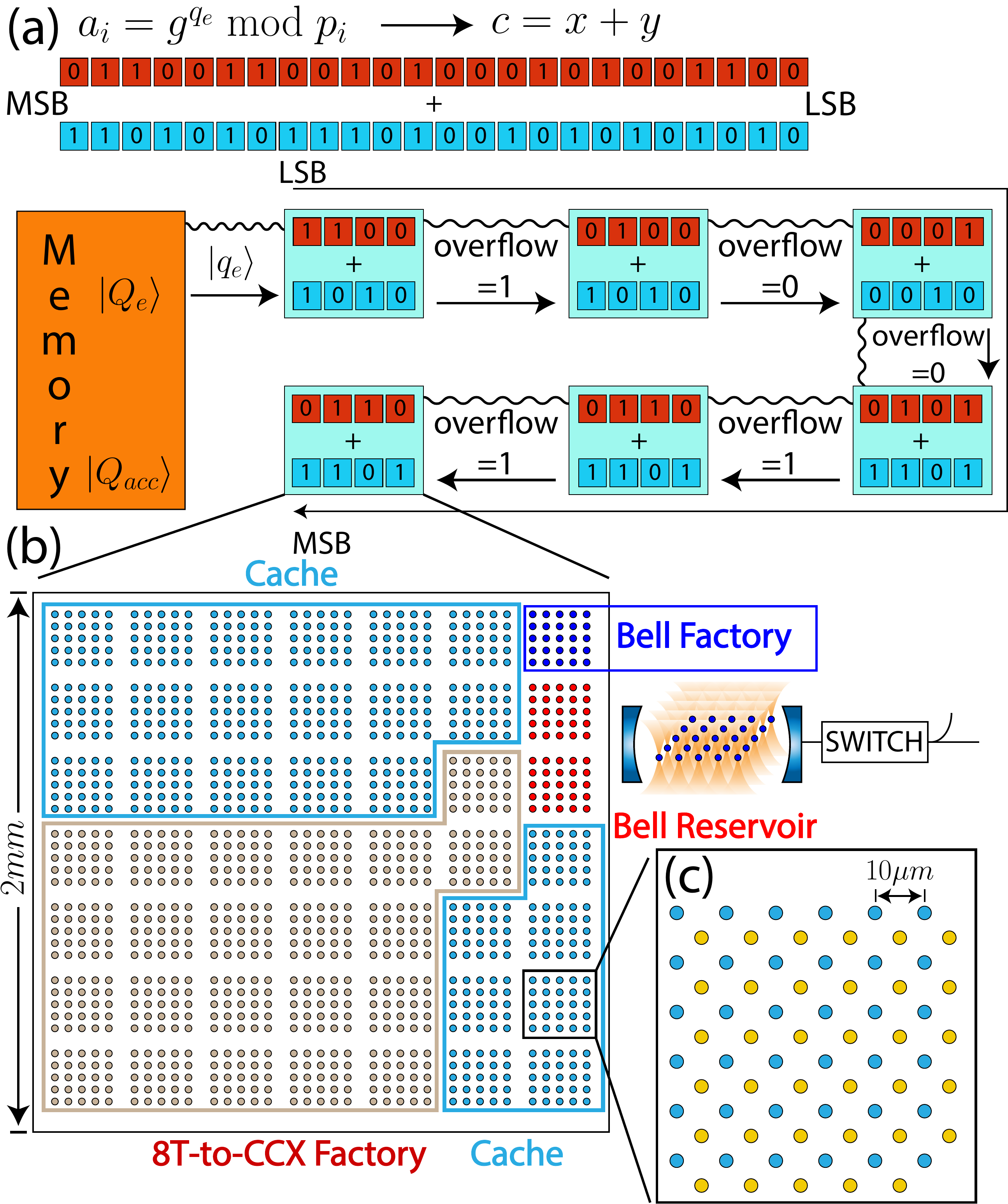}
	\caption{
        \textbf{Architecture of the DShor algorithm}. The architecture of the modular QPU is designed around the modular exponentiations.
        (a) The modular exponentiation can be decomposed to modular additions and distributed to different QPUs. The addition between two 24-bit integers (red squares and blue squares) is distributed to six modules. \red{The most significant bit (MSB) and least significant bit (LSB) are shown on the left and right, respectively.} The additions only need local communications to ripple the overflow information to the neighboring module. (b) Each QPU in the modular processor is divided into caches, the CCX factory, and the communication zone. Temporary results of Shor's algorithm are stored in caches and non-Clifford gates are distilled from the CCX factory. Inter-module communications are created by the communication atoms (blue dots) in the cavity, which connects to the neighboring modules, and are stored in the Bell reservoir (red dots). (c) Logical qubits are stored in the surface code patches where light blue dots represent QEC data qubits and yellow dots represent QEC ancilla qubits. The separation between data qubits is $10$ $\mu$m, and the separation between data and ancilla qubits is $5$ $\mu$m.
    }
    \phantomsection
    \label{fig:DShor Structure}
\end{figure}
\endgroup

\section{Introduction}
\label{Chapter: Introduction}

Shor's algorithm is one of the few quantum algorithms with a promised quantum advantage, where the time complexity of factoring $n$-bit integers is reduced from \red{the heuristic classical scaling $\exp\!\bigl[O(n^{1/3}(\log n)^{2/3})\bigr]$ to $O(poly(n))$~\cite{Babbush2026, Gidney2025, Gidney2021, Ekera2017, Shor1997,Shor1994,Zhou2025,Jones2012,Lanes2025, Huang2025Qadv, Shor2022,Fowler2012,Gorman2017, Gheorghiu2019, Chevignard2024, Buhler1993}}. This potential utility combined with its highly specific \textcolor{revisionred}{goal makes it} a unique benchmark for designing processor architectures and for resource requirement estimation. The estimated physical qubit count required for RSA 2048 has decreased from $\sim10^9$ in 2010 to $\sim10^6$ in 2025~\cite{ Gidney2025, Gidney2021,Zhou2025,Jones2012,Fowler2012,Gorman2017, Gheorghiu2019}. This decrease was largely fueled by advances in magic state cultivation, windowed quantum arithmetic, \red{approximate modular exponentiation with residue arithmetic~\cite{Chevignard2024},} and the replacement
of long modular exponential with short discrete logarithms~\cite{Gidney2025, Gidney2024, Ekera2017}. These rapid developments bring Shor's algorithm to the near future, and in this paper, we will present an end-to-end compilation of Shor's algorithm on the neutral atom hardware platform. 

 In the past several months, further reductions have been proposed, based on quantum error correction codes such as the quantum low density parity check (qLDPC) codes that have a much higher encoding rate than traditional CSS codes such as the surface code or Steane code~\cite{Webster2026, Cain2026}. The high-rate code can further reduce the physical qubits required to factor 2048-bit RSA integers from $\sim10^6$ to $\lesssim10^5$.
 However, no demonstrations of even the most basic qLDPC operations have been made to date, and the non-local stabilizer checks in the QEC cycles impose a greater challenge to the moving of atoms~\cite{Xu2024,Pecorari2025,Poole2025}. Although the logical operations in qLDPC patches can be accelerated via lattice surgery~\cite{Cross2025, Xu2025,Cowtan2025, Webster2025}\red{,} the recent reports of lattice surgery via the surface code on the neutral atom platform \textcolor{revisionred}{also demonstrate that} logical information is lost quickly during each round of the measurement in the lattice surgery cycle~\cite{Bluvstein2025}. 
 The classical decoding time ($\sim 30 - 100$ ms) of qLDPC also significantly slows the circuit time~\cite{Wang2026, Hillmann2025, Xu2023}. Although these challenges are likely to introduce substantial time overheads, these high-rate codes suggest that \textcolor{revisionred}{future quantum computers capable of tackling RSA 2048 could operate} with $\lesssim10^5$ physical qubits, and we anticipate that better decoding schemes and atom control techniques may reduce the time of logical operations in qLDPC codes to the ms level in the future.

Under the current state-of-the-art for hardware development and quantum error correction, we believe that a balance between the time and space overhead tradeoff is the most promising in the near term to execute utility-scale algorithms with meaningful runtimes. Such a `hybrid’ architecture could, for instance, use qLDPC for memory operations that require ultra-sparse logic, and use surface codes with transversal operations combined with magic state cultivation for processing. The fast decoding ($\sim 1$ ms per classical CPU core) and local connectivity of the surface codes allow a much faster QEC cycle\red{~\cite{Higgott2025,Cain2025Fast,Turner2026Transversal,SerraPeralta2026Transversal}}. As such, we believe that it is crucial to architect a balanced quantum processor to factor 2048-bit RSA integers at the requisite scale of several $100,000$ physical qubits.

At this scale of physical resources, we believe that modularity of some kind will be required. We envision four possible flavors for such modularity: 1) photonic interconnects between modules in separate vacuum chambers~\cite{Reiserer2015,Brekenfeld2020,Daiss2021,Covey2023,LiThompson2024,Li2025}, 2) many separate microscope objective-defined modules in the same vacuum chamber that are connected with transportable atom arrays, 3) many layers in a stack of two-dimensional modules that are integrated into a three-dimensional structure~\cite{Barredo2018,Lu2025,Picard2025} that could involve optical lattices~\cite{Nelson2007,Wu2019,Norcia2024,Gyger2024}, and 4) many ``tiled” regions that define sub-modules within a very large field of view of a single microscope objective system~\cite{Manetsch2024}. It is also quite likely that some combination of these four approaches will be used simultaneously, such as, e.g., many objective-defined modules in which each module could operate on two layers and use a tiled layout.

Regardless of the exact physical implementation, the key figures of merit are the number and functionality of the modules and the communication rate between the zones compared to the clock rate of the overall processor. This abstract picture invites a comparison with classical computing architectures. \red{We investigate two architectures by analogy with classical CPUs and GPUs: a QCPU has a small number of processing modules, each capable of multi-bit arithmetic with a local non-Clifford-state factory, whereas a QGPU uses many physically smaller modules to execute independent modular exponentiations in parallel.  In this paper, we use QPU (quantum processing unit) to represent one computation unit in the QCPU/QGPU architecture.} Naturally, the relative cost of intra- versus inter-module operations is a key parameter in these architectures. \red{We therefore focus on the QCPU architecture as a near-term design target. For photonic interconnects, we use a physical communication rate of $10^5$ physical Bell pairs per second, motivated by cavity-based proposals~\cite{LiThompson2024, Sinclair2025, Ramette2023}. Reducing the code distance to $d = 9$ could enable more processing modules within a fixed physical-qubit budget among modules, making the QGPU architecture attractive if communication and memory access can support the additional parallelism.} We also consider this architecture for $d=9$ in anticipation of a future with physical error rates well below $\approx10^{-4}$ and/or with advanced, co-designed decoders.


Here, we demonstrate an architecture for modular neutral atom hardware to perform factoring of 2048 bit RSA integers. We present an end-to-end compilation \blue{of distributed Shor's algorithm (DShor)} and optimization strategy \red{for the full circuit of Shor's algorithm with Chinese remainder theorem (CRT) arithmetic} that focuses on the interplay between the inter-module communication and the intra-module clock rate. \red{Unlike schemes that retain the full arithmetic workspace within each active processor~\cite{Xiao2022,Xiao2023,Schmidt2025}, we distribute that workspace across QPUs with comparable capacities.} Specifically, we decompose Shor's algorithm to fundamental 
operations and we distribute all fundamental operations across the modules with mostly nearest-neighbor interactions. For instance, 
the addition between two integers is distributed across many modules and only needs local communications to ripple the overflow information to the neighboring module [see Fig.~\ref{fig:DShor Structure}(a)]\red{~\cite{Yimsiriwattana2004,VanMeter2006,VanMeter2010,Pattison2024}}. Additionally, we include a dynamic qubit reallocation protocol via atom movement, which helps balance dynamic resource loads associated with communication buffering and code cultivation (see Appendix\red{.}~\ref{Appendix: Reallocation}).
Our architecture is based on one memory module with qLDPC-based logical encoding, and $N_{\text{QPU}}$ (QPU) modules with surface code-based logical encoding. We consider several different physical error rates and concomitant code distances for logical encoding, and we focus on module counts ranging between 1+2 and 1+12 with emphasis on the 1+6 case. Although our modular vision is largely agnostic to the exact physical implementation, for concreteness we focus on 1) photonic interconnects and 2) transportable super-arrays. In general, we find that realistic estimates for communication rates in both approaches are sufficient to cause only minimal increases in runtime compared to a monolithic architecture. We believe the end-to-end compilation of factoring algorithm on the neutral atom platform, particularly \textcolor{revisionred}{focusing} on the modular atomic hardware, provides a blueprint for the future design of large-scale algorithms.

\section{Hardware Assumptions and Architectures}

\label{Chapter: Hardware and Architecture}
The QPU architecture and the compilation of the distributed Shor's algorithm are designed closely around the hardware of the neutral atom array platform. Neutral atom arrays are a leading candidate for DShor due to their relative scalability to 6,000 qubits and counting~\cite{Manetsch2024,Chiu2025}, their access to efficient optical interconnects via optical cavities~\cite{Reiserer2015,Brekenfeld2020,Daiss2021,Covey2023,LiThompson2024,Shaw2025}, their versatile geometry~\cite{Barredo2018,Kaufman2021,Ebadi2021}, and their robust transportability for intra- and inter-module connectivity~\cite{Beugnon2007,Dordevic2021,Bluvstein2022}. In this section, we will introduce the hardware assumptions and the corresponding QPU architecture used in DShor.

DShor is designed based on the ability to perform disparate tasks simultaneously. Specifically, dissipative operations such as readout and network attempting must be performed in parallel with coherent operations. For simplicity, we assume a ``dual species" atom array system as illustrated in Fig.~\ref{fig:DShor Structure}(c), where the atoms of the two species separately serve as data (light blue dots) and ancilla (yellow dots) qubits. Here, ``dual species" is broadly defined, and could be two actual atomic species~\cite{Singh2021,Singh2022}, the `omg' architecture with a single alkaline earth(-like) species~\cite{Chen2022,Lis2023,Ma2023,Li2025,Senoo2025}, or a zoned single-species system~\cite{Bluvstein2022,Deist2022b,Bluvstein2023} that utilizes `shielding' via local light shifts~\cite{Norcia2023,Hu2024,Bluvstein2025}. The separation between atoms of the same species is assumed to be 10 $\mu$m, and the separation between atoms of different species is 5 $\mu$m. Therefore, one logical patch with the code distance $d$ has the dimension $10d\times 10d$ $\mu$m$^2$. We assume atoms can move at the speed of 1 m/s, and atoms are entangled by Rydberg gates. For simplicity, we assume that Rydberg gates can be applied anywhere in the workspace. We note that it remains an open question how to optimally perform stabilizer measurements in zoned architectures.
Logical CNOTs are applied by moving the control logical patch close to the target logical patch followed by the transversal Rydberg gates~\cite{Bluvstein2023,Bluvstein2025}. Since Rydberg gates are fast ($\approx300$ ns)~\cite{Evered2023,Ma2023}, the gate time of the logical CNOTs is dominated by the moving time of atoms, and we approximate the gate time of logical CNOTs as \red{$10d$} $\mu$s \red{regardless of the range of CNOTs}. This gate time is defined as the ``clock time" in DShor and used to calibrate the gate time of all Clifford gates. {\color{revisionred}The additional timing overheads from long-range CNOTs and atom refilling are discussed in Appendix.~\ref{Appendix: Machine code}. We estimate an approximately $50\%$ increase in the logical clock period when these effects are included.} All single-qubit Pauli gates are ignored as they can be applied without any cost by changing the Pauli frame~\cite{Chamberland2018Pauli, Riesebos2017}, \red{and the $\mathrm{H}$ and $\mathrm{S}$ gates are implemented by the fold-transversal gates~\cite{Zhou2025Transversal}}. We assume a QEC cycle is applied \textcolor{revisionred}{after each logical gate} although QEC cycles can be skipped if the logical error per round is lower than the target logical error~\cite{Sunami2025}. 

As the QEC cycle of the surface code only requires local interactions, the cost of one QEC cycle is dominated by the measurement time~\cite{Kovalev2013, Google2023, Geher2024}. Measurement remains a major bottleneck in the neutral atom community, and we thus consider two different measurement times for full generality. The bottleneck stems from the slow laser cooling rate for tweezer-confined atoms, owing in part to the low trapping frequency ($\approx10-100$ kHz). Measurement that retains the atom with high fidelity and keeps the atom cold enough to immediately be reused takes $\approx1-10$ ms~\cite{Lis2023,Norcia2023}. Measurement that either does not retain the atom or does not keep the atom cold can be performed in $\approx20$ $\mu$s~\cite{Ma2023,Scholl2023,Senoo2025,Falconi2025}, but then recooling or obtaining a new atom from a reservoir array is required, which may take $\approx1$ ms. \textcolor{revisionred}{Here, we consider the measurement times of both options: $t_{\mathrm{mea}}\approx250$ $\mu$s and $t_{\mathrm{mea}} = 1~\mathrm{ms}$.} \red{Please note that the timing parameter $t_{\text{mea}}$ also includes latency from classical experimental control.} Although the latter is realistic now, we also consider the former because improvements in qubit readout are expected, based on innovations such as optical cavity arrays~\cite{Menon2024,Shaw2025,Ding2026} or confocal cavity QED microscopes~\cite{Kroeze2023,Marsh2025,Kroeze2025}.

\red{We emphasize that this study is not meant to accurately predict the required spacetime volume to run RSA 2048. There are simply too many choices for assumptions at a very dynamic time when both hardware capabilities and architectures as well as error correction strategies are improving on a monthly basis. Nevertheless, we are careful to consider nearly all space-like and time-like overheads given our choice of physical error rate, QEC cycle time, and algorithmic architecture. \red{In this paper, we assume a common physical error probability $p_p$ for single-qubit depolarization, two-qubit depolarization, \textcolor{revisionred}{measurement-outcome classification errors}, and reset errors, with $p_p=10^{-3}$ as the baseline and $p_p=10^{-4}$ as the higher-fidelity comparison with two QEC cycle times $\sim0.25$ and $\sim1$ ms at each $p_p$. However, we do not rigorously account for decoherence during atom transport or idle periods.}
 
The main space-like overhead that we do not rigorously consider is reservoir arrays to address atom loss. We estimate the overhead due to this in the outlook. In terms of time-like overhead, we believe that the thoroughness of our estimates, particularly in the exact circuit of \red{$8\mathrm{T}$-to-CCX factory} as well as additions and loadings, presents a more accurate picture than several past works and naturally leads to an inflated spacetime volume~\cite{Jones2013, Cain2026, Webster2025}\textcolor{revisionred}{.} \textcolor{revisionred}{\hypersetup{allcolors=revisionred}In the normalized circuit-volume comparison with Ref.~\cite{Gidney2025}, sequential CNOT fanout accounts for a factor of $2.81$. The remaining factor of $1.23$ may come from the combined effect of the longer assumed $\mathrm{T}$-state cultivation time and more detailed gate accounting. Please see Appendix.~\ref{Appendix: Comparison with other works} for more details. In the result of the main text, we also ignore the additional time overhead from the long-range atom transport and the atom refilling to compensate the atom loss. In Appendix.~\ref{Appendix: Machine code}, we estimate the additional time overhead from the long-range atom transport based on the exact atom schedule of the subroutine in the DShor. We also estimate the time overhead based on the $\textcolor{revisionred}{10^{-3}}$ atom loss rate and $10^6 \mathrm{atoms/s}$ refilling rate using the error-detecting QEC circuit in Ref.~\cite{Baranes2026Loss}}.}


The layout of the processor in the QCPU architecture is shown in Fig.~\ref{fig:DShor Structure}(b). Each QPU can execute complicated tasks such as multi-bit additions/loadings, so $8$ ($+3$ inputs) logical qubits are assigned to the \red{$8\mathrm{T}$-to-CCX factory} to distill non-Clifford gates. As mentioned above, we consider two different inter-module connections: photonic connections and connections via transportable arrays. 
With photonic connections, one logical qubit patch is used to generate inter-module Bell pairs, which are saved in the Bell reservoirs with two logical qubits to buffer inter-module communication delay. The rest of the qubits are used as caches to store all intermediate results. To reduce the space used by DShor and increase the parallelism within modules, we allow qubits to be dynamically allocated between the \red{$8\mathrm{T}$-to-CCX factory} and the cache. Qubits are moved from the factory to the cache at intensive cache use to reduce the space usage, and qubits are moved from cache to factory on light cache use to boost non-Clifford gate production. More details of re-allocation of qubits are shown in Appendix\red{.}~\ref{Appendix: Reallocation}.

With the hardware assumptions based on the neutral atom platform and the QCPU modular architecture, DShor can factor $2048$-bit RSA integers with a half-million modular processor with $6$ QPUs and $1$ quantum memory. As computations only happen to a limited number of qubits at a given time (hot qubits), all idle qubits are stored in qLDPC in the memory, and  hot qubits are stored in the surface code, for which we consider a few values of $d$, to reach the target logical error.

\begin{table*}[t]
\centering
\begin{tabular}{c|c|c|c|c|c}
\hline
Qubit function&Qubit name & $N_{L}$& $N_L$(RSA 2048)&$N_p$ (RSA 2048) & Distribution in DShor \\
\hline
Superposition of all exponents &$Q_{exponent}$ & $n/s+n/2$&$1280$ &\red{$81,600$} & in memory \\
Exponent in short logarithm&$q_e$ & $|p_i|$&$24$ &$30,000$ & $|p_i|/N_{\text{QPU}}$ \\
Result of approximate exponentiation&$Q_{acc}$ & $f$&$36$&\red{$3,060$} & in memory\\
Result of exponentiation of single $p_i$&$q_r$ & $|p_i|+1$&25&$31,250$&$(|p_i|+1)/N_{\text{QPU}}$\\
Intermediate result of multiplication & $q_h$ & $|p_i|+1$&25&$31,250$&$(|p_i|+1)/N_{\text{QPU}}$\\
Temporary memory of table values& $q_l$ & $|p_i|+1$& $25$&$31,250$&$(|p_i|+1)/N_{\text{QPU}}$\\
\end{tabular}
\caption{\textbf{Qubit allocation in DShor}. $Q_{exponent}$ and $Q_{acc}$ are stored in the memory, and \textcolor{revisionred}{all intermediate results used by DShor are distributed} evenly to all QPUs. $N_L$ is the number of logical qubits to store the result, and $N_p$ is the number of physical qubits required at $d=25$. \red{The exponent ($\ket{Q_{exponent}}$) and accumulator ($Q_{acc}$) use $80$ and $3$ GB blocks, respectively, with $1,020$ physical qubits per block. $12$ surface-code patches are used to access the information from the GB code via lattice surgery, which adds $15,000$ physical qubits to the memory, giving total $99,660$ physical qubits in the memory. We round it up to $110,000$ physical qubits to the memory.}}
\label{Table: Qubit usage in Shor}
\end{table*}

To reduce and balance the communications used by DShor, all intermediate results are distributed evenly to the caches of QPUs. Shor's algorithm can be decomposed to short modular exponentiations and then decomposed further to modular additions. Since the $k$-th bit in the result of additions $r = x + y$ depends only on the current adding bits and the overflow bit: $r_k = x_k \oplus y_k \oplus c_{k-1}$, $c_k = \text{MAJ}(x_k,y_k,c_{k-1})$, where $\text{MAJ}(x_k,y_k,c_{k-1})$ is the majority gate, which returns the majority bit among $(x_k,y_k,c_{k-1})$, additions only require local (nearest-neighbor) communications to propagate the overflow information $c_{k-1}$~\cite{Gidney2018}. In Fig.~\ref{fig:DShor Structure}(a), the exponent $\ket{q_e}$ of the modular exponentiation $\ket{g^{q_e}}$ is read from the quantum memory, and the modular exponentiation is decomposed to additions. All bits involved in the additions are evenly distributed to QPUs, and the overflow information is rippled through all QPUs by a cascade of local communications. More discussion of the distributed operations in DShor can be found in \ref{Chapter: Distributed fundamental operations}.

\section{Theory and Methods}
\subsection{Shor's algorithm with residue arithmetic}
\label{Chapter: residue arithmetic}
In this study, we focus on the Shor's algorithm using the residue arithmetic to reduce the number of qubits used~\cite{Gidney2025}. Shor's algorithm roots at \red{modular exponentiations}: 
\begin{align}
    \sum_{e}\ket{e} \rightarrow \sum_{e}\ket{e} \ket{g^e\bmod N}
\end{align}
    Therefore, Shor's algorithm needs to store all exponents $\ket{e}$ \red{($e\leq 2^{{2n}}$)~\cite{Shor1994,Shor1997} and all results $\ket{g^e \bmod N}$. The recent development of \textcolor{revisionred}{Eker{\aa}--H{\aa}stad}'s factoring algorithm reduces the size of exponents from $2n$ to $m=n/2 + n/s$ with $s$ classical iterations of factoring algorithm \textcolor{revisionred}{using the short discrete logarithm}. For simplicity, we use Shor's factoring scheme to provide the necessary background for the compilation of modular exponentiations using residue arithmetic. The compilation from the Shor's factoring algorithm can be naively transferred to the \textcolor{revisionred}{Eker{\aa}--H{\aa}stad}'s factoring algorithm. Please check Appendix.~\ref{Appendix: Ekera factoring} and Refs.~\cite{Ekera2016,Ekera2017,Ekera2020} for more details of the compilation of \textcolor{revisionred}{Eker{\aa}--H{\aa}stad}'s factoring algorithm}.

As most computations in Shor's algorithm are modular exponentiations, we focus on the simplification and distribution of modular exponentiations in this paper. The modular exponentiation of $g^e\bmod N$ can be decomposed to multiple short modular exponentiations from a prime system \red{$\{p_i |gcd(p_0...p_{K-1}) = 1\}$ using the residue arithmetic \cite{Chevignard2024}.
Write $e=\sum_{j=0}^{2n-1}2^j e_j$, where $e_j\in\{0,1\}$, and define $M_j=g^{2^j}\bmod N$~\cite{Gidney2025}. Then}

\red{\begin{equation}
{\begin{aligned}[b]
    g^e\bmod N
    &=\left[\prod_{j=0}^{2n-1}(g^{2^j}\bmod N)^{e_j}\right]\bmod N\\
    &=\left[\prod_{j=0}^{2n-1}M_j^{e_j}\right]\bmod L\bmod N\\
    &=\left[\sum_i a_iP_ir_i\right]\bmod L\bmod N.
\end{aligned}
\label{eq:main_crt_product}}
\end{equation}}
where \red{$a_i=\bigl[\prod_{j=0}^{2n-1}M_j^{e_j}\bigr]\bmod p_i$, $P_i=L/p_i$, $r_i=P_i^{-1}\bmod p_i$, and $L=\prod_i p_i>N^{2n}$. The product contains at most $2n$ factors smaller than $N$, so $g^e \bmod N = \left[\sum_i a_iP_ir_i \right]\bmod L\bmod N $ when $L\geq N^{2n}$ (the magnitude of L can be smaller after the windowed arithmetic is implemented~\cite{Gidney2025})}.  Both $P_i$ and $r_i$ can be found classically, and only $\{a_i\}$ are calculated on the quantum computers. \red{Although the intermediate results $a_iP_ir_i$ have $\approx nm$ bits, they are compressed to $f = O(\log_2{n})$ bits using the approximate additions~\cite{Gidney2025} (see Appendix.~\ref{Appendix: CRT truncation} for more details). Therefore, the size of the intermediate quantum} results is $|p_i| = \log{n}$, which significantly reduces the space required by the Shor's algorithm. More details and further simplifications of Shor's algorithm using the residue arithmetic can be found in Appendix\red{.~\ref{Appendix: CRT}}. 

\begin{figure*}[t]
    \includegraphics[width=\linewidth]{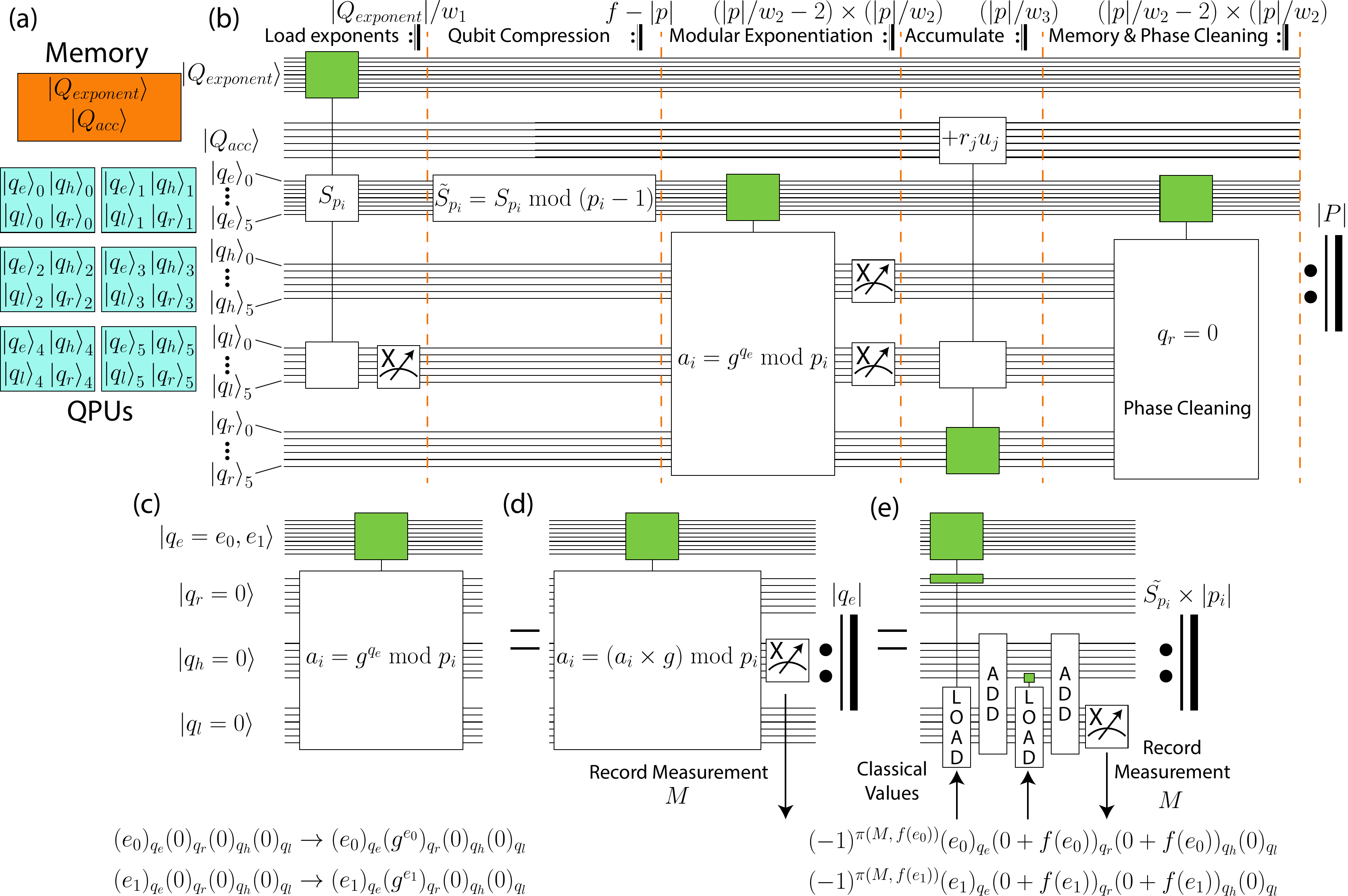}
	\caption{
        \textbf{Algorithmic structure of DShor}. (a) Results of Shor's algorithm are stored in the memory (the orange square), and all intermediate results are distributed evenly to all QPUs (blue squares). (b) Shor's algorithm is decomposed to five steps: 1. loading short logarithms from the memory to the QPUs, 2. compressing the short logarithms, 3. modular exponentiations, 4. accumulating results to the memory, and 5. cleaning all intermediate results and phases. The whole procedure is iterated over all primes in the prime system. (c) The circuit of DShor is simulated and verified by the Shor's simulator. When the modular exponentiation $g^{q_e}$ is implemented, it will be decomposed to modular additions by the compiler. The simulator initializes the limited number of tracked Hilbert trajectories (such as the trajectory with $\ket{q_e = e_0}$). The modular exponentiation is decomposed to (d) $|q_e|$ modular multiplications and then to (e) $|q_e|\times |p_i|$ modular additions. Modular additions are further decomposed to fundamental operations (loading+adding+cleaning). \red{The phase cleaning is skipped for simpler illustration. Please see Appendix.~\ref{Appendix: Distributed addition} and Fig.~\ref{fig:compilation of arithmetic}(e) for the exact circuit}. Each time a fundamental operation is implemented, the front-end circuit simulator will evolve the circuit, and the back-end simulator will update the Hilbert trajectories. At the end of the modular exponentiation, the simulator will check if all trajectories are $\ket{q_e}\otimes\ket{g^{q_e}}$. The final circuit after the verification is used for the resource estimation. 
    }
    \phantomsection
    \label{fig:Shor's algorithm}
\end{figure*}

In DShor, qubits storing exponents $\ket{Q_{exponent}}$ and the \red{compressed result $\ket{Q_{acc}} \approx \ket{\sum a_iP_ir_i} \approx \ket{g^e\bmod N}$} are stored in the quantum memory, and all qubits required for the modular exponentiations \red{$\ket{a_i}$} are evenly distributed to all QPUs as shown in Fig.~\ref{fig:Shor's algorithm}(a). \red{Qubits in the quantum memory are stored in $[[510, 16, 24]]$ generalized bicycle (GB) code with logical error $\simeq 4\times 10^{-16}$ at $p_p = \textcolor{revisionred}{10^{-3}}$~\cite{Webster2026}. \textcolor{revisionred}{Two additional banks} of $w_1$ surface-code access patches \textcolor{revisionred}{are also included in the memory} for the general lattice surgery~\cite{Lin2025Decoding,Kovalev2013,Webster2025, Webster2026}.} In DShor, we assume all qubits are stored in a single-module memory, but the memory can be split into multiple modules as needed based on the hardware implementation. 
Splitting the memory into more modules has a negligible effect on the DShor due to the infrequent usage of qubits in the memory. \red{To factor $2048$ bit RSA integers, $83$ GB code patches and $12$ surface-code ancilla patches are required, \textcolor{revisionred}{which require $99,660$ physical qubits}. We allocate $110,000$ physical qubits to the memory to allow additional communication and surgery workspace.} The exact space usage of DShor to factor general RSA integers and 2048 bit RSA integers \textcolor{revisionred}{in the $d=25$ surface code patches} is shown in Table.~\ref{Table: Qubit usage in Shor}\textcolor{revisionred}{.}

The Shor's algorithm with the residue arithmetic can be constructed with the following five subroutines: loading exponents, qubit compression, modular exponentiation, accumulating results to the memory, and memory $\&$ phase cleaning as shown in Fig.~\ref{fig:Shor's algorithm}(b)~\cite{Gidney2025}. In DShor, we further decompose the five subroutines to three fundamental operations: loading, adding, and cleaning. Loading transfers classical information $f(e)$ to quantum registers depending on the quantum address $\ket{e}$:
\begin{align}
    \ket{e}\ket{0} \rightarrow \ket{e}\ket{f(e)}
\end{align}
Adding applies a direct addition on two quantum registers:
\begin{align}
    \ket{a}\ket{b} \rightarrow \ket{a+b}\ket{b}
\end{align}
And cleaning resets all temporary quantum memories for the future reuse:
\begin{align}
    \ket{a+b}\ket{b} \rightarrow \ket{a+b}\ket{0}
\end{align}
A naive decomposition of modular exponentiation to fundamental operations is in Fig.~\ref{fig:Shor's algorithm}(c,d,e), where one modular exponentiation $g^{q_e}$ is decomposed to $|q_e|$ modular multiplications and further to $|q_e|\times |p_i|$ modular additions with two additions and two loadings. The detailed decomposition of Shor's algorithm to fundamental operations can be found in Appendix\red{.}~\ref{Appendix: Windowed Arithmetic}, \ref{Appendix B: Compilation of arithmetic}, \ref{Appendix: Modular addition}, and \ref{Appendix: Verification of DShor}. 

\subsection{Distributed fundamental operations}
\label{Chapter: Distributed fundamental operations}
As noted in the introduction, Shor's algorithm needs to be distributed over modules at the current stage of quantum computers. Shor's algorithm can be decomposed to modular additions and further to fundamental operations as shown in~\ref{Chapter: residue arithmetic}, so the distribution of Shor's algorithm can be simplified to the distribution of fundamental operations.

\begin{figure}[t]
    \includegraphics[width=\linewidth]{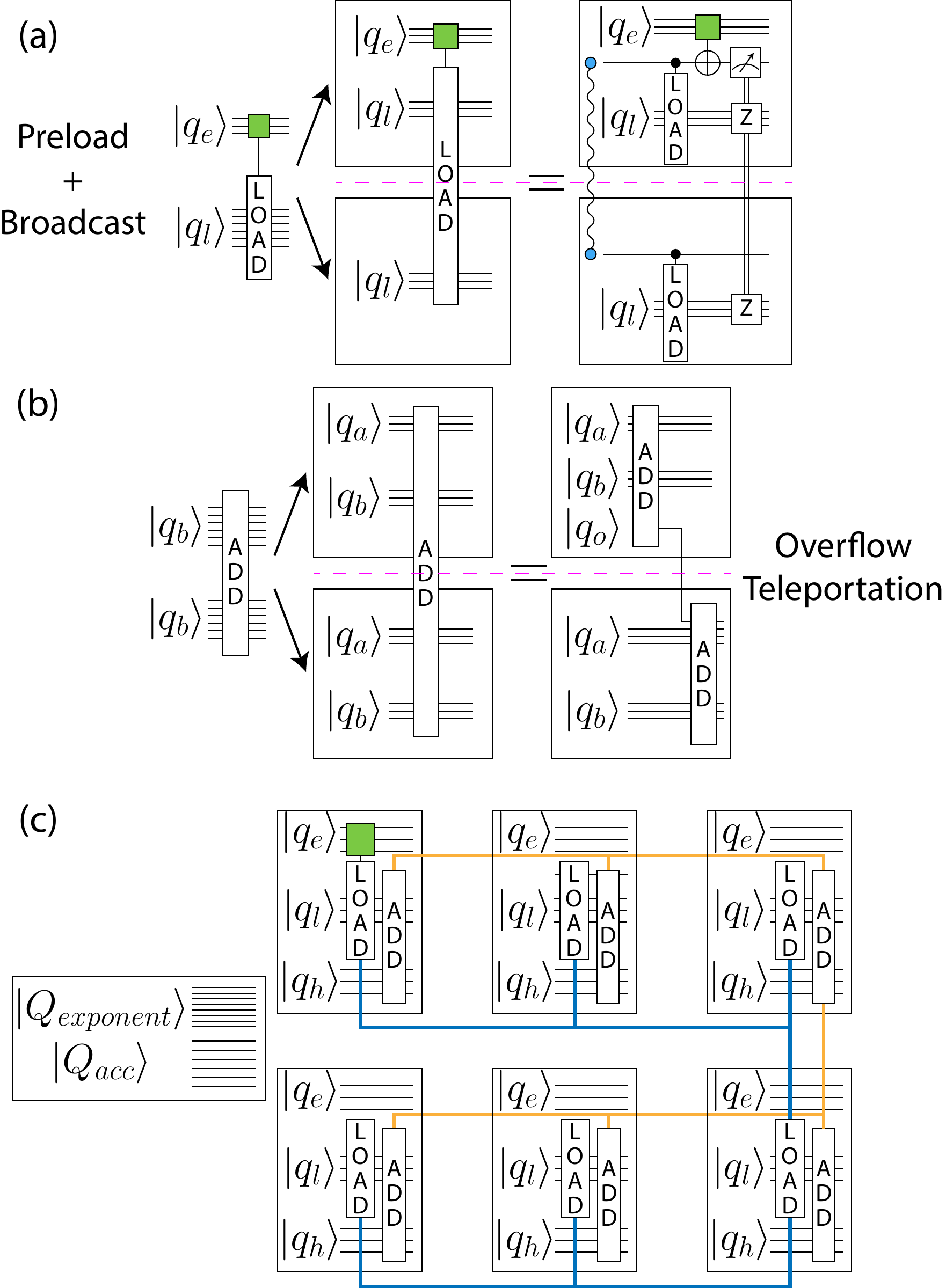}
	\caption{
        \textbf{Distributed fundamental operations of DShor}. (a) Loadings transfer classical information to the quantum registers depending on the quantum address (green square) $\ket{q_e}$. Loading is distributed by $N_{\text{QPU}}$ intra-module pre-loadings, and the quantum address is broadcast to all modules through one inter-module GHZ state. (b) Adding sums two quantum registers, and it is distributed by teleporting the overflow information to the neighbor module. Un-Adding is skipped in (b) for the cleaner illustration.
        (c) \textcolor{revisionred}{A modular addition can be decomposed into two loadings and two additions}, which can be distributed with only local (nearest-neighbor) communications. Each loading requires one GHZ state across modules as labeled by the blue lines, and the addition requires the ripple teleportation of the overflow information, which is labeled as yellow lines.
    }
    \phantomsection
    \label{fig:modular fundamental operations}
\end{figure}

All fundamental operations in DShor can be efficiently distributed using qubit and gate teleportation, which are transversal operations at the \red{physical} level and thus each consume $d^2$ physical Bell pairs. In Fig.~\ref{fig:modular fundamental operations}, we demonstrate the distribution of loading and adding, and the distribution of cleaning can be found in Appendix\red{.}~\ref{Appendix: Modular addition}. In Fig.~\ref{fig:modular fundamental operations}(a), a loading gate is distributed over 2 QPUs, and the naive compilation of loading requires $|q_l|/2$ communications. We decompose one loading gate (one cascade of CNOTs) with two pre-loading gates on two qubits of a Greenberger-Horne-Zeilinger (GHZ) state. \red{The address information is encoded in one qubit of a shared GHZ state and broadcast coherently to all QPUs via gate teleportation~\cite{Zhou2025}.} The distributed loading reduces the communication from $|q_l|$ non-local communications to $1$ GHZ state, or \textcolor{revisionred}{$1$ local communication} between the neighboring QPUs.

Addition ladders have exhibited good distributivity as mentioned in Section\red{.}~\ref{Chapter: Hardware and Architecture}, where $3$ local communications are required for both the addition and the un-addition (cleaning of all temporary memories)~\cite{Gidney2018}. We replace $3$ local communications with \textcolor{revisionred}{$1$ local communication} by the qubit teleportation as shown in Fig.~\ref{fig:modular fundamental operations}(b). The exact circuit of distributed loading and addition can be found in Appendix\red{.}~\ref{Appendix: Modular addition}.

\begin{figure}[t!]
    \includegraphics[width=\linewidth]{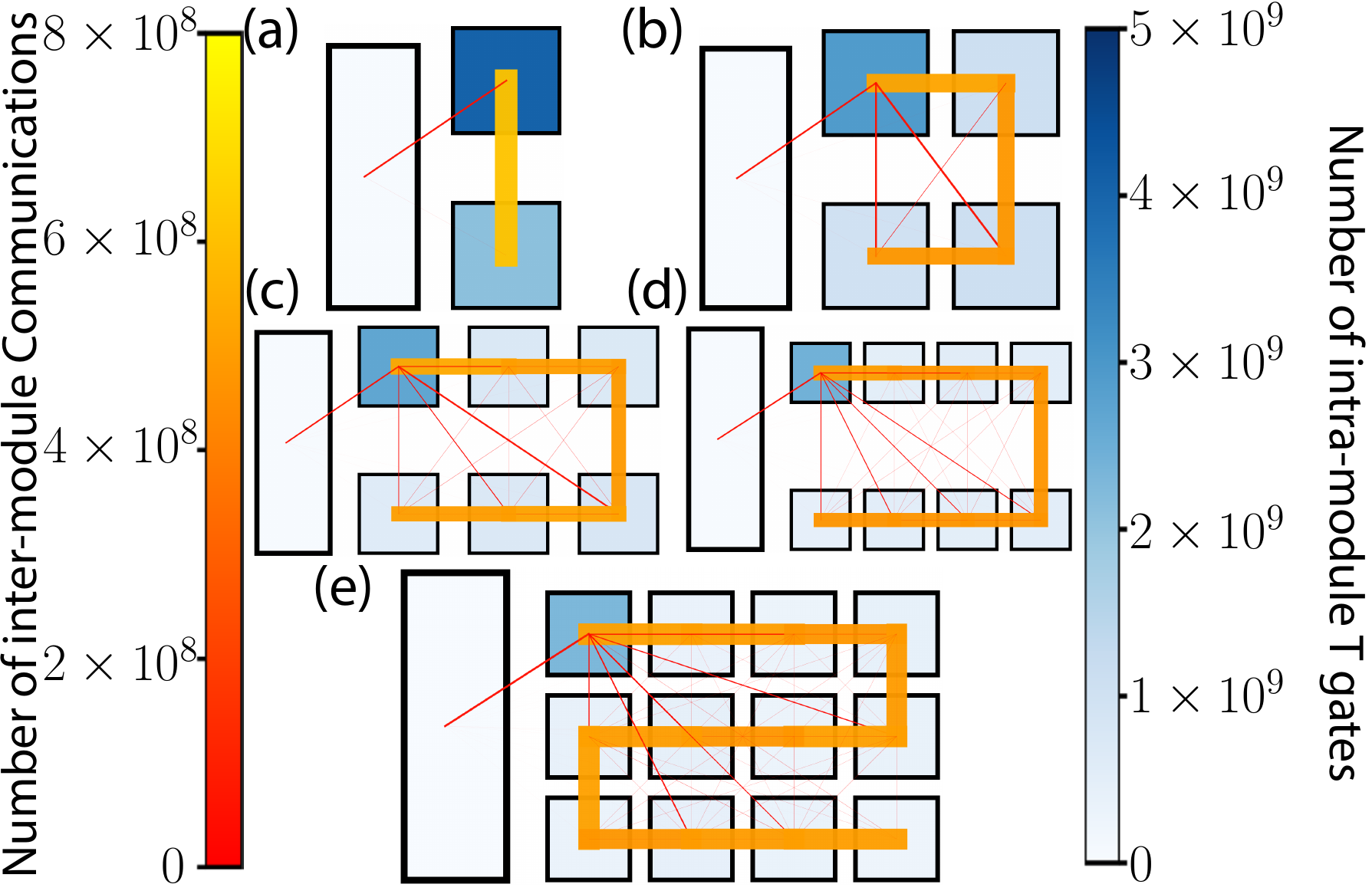}
	\caption{
        \red{\textbf{Inter-module communications and the $\mathrm{T}$ gate count in DShor}}. We simulate the DShor to factor 2048-bit RSA integers with $2,4,6,8,12$ QPUs in (a)-(e), respectively. The rectangle on the left is the quantum memory, and the squares on the right are QPUs. The white-to-blue color scale (right) of the rectangles and squares represents the total number of $\mathrm{T}$ gates used in each Shor's shot, and the red-to-yellow color scale (left) and \textcolor{revisionred}{thickness of lines represent} the communications between modules. Except the 2-QPU DShor, the communications between modules are roughly unchanged with the number of QPUs in DShor, but the number of intra-module $\mathrm{T}$ gates per module decreases as DShor is distributed to more QPUs. The QPU with significantly more $\mathrm{T}$ gates is the router QPU, which reads and loads the quantum registers from the memory. Most connections in DShor are local communications as the fundamental operations only require local communications. 
    }
    \phantomsection
    \label{fig:QPU_connections}
\end{figure}

As the fundamental operations can be distributed with local communications, modular additions can also be distributed with only local communications as shown in Fig.~\ref{fig:modular fundamental operations}(c), where the blue lines label $1$ GHZ state consumed by the loading and the orange lines label the trajectory of qubit teleportations of one addition. In Fig.~\ref{fig:QPU_connections}, we show that the vast majority of communications needed for DShor are local (nearest-neighbor). The widths and colors of lines connecting all QPUs indicate the number of communications between QPUs. If address qubits of loadings are not in the neighbor modules, a non-local qubit teleportation is used to teleport all address qubits into the same module, which leads to few non-local communications in DShor. Here, we assume all-to-all connections among QPUs. However, to factor RSA $2048$ integers with $6$ QPUs, only \red{$6.10\%$ of \textcolor{revisionred}{communications are non-local}}, which has a negligible effect on DShor if all non-local connections are distilled by GHZ states created by local connections~\cite{de_Bone2020}. Most of the non-local connections are from the address teleportations during the modular multiplications. 
We will elaborate on the amount of inter-module communications and intra-module $\mathrm{T}$ gates in Section\red{.}~\ref{Chapter: Result}.

\subsection{Simulation method}
To verify our decomposition of Shor's algorithm to fundamental operations, we invent a Shor's simulator that can simultaneously evolve the front-end circuit and track the back-end Hilbert trajectories. To simulate Shor's algorithm in large systems, we only track a limited number of trajectories in the Hilbert space \textcolor{revisionred}{similar to Ref.}~\cite{Gidney2025}. In the naive decomposition of modular exponentiation shown in Fig.~\ref{fig:Shor's algorithm}(c,d,e), two Hilbert trajectories with different exponents $\ket{q_e}$ are tracked, and the modular exponentiation \red{$\ket{g^{q_e}\bmod p_i}$} is calculated on $\ket{q_r}$. All classical information is loaded on $\ket{q_l}$ and intermediate results are saved in $\ket{q_h}$. All registers except those storing $\ket{q_e}$ are initialized to $\ket{0}$. More details of the resource estimation methods used in the Shor's simulator are in Appendix.~\ref{Appendix: Estimation method}\textcolor{revisionred}{.} 

 In our resource estimation of Shor's algorithm of RSA $2048$, $32$ random Hilbert trajectories are tracked. Modular exponentiations in the DShor are decomposed to fundamental operations, which evolve the circuit and the backend trajectories. For example, the modular additions will be decomposed to two additions, two loadings, and one cleaning (Fig.~\ref{fig:Shor's algorithm}(e)). Additions and loadings will evolve the Hilbert trajectory to $\ket{e}_{q_e}\ket{0+f(e)}_{q_r}\ket{0+f(e)}_{q_h}\ket{0+f(e)}_{q_l}$. The classical values of $\ket{q_l}$ are removed by transversal X measurements. Each trajectory will get a phase error depending on the measurement result and the loaded classical values, which will be cleaned later \red{(see Appendix.~\ref{Appendix B: Compilation of arithmetic} and Fig.~\ref{fig:compilation of arithmetic} for the exact circuit)}. Therefore, after one modular addition, the Hilbert trajectories will evolve to \red{$(-1)^{\pi(f(e),M)}\ket{e}_{q_e}\ket{0+f(e)}_{q_r}\ket{0+f(e)}_{q_h}\ket{0}_{q_l}$}, and the circuit of the modular addition is also recorded. \red{\textcolor{revisionred}{Here,} $M$ is the measurement result and $\pi(x,y)\equiv\left[\sum_b x_b y_b\right]\bmod 2$.} Accordingly, the circuit of modular exponentiations can be verified by checking if all evolved trajectories reach $\ket{e}\ket{g^e\bmod p_i}$ without phase errors. A detailed implementation of Shor's simulator on the modular exponentiation is shown in Appendix.~\ref{Appendix: Verification of DShor}\textcolor{revisionred}{.}

\section{DShor on the arrays with photonic interconnects}
\label{Chapter: Result}
\subsubsection{Choice of parameters}
We simulate DShor on the canonical RSA $2048$ challenge and specially design the parameters suitable to DShor as shown in Table~\ref{Table: Qubit usage in Shor}. In DShor, each QPU requires at least $4|p_i|/N_{\text{QPU}}$ qubits to store all intermediate results. As the optimal size of primes is $O(\log{(n)})$, we choose $|p_i| = 24$ for the QCPU modular architecture. \red{We \textcolor{revisionred}{use} the loading window $w_1 = 6$ and multiplication windows $w_e = w_m = 3$ for the benchmark configurations.} More data of DShor at different window sizes can be found in Appendix\red{.}~\ref{Appendix: More parameters}.

\subsubsection{QPU connectivity, strong scaling analysis}
\begin{figure*}[t!]
    \includegraphics[width=0.8\linewidth]{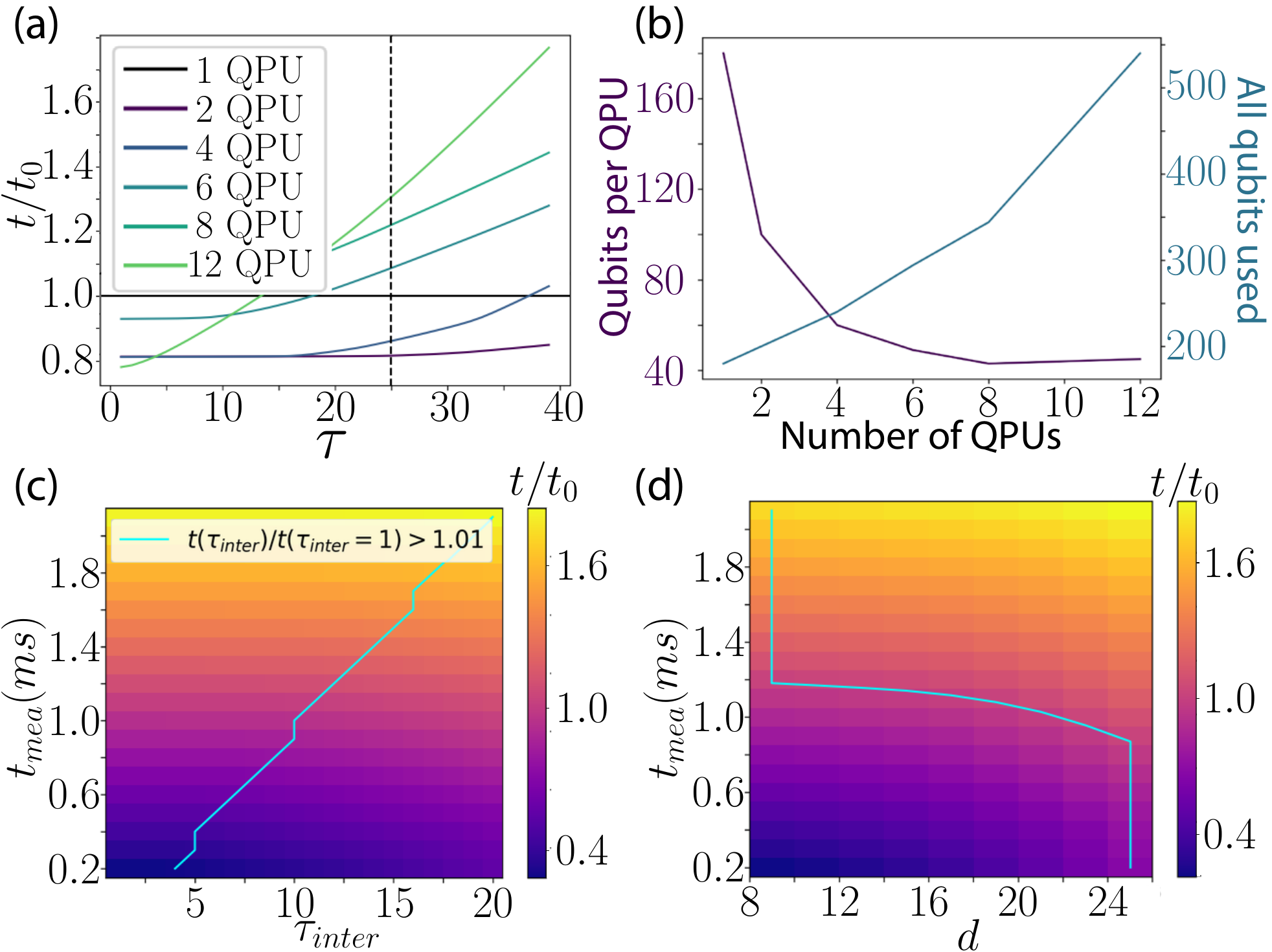}
	\caption{
        \textbf{Results of DShor}. We estimate the space and time usage of the DShor with $2,4,6,8,12$ QPUs to factor RSA 2048-bit integers. The circuit time is benchmarked by the circuit time of single-core Shor's algorithm with $180$ qubits: \red{$t_0 = 190$} days. (a) The circuit time of DShor with $2,4,6,8,12$ QPUs at different costs of inter-module communications, where the communication cost is benchmarked by the cost of logical intra-module CNOTs (without QEC) time: $\tau = \frac{t_{\text{inter}}}{t_{\text{intra}}}$ and $t_{\text{intra}} = 250\mu s$. $\tau = 25$ corresponds to $10^5$ physical inter-module Bell pairs generated per second as labeled by the dashed black line. DShor gains small advantage at low communication cost, and the delay from the inter-module communications is more significant as more QPUs are involved in the DShor. The circuit time of 6-QPU DShor with communication rate $10^5$ Bell pairs/sec is \red{approximately $8.6\%$ longer} than the single-module Shor. (b) The strong scaling of DShor at different number of QPUs (excluding memory). The purple line represents the qubits per QPU, and the blue line represents the total number of qubits used. \textcolor{revisionred}{The qubits used per QPU decrease} with a diminishing slope. \textcolor{revisionred}{The qubits per QPU reach} the minimum \textcolor{revisionred}{at $8$ QPUs and increase at $12$ QPUs} as more qubits are required to accommodate the frequent communications and teleportations in the DShor. The total number of qubits used \textcolor{revisionred}{increases roughly linearly with} the number of QPUs with a sudden increase at $12$ QPUs. (c) The circuit time of 6-QPU DShor at different communication cost and measurement time. \red{The vertical axis gives the measurement time $t_{\text{mea}}$ in ms; its ratio to the logical intra-module CNOT duration (without QEC) is $\tau_{mea} = t_{\text{mea}}/t_{\text{intra}}$.} The distribution of Shor's algorithm is not delayed by the inter-module communications to the left of the \textcolor{black}{blue} line (``zero" communication overhead), where $t(\tau_{\text{inter}})/t(\tau_{\text{inter}}=1) < 1.01$. The threshold of ``zero" communication overhead also increases with a higher measurement time since more operations can happen in parallel behind the measurement. \red{(d) The circuit time of 6-QPU DShor at different code distance and measurement time. The shorter code distance leads to fewer communications required by the inter-module logical operations, and the delay from the inter-module communications is less significant.}
    }
    \phantomsection
    \label{fig:Result}
\end{figure*}

We apply DShor on $2,4,6,8,12$ QPUs to factor 2048 bit RSA integers using the parameters mentioned above and estimate the space and time usage from the circuit evolved by the Shor's simulator as shown in Fig.~\ref{fig:Result}. We assume that the physical error is $10^{-3}$ and benchmark our gate time to $250\, \mu s$ as mentioned in Section\red{.}~\ref{Chapter: Hardware and Architecture}. \textcolor{revisionred}{We also} assume all logical operations are followed by one QEC cycle. The QEC cycle time is estimated as the measurement time, as discussed in Section\red{.}~\ref{Chapter: Hardware and Architecture}. Although increasing the number of qubits per module can accelerate the algorithm (See Appendix.~\ref{Appendix: More parameters}), we use the minimal \textcolor{revisionred}{spatial resources} in the QPU (only one CCX factory) for our resource estimation. The delay $\tau$ from inter-module communications is benchmarked by the clock time: $\tau \equiv \frac{t_{\text{inter}}}{t_{\text{intra}}}$. The circuit time is benchmarked by the single-module Shor's algorithm with $180$ hot qubits, which has  circuit depth of \red{$6.56\times10^{10}$ and circuit time of $190$} days per shot. 

We estimate the inter-/intra-module resources used by DShor with $2,4,6,8,12$ QPUs to factor 2048 bit RSA integers in Fig.~\ref{fig:QPU_connections}. The rectangles on the left are the quantum memories, and the squares on the right are QPUs. Memory is only accessed during two steps of DShor: \textit{load exponents} and \textit{accumulate results}. During \textit{load exponents}, $w_1 = 6$ qubits from the exponents $\ket{Q_e}$ are read to the buffer surface code patches by the lattice surgery, and the buffer patches are teleported to the ``router" QPU. Short logarithms $\ket{q_e}$ are then computed based on these six buffer qubits in the router QPU, and the short logarithms $\ket{q_e}$ are used for modular exponentiations. When the modular exponentiation is finished, the result of approximate exponentiation $\ket{Q_{acc}}$ is teleported to all QPUs to record the current result of modular exponentiation $\ket{q_r}$. Therefore,  all calculations happen in the QPUs, which leads to $0$ $\mathrm{T}$ gates in the memory. More details about memory access in DShor can be found in Appendix\red{.}~\ref{Appendix: Estimation method}.

Except the router QPU (the upper-left QPU with more $\mathrm{T}$ gates), each QPU uses a similar number of $\mathrm{T}$ gates, and the number of $\mathrm{T}$ gates per QPU decreases as the number of QPUs used in DShor increases. The ``router" QPU has more $\mathrm{T}$ gates as all information from the memory is processed by the ``router" QPU, and most of the communications are local as discussed in Section\red{.}~\ref{Chapter: Distributed fundamental operations}.

We also estimate the strong scaling of RSA 2048 DShor as shown in Fig.~\ref{fig:Result}, and the circuit time is benchmarked by the circuit time of the single-core Shor's algorithm with $180$ qubits and $1$ ms measurement time: \red{$t_0 = 190 \text{ days}$}. \red{Figure~\ref{fig:Result}(a) shows that the circuit time of DShor increases with the communication delay $\tau$, and the circuit time of DShor with $6$ QPUs is approximately the same as the single-core Shor's algorithm at $\tau \approx 18.0$, \textcolor{revisionred}{which we refer to as} ``free modularization".} The delay of DShor is more significant as the cost of communications increases since the longer inter-module communications delay more subroutines. The space used by DShor is illustrated in Fig.~\ref{fig:Result}(b), where the purple curve represents the number of qubits within one QPU, and the blue curve represents the total number of qubits in the modular processor. The number of qubits within one QPU decreases with a diminishing slope. The number of qubits per QPU increases when DShor is distributed to $12$ QPUs as each QPU requires more qubits to buffer the frequent qubit teleportations in the DShor. We focus on the analysis of 6 QPU DShor to factor 2048-bit RSA integers with $p_p = 10^{-3}$, so each QPU has $7\times 7$ logical qubits and $61,250$ physical qubits in a $2\times 2$ mm$^2$ atom array workspace, which we believe is a realistic field of view based on the current state-of-the-art for microscope objectives~\cite{Manetsch2024,Chiu2025}. \red{Using the architecture defined above, the total number of physical qubits (including memory) used in DShor to factor 2048 bit RSA integers with $6$ QPUs is $110,000 + 61,250\times 6 = 477,500$. \textcolor{revisionred}{We round it up to half a million throughout the work.}} 

\begin{table*}[t]
\centering
\begin{tabular}{c|c|c|c|c|c|c|c|c|c}
\hline
&Code distance&$N_L$&$N_{p}$&\makecell{Total qubits used\\(excluding memory)}&\makecell{Clock time\\$10d$ $\mu$s}&$t_{mea}$ (ms)&\makecell{Comm. rate\\(ebits/s)}&Time (one shot)&$t_0$ (days)\\
\hline
\makecell{RSA 1024 \\ ($p_p=10^{-3}$)} & 23 & \red{41} & \red{43,378} & \red{260,268} & 230 & 1& $10^5$ & \red{64.6} days & \red{24.2} \\
\makecell{RSA 1024\\ ($p_p=10^{-4}$)}& 14 & \red{41} & \red{16,072} & \red{96,432} & 140 & 1 & $10^5$ & \red{61.5} days & \red{22.5} \\
\makecell{RSA 2048 \\ ($p_p=10^{-3}$)} & 25 & 49 & 61,250 &367,500& 250 & 1 & $10^5$ & \red{206} days & \red{190} \\
\makecell{RSA 2048 \\ ($p_p=10^{-3}$)} & 25 & 49 & 61,250 &367,500& 250 & 0.25 & $10^5$ & \red{129} days & \red{74.0} \\
\makecell{RSA 2048\\ ($p_p=10^{-4}$)} & 15 & 49 & 22,050 & 132,300&150 & 1 & $10^5$ & \red{169} days & \red{176} \\
\makecell{RSA 2048 \\ ($p_p=10^{-4}$)} & 15 & 49 & 22,050 &132,300& 150 & 0.25 & $10^5$ & \red{64.7} days & \red{59.9} \\
\makecell{RSA 4096 \\ ($p_p=10^{-3}$)}& 27 & 49 & 71,442 & 428,652&270 & 1 & $10^5$ & \red{1,280} days & \red{1,170} \\
\makecell{RSA 4096\\ ($p_p=10^{-4}$)}& 16 & 49 & 25,088 & 150,528&160 & 1 & $10^5$ & \red{1,070} days & \red{1,080} \\

\end{tabular}
\caption{\textbf{Resource usage in DShor}. \red{Runtimes are rounded to three significant digits.} We estimate the space and time usage of DShor with $6$ QPUs to factor ($1024, 2048, 4096$ bit\textcolor{revisionred}{) RSA integers} at physical errors $p_p = 10^{-3}$ and $p_p = 10^{-4}$. We record the number of logical qubits ($N_L$) and physical qubits ($N_p$) per module, and we assume the communication rate between modules is $10^5/s$ physical Bell pairs. $t_0$ represents the circuit time of the DShor with $180$ logical qubits with the measurement time $t_{mea}$. In DShor, the time of \textcolor{revisionred}{one logical operation} (CNOT, $\mathrm{H}$, $\mathrm{S}$, etc.) at least needs $t_{\text{mea}} + 10d\mu s$. Therefore, the circuit time of DShor does not change significantly with the code distance given the long measurement time, which becomes a significant parameter influencing the circuit time. The circuit time of single-core Shor's algorithm ($t_0$) suggests this trend, where \red{$t_0 (t_{\text{mea}} = 250~\mu s)/t_0 (t_{\text{mea}} = 1~ ms) \simeq 0.390$}. For $d=25$, the CNOT plus QEC duration decreases from $1.25$ ms to $0.5$ ms, giving a ratio $0.5/1.25=0.4$, close to this result. For the DShor with $6$ QPUs, the circuit time is less sensitive to the measurement time as the latency from the inter-module communications becomes a significant bottleneck at short measurement time as shown in Appendix\red{.}~\ref{Appendix: More parameters}.}
\label{Table: Running time in different RSA}
\end{table*}

We also estimate DShor for different physical error rates \textcolor{revisionred}{with the associated gate times in} different RSA challenges with constant communication rate $10^5$ physical Bell pairs per second, as shown in Table.~\ref{Table: Running time in different RSA}. This communication rate is based on recent predictions for multiplexed and parallelized cavity-based networking protocols~\cite{LiThompson2024,Li2025,Shaw2025}. Here, we only consider the time required for one shot of Shor's algorithm. The expected number of classical shots required to factor RSA $2048$ integers is $9.2$ at $s=8$, and all shots can be collected by 10 identical modular quantum computers working in parallel or by using $6\times 10$ QPUs connected to the shared memory~\cite{Ekera2017, Ekera2017Post, Ekera2026, Ekera2017Post2, Gidney2025}. $s$ is the Eker\aa's parameter to control the space and time tradeoff and Shor's algorithm requires $n/2 + n/s$ qubits in the memory and $\sim s+1$ classical shots. The compilation of DShor in the QGPU architecture ($\sim 100$ modules) can be found in Section\red{.}~\ref{Chapter: QGPU}.

The circuit time of Shor's algorithm strongly depends on the measurement time, as shown in Fig.~\ref{fig:Result}\red{(c)}, since the time of the QEC cycle depends on the measurement time. Fig.~\ref{fig:Result}\red{(c)} shows the circuit time of 6-QPU DShor to factor 2048-bit RSA integers at different measurement time $\tau_{mea}$ and communication delay $\tau_{inter}$. The QEC cycle is longer as the measurement increases, and further increases the running time of DShor. However, the longer measurement time also gives more time for parallelism between intra-module operations and inter-module communications, which increases the threshold of \red{``zero" communication overhead.}  \red{The \textcolor{black}{blue} curve marks the first sampled communication cost for which $t(\tau_{\text{inter}})/t(\tau_{\text{inter}}=1)>1.01$.} The measurement time can also significantly change the compilation of DShor. We assume measurement is much more expensive than physical gates, so logical CNOTs are implemented by transversal CNOTs instead of the lattice surgery. Therefore, the cascade of CNOTs can only be implemented sequentially, which significantly increases the circuit time. This delay can be resolved by a GHZ-empowered CNOT fanout at the price of more qubits per QPU or lattice surgery on platforms with fast measurement (e.g. superconducting circuits)~\cite{Zhou2025, Gidney2025,Chamberland2022}. \textcolor{revisionred}{\hypersetup{allcolors=revisionred}Please see Appendix.~\ref{Appendix: Comparison with other works} for more details about using the sequential CNOTs over the lattice surgery in DShor's compilation.}

The resource and communications used by DShor also strongly depend on the physical error. As the physical error decreases from $10^{-3}$ to $10^{-4}$, the code distance can decrease by $\sim 10$, which can reduce the qubit usage by $\sim 60\%$ and reduce the time of DShor by \red{approximately $4.7\%$ or more}. Table~\ref{Table: Running time in different RSA} illustrates this trend, with resource estimates for both $p_p=10^{-3}$ and $10^{-4}$ at each RSA challenge size. \red{The code distances in Table.~\ref{Table: Running time in different RSA} are \textcolor{revisionred}{extrapolated} from the distance-LER scaling from Ref.~\cite{Gu2026Neural} so that $p_{\mathrm{logical}}N_{\mathrm{operations}} \lesssim \textcolor{revisionred}{10^{-2}}$.} Since the exact code distance required \textcolor{revisionred}{at a certain physical error} can fluctuate significantly depending on the physical fidelity of different gates and the decoder~\cite{Higgott2025, OROURKE2024, iOlius2024,Cain2025,Perrin2025}, we use the code distance as a measure of physical error and decoders. We simulate DShor with $6$ QPUs with $10^5~s^{-1}$ inter-module \textcolor{revisionred}{physical Bell pairs} at different measurement time $t_{\text{mea}}$ and code distance $d$. One shot in DShor can reach as low as \red{$40.5\text{ days}$} per shot if $t_{mea} = 0.2$ ms and $d = 9$. The \textcolor{black}{blue} line captures $t/t_0\leq 1$.

\begin{figure}[t]
    \includegraphics[width=\linewidth]{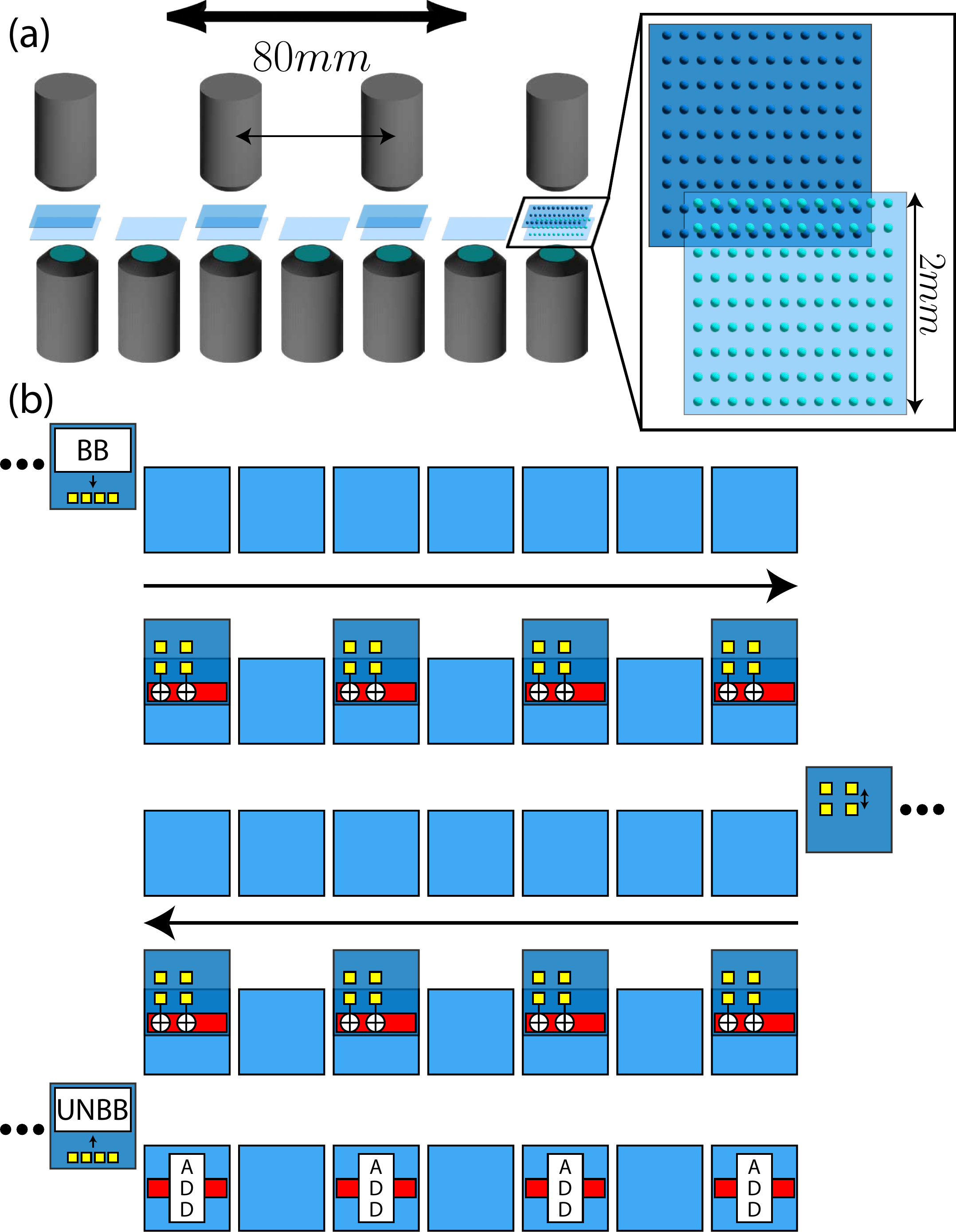}
	\caption{
        \textbf{DShor on the dynamic array}. (a) The modular atomic processors connected by the dynamic arrays. Seven static arrays (six QPUs and one memory) are connected by four dynamic arrays. Each array contains a $2\times 2$ mm$^2$ atom array as illustrated in Fig.~\ref{fig:DShor Structure}. The separation between the static objectives is $40$ mm, and the separation between dynamic arrays is $80$ mm. All dynamic arrays can move collectively and create inter-module connections \textcolor{revisionred}{when the dynamic array overlaps with} the static array. They all share a common focal plane. (b) Compilation of DShor on the dynamic arrays with multi-address loadings. A quantum address is loaded on the dynamic array and processed by the Bucket-Brigade circuit. Classical information is loaded to static QPUs as dynamic arrays sweep through all static QPUs. The address qubits are cleaned by Un-Bucket-Brigade circuit while the quantum addition happens in parallel.
    }
    \phantomsection
    \label{fig:dynamic array illustration}
\end{figure}

\section{DShor with dynamic super-arrays}
\label{Chatper: Dynamic array}
\textcolor{revisionred}{Besides the compilation} of DShor on the static arrays with photonic inter-module connection, we also provide the compilation of DShor on super-arrays connected by dynamic arrays as illustrated in Fig.~\ref{fig:dynamic array illustration}(a).

\begin{figure}[t]
    \includegraphics[width=\linewidth]{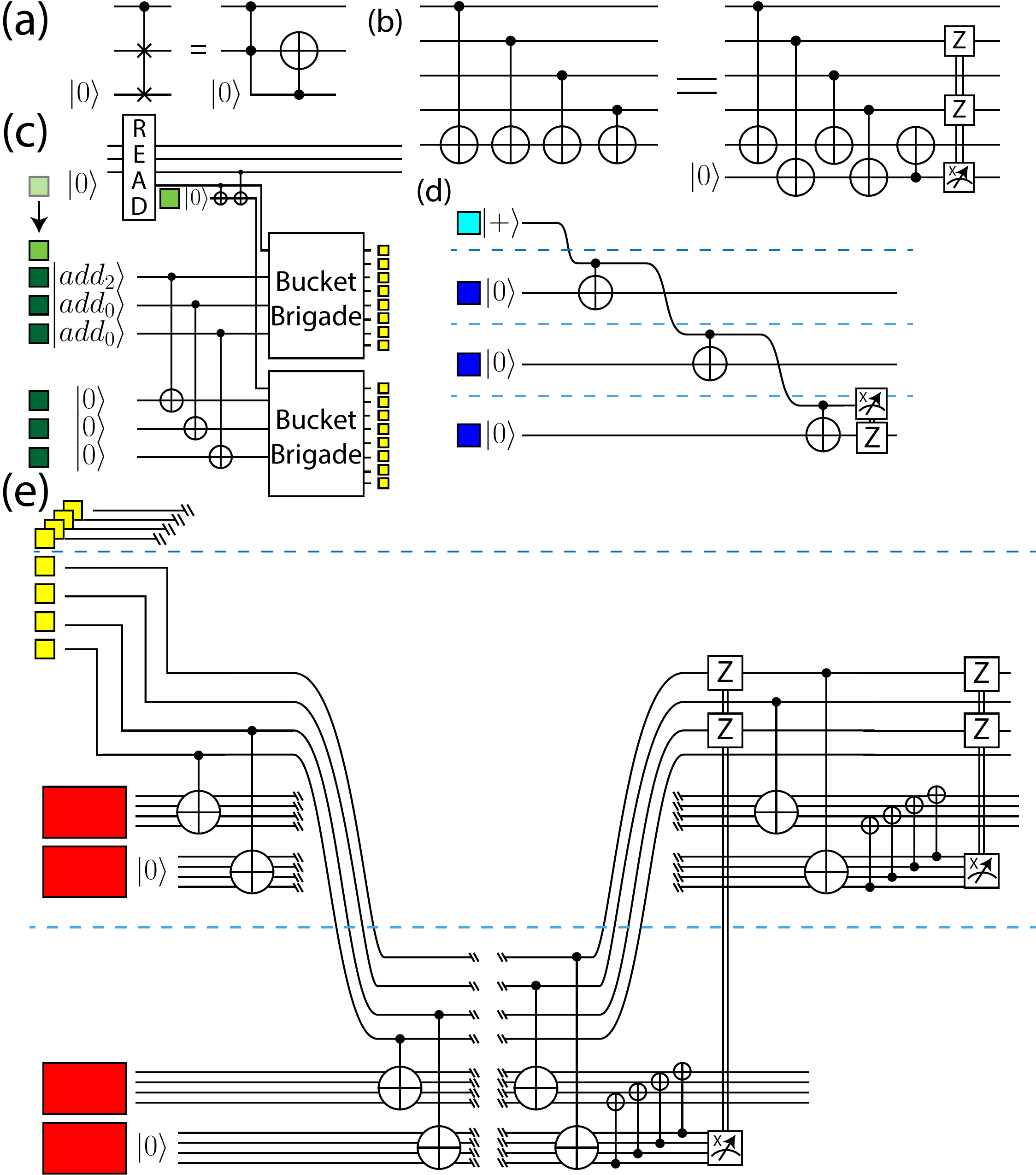}
	\caption{
        \textbf{Circuit of DShor on the dynamic arrays}. (a) The controlled swap gate between one data qubit and one $\ket{0}$ in the Bucket-Brigade circuit can be simplified. (b) One cascade of CNOTs can be split to two shorter cascades of CNOTs with one ancilla qubit. (c) Hybrid quantum loadings on the dynamic array. All quantum addresses are read by unary loadings except the last three quantum addresses (dark green squares). The three quantum addresses are expanded to three Bell pairs and further expanded to $16$ address qubits (yellow squares) by the Bucket-Brigade circuit. (d) \textcolor{revisionred}{A cascade of CNOTs is applied} from the dynamic array to the static array to generate one GHZ state across all static arrays. (e) Quantum information is loaded from the address qubits (yellow squares) to the data registers (red squares). Each data register is accompanied by one ancilla qubit to double the loading bandwidth using the circuit in (b). Classical information is loaded from the address qubits to the data registers during one sweep of the dynamic arrays.
    }
    \phantomsection
    \label{fig:dynamic loading circuit}
\end{figure}

We compile the 6-QPU DShor to factor 2048 bit RSA integers on seven static arrays (6 QPUs and 1 memory) connected by four dynamic arrays. This hardware approach is nascent but provides a method to scale to several 100,000 physical qubits within a single vacuum chamber without the need for photonic interconnects. Anticipating a future in which RSA 2048 can be performed with $\sim10^4-10^5$ qubits via high-rate QEC codes, we may wish to deliberately eschew the photonic interconnects between modules contained in separate vacuum chambers and adopt a single-chamber architecture. To reach the several 100,000 physical qubit scale required for our CSS code-based processor architecture, we envision defining many modules via an array of microscope objectives spaced by $\approx40$ mm.

Communication between these modules is performed by atom arrays that can be transported between them. Dynamic objectives from the opposite side of the vacuum chamber as the static objectives are able to move together laterally via an air bearing linear motion stage on which all four of them are placed. The focal plane in which the atom arrays reside is shared by all objectives. The working principle of the optical system, though not described in detail here, is designed to preserve the full functionality of each array, both static and dynamic. Such linear motion stages are widely used in high-tech manufacturing including in extreme ultraviolet lithography, where wafer and reticle stages move in vacuum with nm-scale precision \textcolor{revisionred}{in two dimensions} with accelerations of $\sim15g$. 

\begin{figure}[t]
    \includegraphics[width=0.8\linewidth]{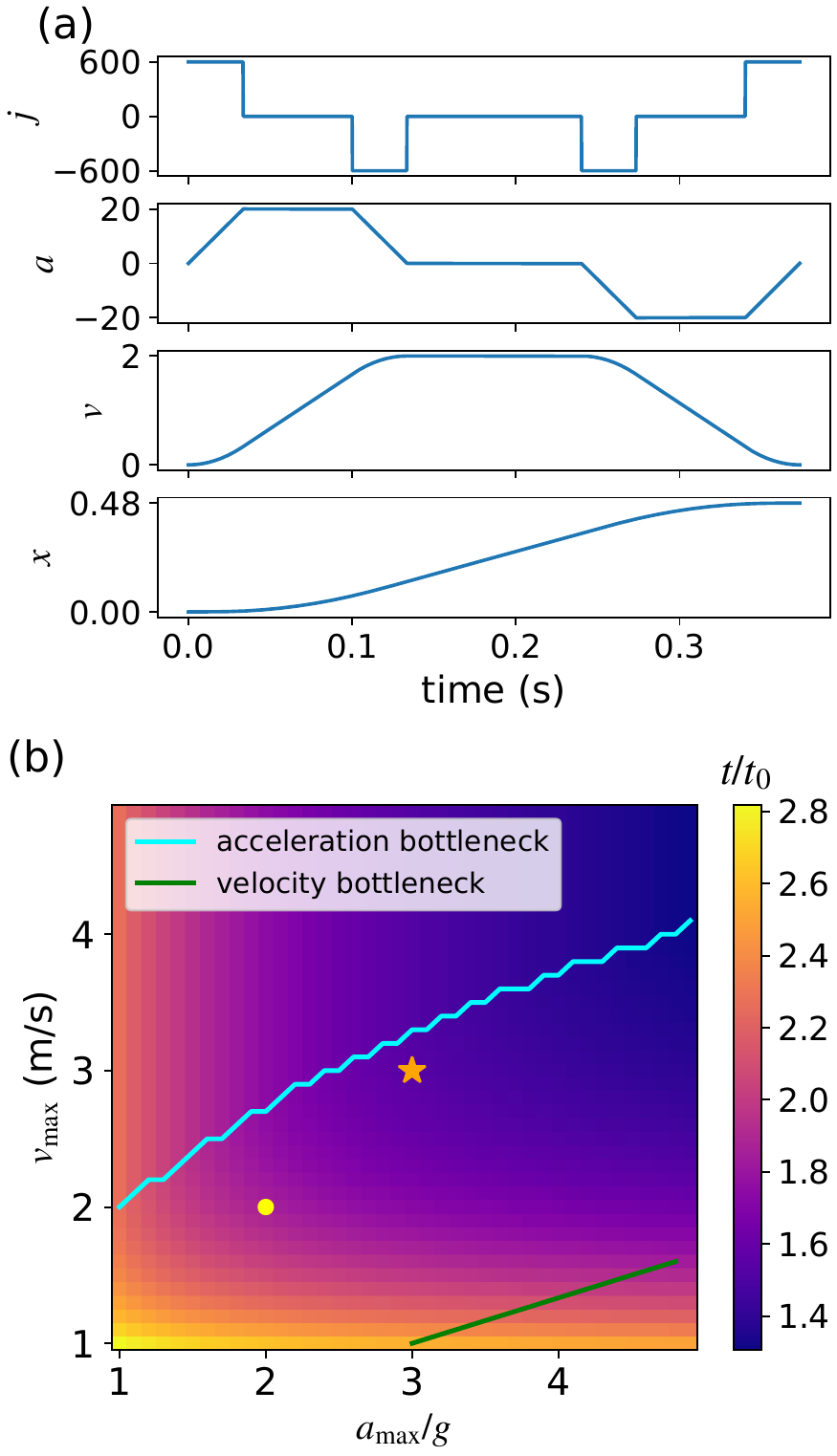}
	\caption{
        \textbf{Dependence on the motion profile of the dynamic arrays}. (a) \textcolor{revisionred}{The moving profile at} $j_{\text{max}} = 600$ m/s$^3$, $a_{\text{max}} = 20$ m/s$^2$, $v_{\text{max}} = 2 $ m/s over a displacement of $480$ mm (one one-way sweep of the dynamic array). The circuit time of DShor with $6$ QPUs to factor 2048 bit RSA integers is $\red{354}$ days, which corresponds to $\tau \approx \red{68.9}$ ($\sim \red{3.63\times 10^4}/s$ \textcolor{revisionred}{physical Bell pairs). (b)} The phase diagram of DShor circuit time to factor 2048 bit RSA integers at different $a_{\text{max}}$ and $v_{\text{max}}$. The time is benchmarked by the circuit time of the single-core with $180$ logical qubits Shor's algorithm: $t_0 = \red{190}$ days. The jerk is $j_{\text{max}} = 30a_{\text{max}}$. The \textcolor{black}{blue} line represents the acceleration bottleneck, above which maximum speed cannot be reached due to the limit of maximum acceleration. The green line represents the speed bottleneck, the right of which the speed reaches its maximum before the acceleration reaches its maximum. The yellow dot labels the circuit time with $a_{\text{max}} = 20$ m/s$^2$ and $v_{\text{max}} = 2$ m/s with circuit time $\red{354}$ days, which corresponds to $\tau \approx \red{68.9}$ ($\sim \red{3.63\times 10^4}/s$ physical Bell pairs). The orange star labels the circuit time with $a_{\text{max}} = 30$ m/s$^2$ and $v_{\text{max}} = 3$ m/s with circuit time $\red{293}$ days, which corresponds to $\tau \approx \red{53.3}$ ($\sim \red{4.69\times 10^4}/s$ physical Bell pairs)\textcolor{revisionred}{.}
    }
    \phantomsection
    \label{fig:dynamic transition}
\end{figure}

For the application presented here, suitable stages with max acceleration of $2g$ and max velocity of 1 m/s are readily available, and stages with $5g$ and 5 m/s are also realistic. Here, we assume that dynamic arrays can move with maximum jerk $j =600$ m/s$^3$, maximum acceleration $20$ m/s$^2$ ($\approx2g$) and maximum speed $2$ m/s. The moving profile of the dynamic arrays is illustrated in Fig.~\red{\ref{fig:dynamic transition}(a)}. We assume both static array and dynamic arrays have dimension $2\times 2$ mm$^2$ with $7\times 7$ logical qubits, the same as in Section\red{.}~\ref{Chapter: Result}.

We assume a periodic motion profile based on zooming back and forth between the two ranges of travel such that, at one end, the fourth dynamic objective overlaps with the first static objective, and at the other end of travel, the first dynamic objective overlaps with the seventh static objective. The range of movement of the dynamic arrays can cover all static arrays as illustrated in Fig.~\ref{fig:dynamic array illustration}(b) and can only move collectively. We assume that `drive-by' Rydberg entangling gates can be performed between the static atoms in the static modules and the dynamic atoms in the dynamic modules, such that the dynamic modules do not need to start and stop at each static module which would enormously inflate the communication time. As noted above, Rydberg-mediated gates require only $\approx300$ ns~\cite{Evered2023,Ma2023}. Assuming the two atoms participating in the gate are separated by $y=5$ $\mu$m and that they have a relative velocity of $v_x=1$ m/s perpendicular to their axis of separation, the effective change to their separation during the 300-ns gate (assuming it is centered on $x=0$) is $\delta r\approx 2$ nm, which is completely negligible. The Doppler shifts associated with the relative motion can likewise be negligible if the Rydberg beam is perpendicular to the direction of motion. Recently, such a drive-by gate was demonstrated with state-of-the-art fidelity at a very low relative velocity of 0.05 m/s, limited by the Doppler sensitivity of the setup~\cite{Lib2026}. We also present the \textcolor{revisionred}{schedule of atom movements for} \textit{loading} using the drive-by gates in Appendix\red{.}~\ref{Appendix: Machine code}.

The dynamic array architecture can provide a very large per-pass bandwidth between modules because of the highly parallelizable transversal drive-by gates between the dynamic array and the static array. If the dynamic array has a complete overlap with the static array, all qubits on the dynamic arrays can be used to generate inter-module Bell pairs. We assume all arrays contain $7\times 7$ $d=25$ surface code patches ($61,250$ physical qubits as shown in Table.~\ref{Table: Running time in different RSA}), and $6$ static arrays are connected by one dynamic array. One transverse move can generate $61,250$ physical $6$-qubit GHZ states among all static arrays, which requires $306,250$ physical Bell pairs between the nearest-neighbor modules and takes \red{$0.373$} s [see Fig.~\ref{fig:dynamic transition}(a)], which corresponds to a bandwidth of \red{$8.20\times 10^5$} physical Bell pairs per second. This exceeds all but the most optimistic estimates for photonic interconnects. 

However, the bandwidth of the inter-module connections heavily depends on the size of the Bell-pair reservoir. In the architecture illustrated in Fig.~\ref{fig:DShor Structure}(b) with 3 logical qubits in the Bell pair reservoir, one transverse move can only generate $1875$ physical $6$-qubit GHZ states (which requires $9375$ physical Bell pairs). Moreover, the latency between passes is long due to the acceleration/\textcolor{revisionred}{deceleration} time of the dynamic array. 


To resolve the competition between the spatial usage of the Bell pair reservoir and the inter-module bandwidth, and to reduce the delay from acceleration/\textcolor{revisionred}{deceleration} of  dynamic arrays, loading in DShor is replaced by multi-address loading as illustrated in Fig.~\ref{fig:dynamic loading circuit}. We apply a hybrid QRAM with the first three qubits read linearly and the latter three qubits read in parallel by the Bucket-Brigade circuit~\cite{Hann2021, Shen2025, Giovannetti2008, Giovannetti2008Architecture, Cesa2025}. The four input qubits are teleported to the dynamic arrays and expanded to $8$ ``hot encoding" address qubits. Quantum information is loaded from the address on the dynamic arrays to the data registers on the static arrays when two arrays overlap using the transversal drive-by gates.
The hybrid Bucket-Brigade circuit is verified by the Shor's simulator, and more details of multi-address loading on the dynamic array can be found in Appendix\red{.}~\ref{Appendix: dynamic array}.

The circuit time of DShor on the dynamic array is estimated by the circuit time of modular additions. The loading in \textcolor{revisionred}{one modular addition} requires $4$ round-trip sweeps ($\red{0.747}$ s each) and one Bucket-Brigade state preparation (circuit depth $\red{2990}$ clock cycles, or circuit time $\red{0.7475}$ s at $t_{\text{mea}} = 1ms$). The motion profile of the dynamic arrays using the assumptions mentioned above ($v_{\text{max}} = 2 \text{ m/s}$, $a_{\text{max}} = 20 \text{ m/s}^2$, $j = 600$ m/s$^3$) is shown in Fig.~\ref{fig:dynamic transition}(a). The addition in the modular addition can be applied in parallel with the state preparation as shown in Fig.~\ref{fig:dynamic array illustration}(b), so the circuit time of one modular addition is $\red{3.73}$ s. In RSA $2048$, there are totally $\sim \red{8.18\times 10^6}$ modular additions, which takes $\red{354}$ days, which corresponds to $\tau \approx \red{68.9}$ ($\sim \red{3.63\times 10^4}/s$ physical Bell pairs between modules) as shown by the yellow dot in Fig.~\ref{fig:dynamic transition}(b). At $v_{\text{max}} = 3\text{ m/s}, a_{\text{max}} = 30 \text{ m/s}^2, j_{\text{max}} = 900 \text{ m/s}^3$, factoring 2048 bit RSA integers needs $\red{293}$ days ($\tau \approx \red{53.3}$ or $\sim \red{4.69\times 10^4}/s$ physical Bell pairs between modules) as shown by the orange star. The compilation with multi-address therefore provides a $4\times$ bandwidth increase compared with the naive compilation of DShor on the dynamic arrays. For additional context, we assumed $\tau\approx25$ ($\sim 10^5$ physical Bell pairs between modules) for our photonic interconnect architecture, such that the run times are within a factor of $\approx \red{1.71}$ between the two approaches. Accordingly, we conclude that the dynamic array architecture could be a viable alternative to the photonic interconnect approach.

\section{DShor in the QGPU architecture}
\label{Chapter: QGPU}

\begin{figure}[t]
    \includegraphics[width=\linewidth]{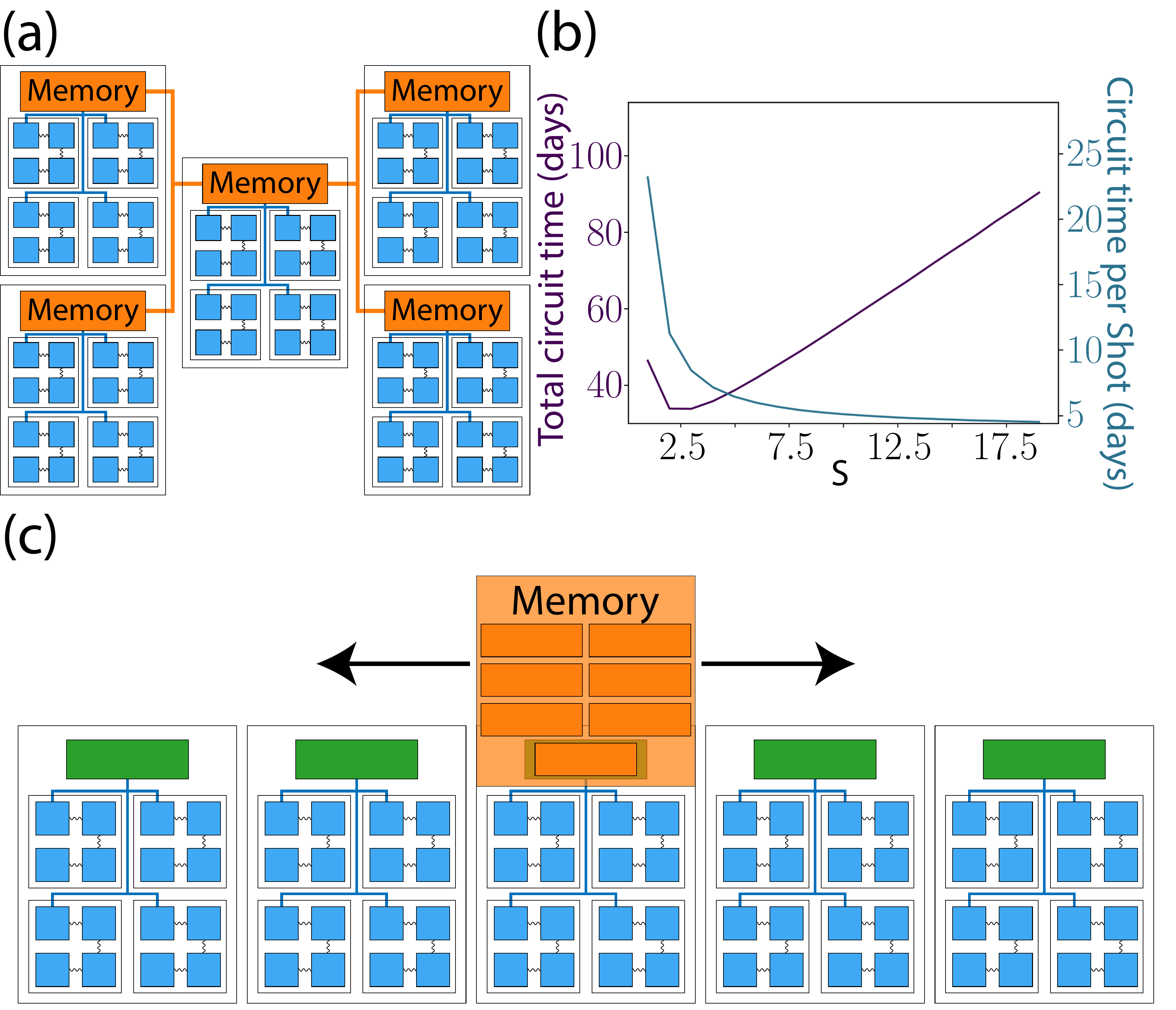}
	\caption{
        \red{\textbf{DShor in the QGPU architecture}. (a) Architecture with photonic interconnects. Each cluster (big white box) contains one memory (orange rectangles) and four modular processors (small white boxes), each with four QPUs (blue squares). Memory qubits use $[[126,12,10]]$ GB codes, with twelve $d=9$ surface-code access patches allocated per modular processor. Blue lines connect memory to the router QPUs; orange lines connect memories. (b) Circuit time at different $s$. The total circuit time (purple curve) reaches $33.8$ days at $s=3$, with $22$ modular processors and $991,188$ physical qubits. The time per shot (blue curve) reaches $4.5$ days at $s=19$. These points use the representative-circuit extrapolation described in the text. (c) Architecture with memories connected by dynamic arrays. Memory states are teleported to the buffer zones (green rectangles) when the patches overlap.}
    }
    \phantomsection
    \label{fig:QGPU}
\end{figure}

With higher gate fidelities that enable drastic reduction in the code distance, we envision that DShor can be implemented with a quantum GPU (``QGPU") architecture ($\approx 100$ modules connected to the shared memory). We estimate that physical error rates of $p_p\approx10^{-5}$ will be needed to run RSA 2048 with $d=9$ using established decoder techniques. However, we anticipate that specialized decoders that are designed with a detailed knowledge of the error channels in the hardware including erasure errors~\cite{Wu2022,Zhang2025} may enable the use of such relatively small code distances with physical error rates of $p_p\approx10^{-4}$. Accordingly, we present a version of our DShor algorithm that is optimized around the dramatically relaxed spacelike resources associated with a $d=9$ logical encoding. 

In Fig.~\ref{fig:QGPU}(a), $20$ modular processors are connected with the shared memory, and each module contains four QPUs with $8\times 8 = 64$ logical qubits encoded in $d=9$ surface code patches. All qubits in the memory are stored in \red{$[[126, 12, 10]]$} GB codes \red{with twelve $d=9$ surface-code access patches allocated per modular processor for lattice surgery.} The communications between the memory and modular processors are labeled by the blue lines, where each memory only needs to communicate with the router QPU in the modular processors. The communications between the memory are labeled by the orange lines, where \textcolor{revisionred}{all memories are connected}. In the QGPU architecture, $N_{\text{processors}}$ modular exponentiations happen in parallel and the circuit time is reduced to $1/N_{\text{processors}}$ of its original circuit time. In Fig.~\ref{fig:QGPU}(c), $20$ modular processors are connected by the memory connected by the dynamic arrays. Memory patches stored in GB codes (\textcolor{revisionred}{orange rectangles}) are teleported to the buffer memory zones (green rectangles) when the two regions overlap. Figure~\ref{fig:QGPU}(c) shows a QGPU architecture based on dynamic arrays.

The memory of Shor's algorithm requires $n/2 + n/s$ qubits to store all exponents at the price of $\sim s+1$ classical runs required. Therefore, the circuit time can be further reduced by reducing Ekera's parameter $s$ at the price of more memory used by DShor. Fig.~\ref{fig:QGPU}(b) shows the circuit time of DShor to factor $2048$-bit RSA integers with $<10^6$ physical qubits  at different Ekera's parameter $s$. We assume the measurement time is $1$ ms and the communication rate is $10^5$ s$^{-1}$. The purple curve is the total circuit time $t_{\text{one shot}}\times (s+1)$, and the blue curve represents the circuit time in one Shor's shot. The total circuit time reaches its minimum \red{$33.8$ days at $s=3$ with $22$ modular processors and $<991,188$ \textcolor{revisionred}{physical qubits in total}.} The circuit time per shot decreases at the higher $s$ as expected, and \red{the circuit time per shot is $4.5$ days at $s=19$. Collecting $20$ required shots on $20$ identical QGPUs would give approximately $4.5$ days with about $20$ million physical qubits.} However, DShor in the QGPU architecture requires more frequent and complicated communications between the modules and the memory, which may become a new factor delaying the circuit time. We leave the more detailed analysis of the memory control of DShor in the QGPU architecture to future work.

\section{Concluding discussion}
We have presented a detailed end-to-end compilation of Shor's algorithm that, while generalizable to any hardware platform, is specifically tailored to modular atomic quantum processors. This work serves to provide a collection and summary of the state-of-the-art algorithmic sub-routines in addition to introducing new methods that further improve the spacetime resource efficiency. These new insights include 1) modular implementation of Shor's algorithm and the concomitant modular architecture, 2) decomposing Shor's algorithm to fundamental operations, 3) qubit re-allocations on the neutral atom platforms, 4) end-to-end simulation of Shor's algorithm and resource estimation based on the concrete circuit. Additionally, a major contribution of this work is to present a detailed strategy for modularizing Shor's algorithm that maximizes the compilation flexibility with respect to the buffering of communication. For the foreseeable future in which communication is more expensive than local operations, our methods for splitting the algorithm across roughly ten modules will be invaluable.

We reiterate that this work is not meant to provide a detailed resource estimate for building a quantum processor today. Rapid advances in atomic hardware, quantum error correction, and algorithm design would render such an estimate obsolete within months. On the atomic hardware side, we anticipate the possibility to reduce measurement time via, e.g., optical cavity arrays~\cite{Menon2024,Shaw2025,Kroeze2023,Marsh2025,Kroeze2025,Ding2026}. We also anticipate the possibility to mitigate the bottleneck of acceleration/deceleration due to atomic inertia during transport between disparate zones through the use of \textit{constant velocity} protocols and drive-by gates, inspired by the transportable array architecture discussed above and recently demonstrated for a low velocity of 0.05 m/s~\cite{Lib2026}. On the error correction side, we anticipate improved strategies for stabilizer measurement and classical decoding of qLDPC codes, improved strategies for code switching and lattice surgery between memory and logic, and even the possibility of transversal logic in the qLDPC code block that could obviate the need for code switching. 
These advances are expected to drastically improve the spacetime volume estimates, but are unlikely to eliminate the value of modular architectures in the near term. \red{Moreover, the efficiency of inter-module communications can be further improved by the cultivation-inspired Bell pairs preparation~\cite{Sunami2026}, qLDPC distilled Bell pair~\cite{Ataides2025} and repeated concatenated distillation of Bell pairs~\cite{Pattison2024}.}

As mentioned in the introduction, we have neglected atom loss in our compiler and resource estimates. The probability of loss during essentially all of the key primitives (initialization, cooling, one- and two-qubit gates, and slow measurement) is at the $\sim\textcolor{revisionred}{10^{-3}}$ scale in state-of-the-art systems (see, e.g., refs.~\cite{Lis2023,Norcia2023,Scholl2023,Ma2023,Evered2026}). Therefore, in principle, the space overhead associated with loss is minimal. The time overhead is potentially more substantial because new atoms must be transported from the reservoir array~\cite{Norcia2024,Gyger2024} into the memory or QPU code blocks, and because the logical encoding must be dynamic as new atoms are cycled in~\cite{Eickbusch2025}. However, we note that our compiler can straightforwardly address this latency. As the refilling of data (ancilla) qubits cannot happen in parallel with transversal gates (QEC cycle), the latency from refilling can be included by increasing the clock time and measurement time. Finally, we note that thermal errors and atom recooling during the computation will be required, and remain largely outstanding tasks in the community. \textcolor{revisionred}{Appendix~\ref{Appendix: Machine code} heuristically estimates the additional atom-refilling latency excluded from our compiled runtimes. Recent atom-loss-aware QEC protocols~\cite{Baranes2026Loss,AtomComputing2026Toric} may help mitigate this additional cost.} Finally, we note that it may be fruitful to design the hardware layer around atom loss, potentially by leveraging \textit{constant velocity} operations including measurement. Such strategies might streamline atom replenishing and enable the use of fast, high-loss atom readout for stabilizer measurement and logical operations.\\

\textit{Acknowledgments}.---We acknowledge the Covey Lab for stimulating discussions. J.P.C. acknowledges funding from the AFOSR Young Investigator Program (AFOSR award FA9550-23-1-0059); the NSF PHY Division (NSF award 2339487); the NSF Quantum Interconnects
Challenge for Transformational Advances in Quantum
Systems (NSF award 2137642); the NSF QLCI for Hybrid Quantum Architectures and Networks (NSF award 2016136); the U.S. Department of Energy, Office of Science, National Quantum Information Science Research Centers; and the Army Research Office (ARO awards W911NF-25-1-0156 and W911NF-25-1-0214). 

{\color{revisionred}\textit{Acknowledgments}.--
The code of DShor simulator and the architecture/compilation of DShor were invented without the assistance of AI. T.X. acknowledges the use of ChatGPT 6.1 sol for assistance in the numerical simulation and the review of the manuscript.}

\setcounter{section}{0}



\appendix
\renewcommand\appendixname{APPENDIX}
\renewcommand\thesection{\Alph{section}}
\renewcommand\thesubsection{\arabic{subsection}}
\makeatletter
\renewcommand\p@subsection{\thesection.}
\makeatother



\setcounter{figure}{0}
\renewcommand{\thefigure}{S\arabic{figure}}

\clearpage
\section{Gate time benchmarking}
\label{Appendix: gate time benchmarking}

\begin{figure}[t]
    \includegraphics[width=\linewidth]{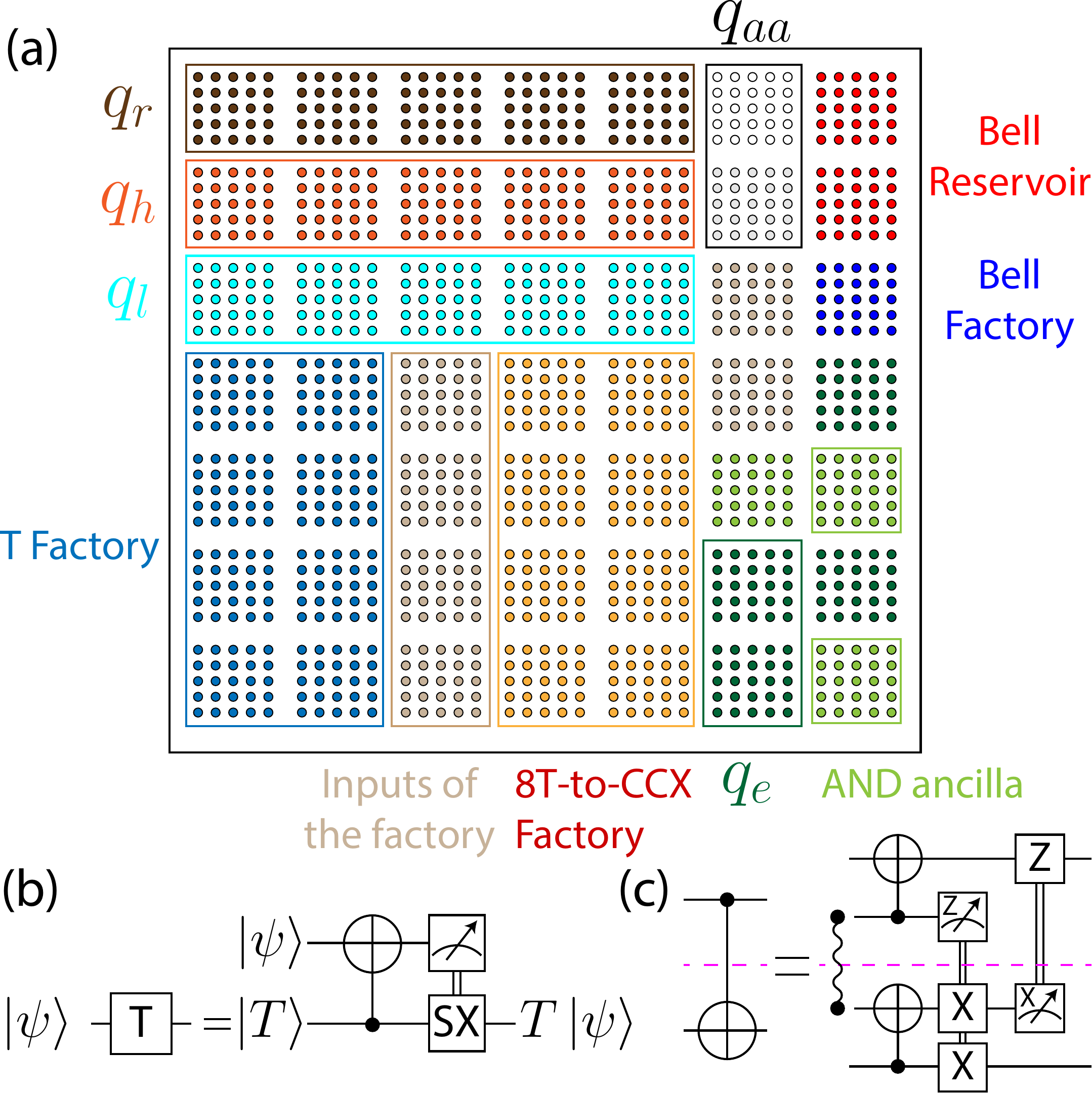}
	\caption{
        \textbf{Gate time Benchmarking}.  (a) The detailed qubit allocation in each QPU. $q_{aa}$ is the antenna ancilla qubits used to generate the inter-module GHZ state, and $q_r, q_h$, $q_{l}, q_e$ are the intermediate results as shown in Table.~\ref{Table: Qubit usage in Shor}. The \red{$8\mathrm{T}$-to-CCX factory} on the bottom left corner of the QPU is further split to $\mathrm{T}$ factory, inputs of the factory and $8$ ancilla qubits used in the \red{$8\mathrm{T}$-to-CCX factory}. (b) The $\mathrm{T}$ gate is implemented by the magic state injection from the magic state $\ket{T}$ from the code cultivation. (c) The inter-module CNOTs are implemented by the gate teleportation using the inter-module Bell pairs stored in the Bell reservoir.
    }
    \phantomsection
    \label{fig:Gate benchmark}
\end{figure} 

In the DShor algorithm, we use the following universal gate set:
X, Y, Z, CNOT(\red{CZ}), $\mathrm{H}$, $\mathrm{S}$, $\mathrm{T}$ with the detailed implementations as follows:
\begin{itemize}
    \item X/Y/Z: 
Pauli \red{corrections are tracked classically in the simulator instead of being assigned separate gate time. The tracked frame determines subsequent correction and measurement choices, including at non-Clifford gates}~\cite{Chamberland2018Pauli, Riesebos2017}.
    \item CNOT(CZ): 
\red{The logical CNOT(CZ) is implemented using transversal Rydberg gates~\cite{Zhou2025Transversal,Cain2025Fast,Turner2026Transversal,SerraPeralta2026Transversal}. Its duration defines the clock time used in this paper.}
    \item $\mathrm{H}$/$\mathrm{S}$: 
\red{The $\mathrm{H}$ gate uses transversal Hadamard gates followed by code rotation, and the $\mathrm{S}$ gate uses a fold-transversal implementation~\cite{Zhou2025Transversal}. We model both intrinsic gate durations as three CNOT clock periods, with the following QEC cycle charged separately, assuming that the required atomic motion remains within a patch~\cite{Chen2024RP2}.}
    \item $\mathrm{T}$:
    \red{A $\mathrm{T}$ gate is implemented by injecting a cultivated magic state into the data register, as in Fig.~\ref{fig:Gate benchmark}(b). We use fold-transversal surface code cultivation at $d=5$, as specified in Appendix.~C and Table~A5 of Ref.~\cite{Sahay2026}. The preparation has one measurement layer after injection, two at $\mathrm{ROT}(5)$, and five during escape. For a completed cultivation attempt, the measurement contribution is}
    \red{\begin{equation}
        {t_{\mathrm{meas}}=(1+2+5)t_{\mathrm{mea}}=8t_{\mathrm{mea}}.
        \label{eq:cultivation-measurement-time}}
    \end{equation}}
    \red{We set $t_T=10t_{\mathrm{mea}}$ for slack. At $p_p=10^{-3}$, we use postselection with a rounded complementary-gap cutoff of $91$ dB in the released Clifford-state simulation data, giving $4.45$ expected cultivation attempts per accepted state and infidelity $p_T\simeq8.93\times10^{-8}$~\cite{SahayCultivationData2026}. After the $\ket{T}$ is prepared, \textcolor{revisionred}{it is injected into} the data qubit with the feedforward $\mathrm{S}$ gate depending on the measurement result.}
    \item \red{CCX:
    The $8\mathrm{T}$-to-CCX factory~\cite{Jones2013} prepares a verified $\mathrm{AND}\ket{++0}$ resource, which is teleported onto the data. The injection includes two CNOTs, two Z-basis measurements, and two conditional CNOT corrections with probability $50\%$~\cite{Gidney2025}.}
\end{itemize}

\begin{table*}[t]
\red{
\centering
\small
\setlength{\tabcolsep}{5pt}
\renewcommand{\arraystretch}{1.5}
\begin{tabular*}{\textwidth}{@{\extracolsep{\fill}}cccccc@{}}
\toprule
\makecell{Type of\\gates} & \makecell{Number of\\gates} & \makecell{Target error\\per shot} & \makecell{Target logical\\error per gate} & \makecell{Required\\physical error} & \makecell{Decoding\\time ($\mu\mathrm{s}$)} \\
\midrule
\makecell{Intra-module\\CNOT/CZ} & $5.92\times10^{10}$ & $<\textcolor{revisionred}{10^{-2}}$ & $<\textcolor{revisionred}{1.68\times10^{-13}}$ & \makecell{$\textcolor{revisionred}{10^{-3}}$~\cite{Gu2026Neural}} & $34.16\pm0.92$ \\
\makecell{Inter-module\\CNOT/CZ} & $2.74\times10^9$ & $<\textcolor{revisionred}{10^{-3}}$ & $<3.64\times10^{-13}$ & \makecell{$\textcolor{revisionred}{10^{-3}}$~\cite{Gu2026Neural}} & --- \\
$\mathrm{H}$ & $1.85\times10^9$ & \makecell{$<\textcolor{revisionred}{10^{-3}}$} & \makecell{$<5.41\times10^{-13}$} & $<\textcolor{revisionred}{10^{-3}}$ & $19.08\pm0.41$ \\
$\mathrm{S}$ & \makecell{$3.07\times10^9$} & $<\textcolor{revisionred}{10^{-3}}$ & \makecell{$<3.25\times10^{-13}$} & \makecell{$<\textcolor{revisionred}{10^{-3}}$~\cite{Chen2024RP2}} & $58.35\pm0.52$ \\
$\mathrm{T}$ & $6.09\times10^9$ & \makecell{---} & $<9.68\times10^{-8}$ & \makecell{$<\textcolor{revisionred}{10^{-3}}$~\cite{Sahay2026,SahayCultivationData2026}\\$\sim 4.45$ attempts} & $(122.131\pm0.037)^{\dagger}$ \\
CCX & $7.61\times10^8$ & $<\textcolor{revisionred}{2\times10^{-4}}$ & $<2.62\times10^{-13}$ & --- & --- \\
\makecell{Memory\\access} & $2.33\times10^7$ & $<\textcolor{revisionred}{10^{-3}}$ & $\textcolor{revisionred}{<2.77\times10^{-11}}$ & $10^{-3}$~\cite{Webster2026} & \makecell{$\textcolor{revisionred}{<84{,}885}^*$} \\
\bottomrule
\end{tabular*}
\caption{\textbf{QEC budget in DShor}. Counts and error budgets per RSA-2048 shot with six QPUs. The two-qubit rows include CNOT and CZ operations and their conditional corrections during the CCX gate teleportation. \textcolor{revisionred}{At $p_p=10^{-3}$, each operation category is allocated a per-shot error budget of at most $10^{-3}$, except intra-module CNOT/CZ, which is allocated $10^{-2}$.} \textcolor{revisionred}{The total allocated logical-error budget is $1.42\times10^{-2}$ per shot.} The decoding time is derived from $10^6$ decoder samples on an AMD Ryzen 9 6900HS laptop. $^\dagger$ The decoding time of $\mathrm{T}$ state cultivation measures total cultivation and escape decoding. $^*$The memory entry is the minimum time allowed in the DShor for one round of lattice surgery between the memory and six $d=25$ surface code patches.}
\label{tab:logical-operation-error-targets}
}
\end{table*}

\red{
\section{QEC budget in DShor}
\label{Appendix: QEC budget}

In this section, we will explain the QEC budget of all operations mentioned in Appendix~\ref{Appendix: gate time benchmarking}.

For DShor to factor 2048 bit RSA integers with six processing QPUs in Table~\ref{Table: Running time in different RSA}, Table~\ref{tab:logical-operation-error-targets} gives the estimated operation counts per shot and the target logical error for each gate. In this paper, we assume the common error $p_p$ of single-qubit depolarization, two-qubit depolarization, \textcolor{revisionred}{measurement-outcome classification errors}, and reset errors. We use $p_p=10^{-3}$ as the baseline and $p_p=10^{-4}$ as the higher-fidelity comparison. Since our target logical error is $\sim 10^{-13}$ for most logical operations, an accurate analysis of the logical error rate (LER) is computationally expensive, so all derived physical/logical errors are extrapolated from the data in other papers or local small-scale simulations. \textcolor{revisionred}{The total allocated logical-error budget is $1.42\times10^{-2}$ per shot.} In DShor, we categorize the logical operations in the following eight categories: Pauli gates (virtually implemented), inter/intra-module CNOT/CZ operations, $\mathrm{H}$, $\mathrm{S}$, $\mathrm{T}$, CCX, and memory access gate. The definition, error budgets and the source of extrapolation of physical/logical errors for each type of gate are as follows.

\begin{itemize}
    \item X/Y/Z: Pauli corrections are tracked classically, as described in the Appendix.~\ref{Appendix: gate time benchmarking}, so the LER analysis is skipped in Table~\ref{tab:logical-operation-error-targets}.

    \item inter/intra-module CNOT/CZ:
In DShor, we assume the intra-module CNOT/CZs are implemented by the transversal gates~\cite{Zhou2025Transversal,Cain2025Fast,Turner2026Transversal,SerraPeralta2026Transversal}. The inter-module CNOT/CZ is applied by gate teleportation from the logical Bell pair derived from $d^2$ physical Bell pairs.  For the resource estimate, we assume a common effective physical-error model and assign the same logical-error estimate to inter- and intra-module CNOT/CZs. Although inter-module gates have lower fidelity than intra-module gates, methods such as physical Bell-pair distillation or cultivation-inspired inter-module Bell-pair preparation can suppress the logical Bell-pair error rate to the required level~\cite{Pattison2024,Sunami2026}.

For the CNOT/CZ error budget, we use the surface-code memory LER of Ref.~\cite{Gu2026Neural} as the \textcolor{revisionred}{estimated error, similarly to} Ref.~\cite{Gidney2025}.   At $p_p=\textcolor{revisionred}{2\times10^{-3}}$, extrapolating the correlated-MWPM fit in Fig.~3(a) of Ref.~\cite{Gu2026Neural} to $d=25$ gives approximately $1.5\times10^{-13}$ per surface-code memory round. To extrapolate from $p_p=\textcolor{revisionred}{2\times10^{-3}}$ to $\textcolor{revisionred}{10^{-3}}$, we additionally assume the below-threshold scaling $p_L(d,p_p)\propto p_p^{(d+1)/2}$. At $d=25$, this gives $p_L\simeq(1.5\times10^{-13})/2^{13}\simeq1.8\times10^{-17}$, corresponding to estimated per-shot contributions of $\textcolor{revisionred}{1.1\times10^{-6}}$ for intra-module CNOT/CZ operations and $\textcolor{revisionred}{5.0\times10^{-8}}$ for inter-module CNOT/CZ operations. \textcolor{revisionred}{\hypersetup{allcolors=revisionred}We conservatively round the intra- and inter-module contributions up to $10^{-2}$ and $10^{-3}$ per-shot error budgets, respectively, in Table~\ref{tab:logical-operation-error-targets} for slack. The intra-module CNOT/CZ budget corresponds to a per-gate target of $1.68\times10^{-13}$, rounded down from $10^{-2}/(5.92\times10^{10})$.} \textcolor{revisionred}{\hypersetup{allcolors=revisionred}Other studies report logical errors of $\sim5\times10^{-14}$ (lower than our target logical error)~\cite{Ismail2026Transversal} and $\sim3.3\times10^{-13}$ (higher than our target logical error)~\cite{Cain2025Fast} under different error models.}

    \item $\mathrm{H}$/$\mathrm{S}$:
For $\mathrm{H}$ gates, we use correlated Blossom MWPM decoding~\cite{Higgott2025} with rotated surface-code circuits and the public QEC schedule of Ref.~\cite{SerraPeralta2026Transversal}. Transversal $\mathrm{H}$ is followed by a classical relabeling of the X/Z stabilizers and logical operators. We add one layer of $10^{-3}$ single-qubit depolarization to simulate decoherence during patch rotation. At $p_p=10^{-3}$, an exponential fit to $10^7$ trials per input basis at each distance $d=5,7,9,11,13$, each containing $d$ repeated $\mathrm{H}$-plus-QEC cycles with ideal encoded input and final readout, gives an effective per-cycle error $p_{\text{H}}\simeq2.56\times10^{-14}$ at $d=25$, taking the larger of the X- and Z-input estimates. With $N_{\text{H}}=1.85\times10^9$, the estimated contribution is $N_{\text{H}}p_{\text{H}}\simeq\textcolor{revisionred}{4.73\times10^{-5}}$ per shot. We allocate $\textcolor{revisionred}{10^{-3}}$ for slack, as in Table~\ref{tab:logical-operation-error-targets}.

$\mathrm{S}$ gates are only used to correct the $\mathrm{T}$ gate injection in the DShor. We use the Z-then-X MWPM decoder of Ref.~\cite{Chen2024RP2}. Extrapolating its Fig.~4 two-$\mathrm{S}$ benchmark gives an error of $3.03\times10^{-13}$ at $p_p=10^{-3}$ and $d=25$. The compiled circuit has $N_{\text{S}}\simeq3.07\times10^9$ $\mathrm{S}$ gates, approximately $N_{\text{T}}/2$ as expected. The estimated contribution is $N_{\text{S}}p_{\text{S}}\simeq\textcolor{revisionred}{9.3\times10^{-4}}$ per shot. We round it up to $\textcolor{revisionred}{10^{-3}}$ for slack.

    \item $\mathrm{T}$:
DShor uses $\mathrm{T}$ gates only in the error-detecting $8\mathrm{T}$-to-CCX factory~\cite{Jones2013}. We use the cultivation scheme of Ref.~\cite{Sahay2026}, with cultivation at code distances $3$ and $5$, followed by five stabilizer rounds at $d=11$ and unitary growth to $d=25$. At $p_p=10^{-3}$, the selected Clifford-state simulation data give an infidelity $p_T\simeq8.93\times10^{-8}$ and $4.45$ expected attempts per accepted state~\cite{SahayCultivationData2026}. We use the reported cultivated-state infidelity to estimate the $\mathrm{T}$-state error after growth to $d=25$, assuming that the final growth adds a negligible logical-error contribution as analyzed by Ref.~\cite{Sahay2026}.

Each $\mathrm{T}$ input to the CCX factory reserves a distance-25 patch with $2d^2-1=1249$ physical qubits. Using the peak footprints of $54$ qubits per initial candidate and $241$ per $d=11$ escape candidate in Table~A5 of Ref.~\cite{Sahay2026}, this allows $20$ parallel initial candidates and four survivors promoted to escape. Under the independent-candidate model, the probability of at least one accepted state per batch is $\simeq99.16\%$. Given this high success probability of T gate cultivation, we ignore the additional time overhead from the failed T gate cultivation.

    \item CCX:
Toffoli gates are distilled by the $8\mathrm{T}$-to-CCX factory for higher fidelity~\cite{Jones2013}. According to Ref.~\cite{Jones2013}, \textcolor{revisionred}{the error probability of a CCX gate is}
\begin{equation}
    p_{\mathrm{CCX}}\simeq28p_T^2
    \simeq2.23\times10^{-13}.
    \label{eq:CCX-error-probability}
\end{equation}
With $N_{\mathrm{CCX}}\simeq7.61\times10^8$ CCX gates per shot, the resulting error contribution is
\begin{equation}
    N_{\mathrm{CCX}}p_{\mathrm{CCX}}
    \simeq\textcolor{revisionred}{1.70\times10^{-4}}.
    \label{eq:CCX-error-contribution}
\end{equation}
We round this number up to $\textcolor{revisionred}{2\times10^{-4}}$ \textcolor{revisionred}{for slack.}
    \item Memory access:
Memory access is the operation to transfer the information from the quantum memory to the surface code, which can be achieved by either fanout or qubit teleportation. In DShor's compilation, \textcolor{revisionred}{less than $0.1\%$ of gates are memory-access operations}, where \textcolor{revisionred}{DShor fans out} the exponent qubits from the memory to the buffer surface codes. In this paper, we use the $[[510,16,24]]$ GB code as an example quantum memory, with a logical-measurement error of approximately $4\times10^{-16}$ per logical qubit per logical cycle at $p_p=10^{-3}$, extrapolated from the $d=4,6,10$ data in Fig.~6 of Ref.~\cite{Webster2026}. {\color{revisionred}
Even in the worst case where one syndrome extraction circuit (SEC) is applied to the GB memory every $1$ ms throughout a shot lasting $200$ days, the estimated total logical-error contribution from SECs is $\textcolor{revisionred}{3.53\times10^{-4}}$. We use the same LER analysis for lattice surgery between the GB memory and the surface code. Assuming $2.33\times10^7$ memory accesses per DShor shot, each requiring $26$ lattice-surgery rounds, the total logical-error contribution is approximately $(2.33\times10^7)(4\times10^{-16})=9.3\times10^{-9}$, which is negligible compared with the SEC contribution. We round the total memory-error contribution up to $\textcolor{revisionred}{10^{-3}}$ for slack.
}

Since the arithmetic compilation is independent of the memory-code family, the actual qLDPC code used in the DShor will only change the circuit time very weakly. Therefore, rather than reporting \textcolor{revisionred}{a lattice-surgery time depending} only on the GB code, we report the shortest interval observed between successive memory-loading layers  (\textcolor{revisionred}{consistently} $6$ logical qubits fanout from the memory to the \textcolor{revisionred}{surface codes) in} the sampled circuit: $\Delta_{\min}=2.2070\mathrm{s}$. {\color{revisionred}Assuming $26$ lattice-surgery rounds per fanout layer, with six surgeries happening in parallel, the average decoding time per surgery round on each core must satisfy $t_{\text{surgery}}<\Delta_{\min}/26\simeq84{,}885\,\mu\mathrm{s}$, as shown in Table~\ref{tab:logical-operation-error-targets}.}
\end{itemize}

The measured mean decoding times for $\mathrm{H}$, $\mathrm{S}$, intra-module CNOT, and cultivation at $p_p=10^{-3}$ are below $0.5\,\mathrm{ms}$ (Table~\ref{tab:logical-operation-error-targets}), shorter than both logical clock periods considered in this paper ($1.25\,\mathrm{ms}$ and $0.5\,\mathrm{ms}$). These benchmarks suggest that most logical operations will not be delayed by classical decoding, provided that syndrome-data collection and classical control do not introduce a larger latency.
}

\onecolumngrid
\begingroup

\small
\setlength{\tabcolsep}{6pt}
\renewcommand{\arraystretch}{1.18}
\begin{table}[H]
\centering
\makeatletter
\let\@parboxrestore\@arrayparboxrestore
\makeatother
\begin{tabular}{@{}ll@{}}
\toprule
\red{Symbol} & \red{Definition} \\
\midrule
\multicolumn{2}{l}{\red{\textbf{RSA and short discrete logarithms (Appendix.~\ref{Appendix: Ekera factoring})}}}\\
\parbox[t]{0.25\textwidth}{\strut \red{$N$, $n$}\strut} & \parbox[t]{0.71\textwidth}{\strut \red{Integer to be factored with $n=\lfloor\log_2 N\rfloor+1$ bits.}\strut} \\
\parbox[t]{0.25\textwidth}{\strut \red{$P,Q$; $\nu$}\strut} & \parbox[t]{0.71\textwidth}{\strut \red{Unknown distinct odd primes satisfying $N=PQ$; $\nu=\max\{\lfloor\log_2 P\rfloor,\lfloor\log_2 Q\rfloor\}+1$ is the larger factor bit length.}\strut} \\
\parbox[t]{0.25\textwidth}{\strut \red{$g$; $r_N,r_P,r_Q$}\strut} & \parbox[t]{0.71\textwidth}{\strut \red{$g$ is an integer coprime to $N$. For $R\in\{N,P,Q\}$, $r_R$ is the smallest positive integer satisfying $g^{r_R}\equiv1\pmod R$.}\strut} \\
\parbox[t]{0.25\textwidth}{\strut \red{$m_{\mathrm{log}}$}\strut} & \parbox[t]{0.71\textwidth}{\strut \red{Bound of the short logarithm: $0<d_{\mathrm{log}}<2^{m_{\mathrm{log}}}$. Equal-length factors permit $m_{\mathrm{log}}\ge\nu-1$; the reported resource estimates use $m_{\mathrm{log}}=\nu$.}\strut} \\
\parbox[t]{0.25\textwidth}{\strut \red{$x$, $d_{\mathrm{log}}$}\strut} & \parbox[t]{0.71\textwidth}{\strut \red{$x=g^{(N-1)/2-2^{\nu-1}}\bmod N$ is computed from known quantities. The short logarithm to be recovered is $d_{\mathrm{log}}$, satisfying $g^{d_{\mathrm{log}}}\equiv x\pmod N$ and $0<d_{\mathrm{log}}<2^{m_{\mathrm{log}}}$.}\strut} \\
\parbox[t]{0.25\textwidth}{\strut \red{$s$, $\ell$}\strut} & \parbox[t]{0.71\textwidth}{\strut \red{$s\ge1$ is the Eker{\aa}--H{\aa}stad tradeoff parameter; $\ell=\lceil m_{\mathrm{log}}/s\rceil$. The two exponent registers contain $m_{\mathrm{log}}+\ell$ and $\ell$ qubits.}\strut} \\
\parbox[t]{0.25\textwidth}{\strut \red{$m$}\strut} & \parbox[t]{0.71\textwidth}{\strut \red{Size of the input exponent registers in the  factoring algorithm: $m=m_{\mathrm{log}}+2\ell$ for the Eker{\aa}--H{\aa}stad algorithm, and $m = 2n$ in the naive Shor's algorithm.}\strut} \\
\midrule
\multicolumn{2}{l}{\red{\textbf{Residue arithmetic (Appendices.~\ref{Appendix: CRT} and \ref{Appendix: Approximate addition})}}}\\
\parbox[t]{0.25\textwidth}{\strut \red{$e$, $e_j$}\strut} & \parbox[t]{0.71\textwidth}{\strut \red{The exponent $e$ is represented in the binary form: $e=\sum_{j=0}^{m-1}2^j e_j$, where $e_j\in\{0,1\}$ and $e_0$ is the least significant bit.}\strut} \\
\parbox[t]{0.25\textwidth}{\strut \red{$F(e)$}\strut} & \parbox[t]{0.71\textwidth}{\strut \red{Known integer-valued function in the circuit $\ket e\ket0\mapsto\ket e\ket{g^{F(e)}\bmod N}$; $F(e)=e$ for ordinary order finding.}\strut} \\
\parbox[t]{0.25\textwidth}{\strut \red{$p_i$, $K$}\strut} & \parbox[t]{0.71\textwidth}{\strut \red{$K$ distinct known primes $p_0,\ldots,p_{K-1}$ used as residue moduli.}\strut} \\
\parbox[t]{0.25\textwidth}{\strut \red{$\lambda_i$}\strut} & \parbox[t]{0.71\textwidth}{\strut \red{bit size of the residue: $\lambda_i=\lceil\log_2 p_i\rceil$.}\strut} \\
\midrule
\multicolumn{2}{l}{\red{\textbf{Truncation and masking (Appendix.~\ref{Appendix: Approximate addition})}}}\\
\parbox[t]{0.25\textwidth}{\strut \red{$f$}\strut} & \parbox[t]{0.71\textwidth}{\strut \red{Accumulator width in qubits; each classically reduced $n$-bit contribution is truncated by discarding its $n-f$ least significant bits.}\strut} \\
\bottomrule
\end{tabular}
\red{\caption{Notations used in the Appendix.~\ref{Appendix: Ekera factoring}-\ref{Appendix: Approximate addition}}}
\label{tab:appendix_notation}
\end{table}
\par\endgroup
\twocolumngrid

\red{
\section{{Eker{\aa}--H{\aa}stad's factoring algorithm}}
\label{Appendix: Ekera factoring}

{In Section.~\ref{Chapter: residue arithmetic}, we introduce modular exponentiation using the standard order-finding formulation of Shor's algorithm~\cite{Shor1997},}
\begin{equation}
{\begin{gathered}
\frac{1}{2^n}\sum_{e=0}^{2^{2n}-1}\ket e\ket0
\longrightarrow
\frac{1}{2^n}\sum_{e=0}^{2^{2n}-1}\ket e\ket{g^e\bmod N}.
\end{gathered}}
\end{equation}

{The conventional order-finding circuit uses $2n$ input qubits for an $n$-bit modulus $N$. Eker{\aa} and H{\aa}stad instead reduce the factorization of an RSA modulus to a short discrete-logarithm problem~\cite{Ekera2017,Ekera2016}. For an \({m_{\mathrm{log}}}\)-bit short logarithm and a tradeoff factor \(s\), their quantum algorithm uses two exponent registers of \({m_{\mathrm{log}}}+\ell\) and \(\ell\) qubits, respectively, where}
\begin{equation}
{\ell=\left\lceil\frac{{m_{\mathrm{log}}}}{s}\right\rceil.}
\end{equation}
{The total exponent-register width is therefore}
\begin{equation}
{m=}{{m_{\mathrm{log}}}}{+2\ell
\approx}
{{m_{\mathrm{log}}}}{+\frac{2{m_{\mathrm{log}}}}{s}.}
\end{equation}
{For balanced RSA integers, \({m_{\mathrm{log}}}\approx n/2\), giving a total width of approximately}
\begin{equation}
{\frac{n}{2}+\frac{n}{s}.}
\end{equation}
{Increasing $s$ reduces the exponent-register width in each run at the cost of more classical repetitions of the quantum algorithm to collect enough information for order finding.}

\subsection{{Reduction of RSA factorization to a short discrete logarithm}}

{We will follow the proof from \cite{Ekera2017Post2}. Let $N=PQ$ be an RSA modulus, where \(P\) and \(Q\) are distinct odd primes of the same bit length $\nu$, and select}
\begin{equation}
{g\in\mathbb{Z}_N^*.}
\end{equation}
{Define}
\begin{equation}
{r_N=\operatorname{ord}_N(g),\qquad
r_P=\operatorname{ord}_P(g),\qquad
r_Q=\operatorname{ord}_Q(g).}
\end{equation}
{Since \(P\) and \(Q\) are coprime, the Chinese remainder theorem implies
$g^k\equiv 1\pmod N$
if and only if}
\begin{equation}
{g^k\equiv1\pmod P,
\qquad
g^k\equiv1\pmod Q.}
\end{equation}
{Consequently,}
\begin{equation}
{r_N=\operatorname{lcm}(r_P,r_Q).}
\end{equation}
{By Fermat's little theorem,}
\begin{equation}
{r_P\mid(P-1),
\qquad
r_Q\mid(Q-1),}
\end{equation}
{and hence}
\begin{equation}
{r_N
\mid
\operatorname{lcm}(P-1,Q-1).}
\end{equation}
{Define
$\widetilde P=\frac{P-1}{2}, \widetilde Q=\frac{Q-1}{2}$. Since
$P,Q$ are distinct odd prime integers, $2\mid\operatorname{gcd}(P-1, Q-1)$ and $\operatorname{lcm}(P-1, Q-1)\mid((P-1)(Q-1)/2)$,
which implies}
\begin{equation}
{g^{2\widetilde P\widetilde Q}
\equiv1\pmod N,
\label{eq:ekera_trivial_exponent}}
\end{equation}
{where}
\begin{equation}
{2\widetilde P\widetilde Q = (N-1)/2
-\widetilde P-\widetilde Q.}
\end{equation}
{and}
\begin{align}
{g^{(N-1)/2}
=
g^{2\widetilde P\widetilde Q}
g^{\widetilde P+\widetilde Q}
\equiv
g^{\widetilde P+\widetilde Q}
\pmod N.
\label{eq:ekera_unshifted_log_identity}}
\end{align}
{Define the unshifted short logarithm}
\begin{equation}
{d_0=\widetilde P+\widetilde Q.}
\end{equation}
{Hence, provided $r_N \geq d_0$, the known group element}
\begin{equation}
{x_0\equiv g^{(N-1)/2}\equiv g^{d_0}\pmod N}
\end{equation}
{has short discrete logarithm $d_0$ with respect to $g$.
The factors $P,Q$ can then be recovered from $P+Q=2d_0+2$ and $PQ=N$}.

{Later work~\cite{Ekera2020,Chevignard2024} reduces the bit length of the discrete logarithm by an additional bit by defining}
\begin{align}
    {x} &{\equiv g^{\kappa_N}\pmod N\nonumber}\\
    {{d_{\mathrm{log}}}} &{= \widetilde{P}+\widetilde{Q}-2^{\nu - 1}.}
\end{align}
{where
$
\nu=\operatorname{bitlen}(P)=\operatorname{bitlen}(Q)=\left\lceil\frac n2\right\rceil
$
for equal-length RSA factors, and $\kappa_N=\frac{N-1}{2}-2^{\nu-1}.$

Multiplying both sides of Eq.~\eqref{eq:ekera_unshifted_log_identity}
by $g^{-2^{\nu-1}}$ gives}
\begin{equation}
    {x\equiv g^{\kappa_N}\equiv g^{d_{\mathrm{log}}}\pmod N.}
\end{equation}

{Thus, the known exponent $\kappa_N$ lets us
compute $x$, while the unknown short exponent $d_{\mathrm{log}}$
encodes the factor sum $P+Q$ and is the quantity to be recovered.}

\subsection{{Quantum short-discrete-logarithm procedure}}

{The known inputs are $g$, $x$,
and a bound on the unknown logarithm. 

For equal-length RSA factors, $0<d_{\mathrm{log}}<2^{\nu-1}$. One may choose any integer bound $m_{\mathrm{log}}\ge\nu-1$ with $d_{\mathrm{log}}<2^{m_{\mathrm{log}}}$. For consistency with the reported resource parameters, we use the conservative choice $m_{\mathrm{log}}=\nu$: for RSA 2048 and $s=8$, this gives $\ell=128$ and $m=1024+2(128)=1280$.
Each quantum run uses two input registers storing integers $a$ and $b$.
According to~\cite{Ekera2017,Ekera2020}, for an integer tradeoff parameter $s\ge1$, set
$\ell=\lceil m_{\mathrm{log}}/s\rceil$: the $b$ register contains
$\ell$ qubits, and the $a$ register contains $m_{\mathrm{log}}+\ell$
qubits.

In a single run, prepare uniform superpositions over $a$ and $b$, then
compute $g^a x^{-b}\bmod N$ into the output register. The resulting state is}
\begin{equation}
    {\ket{\psi}=\frac{1}{\sqrt{2^m}}
    \sum_{a=0}^{2^{m_{\mathrm{log}}+\ell}-1}
    \sum_{b=0}^{2^\ell-1}
    \ket a\ket b\ket{g^a x^{-b}\bmod N}.
    \label{eq:ekera_pre_qft}}
\end{equation}
{\textcolor{revisionred}{The }$\pmod N$ notation is suppressed in this section for simplicity.
Applying the QFT to both input registers gives}
\begin{align}
    {\ket{\psi}}
    &{=\frac{1}{2^m}
    \sum_{a,j=0}^{2^{m_{\mathrm{log}}+\ell}-1}
    \sum_{b,k=0}^{2^\ell-1}\nonumber}\\
    &{\quad\times
    e^{\frac{2\pi i(aj+2^{m_{\mathrm{log}}}bk)}{2^{m_{\mathrm{log}}+\ell}}}
    \ket j\ket k\ket{g^a x^{-b}}.}
\end{align}
{For this derivation, assume} 
\begin{equation}
    {r_N\ge2^{m_{\mathrm{log}}+\ell}+2^\ell d_{\mathrm{log}},}
\end{equation}
{as in Ref.~\cite{Ekera2017}.

Measuring the group element then
identifies a unique exponent difference $\xi=a-bd_{\mathrm{log}}$
within the range generated by the input registers.
Define}
\begin{equation}
    {\mathcal B_\xi=\left\{b:\begin{array}{l}
        0\le b<2^\ell,\\
        0\le\xi+bd_{\mathrm{log}}<2^{m_{\mathrm{log}}+\ell}
    \end{array}\right\}.}
\end{equation}
{The resulting unnormalized input state is}
\begin{align}
    {\ket{\psi_\xi}}
    &{=\frac{1}{2^m}
    \sum_{j=0}^{2^{m_{\mathrm{log}}+\ell}-1}
    \sum_{k=0}^{2^\ell-1}
    e^{\frac{2\pi i\xi j}{2^{m_{\mathrm{log}}+\ell}}}\nonumber}\\
    &{\quad\times\left[
    \sum_{b\in\mathcal B_\xi}
    e^{\frac{2\pi ib(d_{\mathrm{log}}j+2^{m_{\mathrm{log}}}k)}{2^{m_{\mathrm{log}}+\ell}}}
    \right]\ket j\ket k.}
\end{align}
{Thus the joint probability of $(j,k)$ and $g^\xi$ is}
\begin{equation}
    {\mathcal P(j,k,\xi)=\frac{1}{2^{2m}}
    \left|\sum_{b\in\mathcal B_\xi}
    e^{\frac{2\pi ib(d_{\mathrm{log}}j+2^{m_{\mathrm{log}}}k)}{2^{m_{\mathrm{log}}+\ell}}}
    \right|^2.}
\end{equation}
{Constructive interference occurs when phase is close to $2z\pi$. \textcolor{revisionred}{Define the deviation} of the phase from $2z\pi$ as $\alpha$:}
\begin{equation}
    {d_{\mathrm{log}}j+2^{m_{\mathrm{log}}}k
    =z2^{m_{\mathrm{log}}+\ell}+\alpha,\qquad z\in\mathbb Z.}
\end{equation}
{The original analysis~\cite{Ekera2017} calls a Fourier pair $(j,k)$
good when $|\alpha|\le2^{m_{\mathrm{log}}-2}$.

Measuring the two input registers produces a Fourier sample $(j,k)$.
Samples from independent quantum runs are combined in classical lattice
post-processing to recover $d_{\mathrm{log}}$.
Increasing $s$ shortens both input registers but requires more samples:
the required number of runs scales approximately with $s$.
For fixed $s$ and large $m_{\mathrm{log}}$, the improved post-processing
analysis estimates $s+1$ runs when lattice enumeration is
avoided~\cite{Ekera2020}. The detailed sample requirements and classical recovery procedure are
given in Refs.~\cite{Ekera2016,Ekera2017,Ekera2020}.}

\section{{Modular exponentiations with residue arithmetic}}
\label{Appendix: CRT}

{We compile modular exponentiation using the residue-arithmetic methods of Refs.~\cite{Chevignard2024,Gidney2025}. The input registers control the evaluation of $g^{F(e)}\bmod N$, followed by the \textcolor{revisionred}{quantum} Fourier transformations. This appendix describes the CRT representation of that arithmetic. Appendix.~\ref{Appendix: Approximate addition} then introduces precomputed exponents, truncated accumulation, and superposition masking.

For the RSA modulus $N=PQ$, applying CRT directly to its factors
would require knowing the solution of the factoring problem. Instead,
choose $K$ known, distinct small primes $p_i$ and the auxiliary modulus}
\begin{equation}
    {L=\prod_{i=0}^{K-1}p_i.}
\end{equation}
{The Chinese remainder theorem gives}
\begin{equation}
    {\mathbb Z_L\cong\prod_{i=0}^{K-1}\mathbb Z_{p_i}.}
\end{equation}
{Define the CRT cofactor and its modular inverse by}
\begin{equation}
    {P_i=L/p_i,\qquad r_i=P_i^{-1}\bmod p_i.}
\end{equation}
{Any integer
$v$ can then be reconstructed modulo $L$ from its residues:}
\begin{equation}
    {v\equiv\sum_{i=0}^{K-1}(v\bmod p_i)P_i r_i\pmod L.}
\end{equation}
{The auxiliary modulus $L$ is selected so that 
\textcolor{revisionred}{the reconstructed value modulo $L$} can be converted into an approximation modulo $N$ as described
below.

Let $e=\sum_{j=0}^{m-1}2^j e_j$ encode the $m$ input bits, and
write the exponent as the bit-linear function}
\begin{equation}
    {F(e)=\sum_{j=0}^{m-1}e_j F(2^j).}
\end{equation}
{Here $m$ denotes the total input width: in Eker{\aa}--H{\aa}stad's factor algorithm \cite{Ekera2016},
$m=m_{\mathrm{log}}+2\ell$ and $e=a+2^{m_{\mathrm{log}}+\ell}b$.
In that case $F(e)=a-\kappa_N b$, so
$g^{F(e)}\equiv g^a x^{-b}\pmod N$; for ordinary order finding,
$F(e)=e$. The multipliers $M_j=g^{F(2^j)}\bmod N$ are therefore known
classically without knowing $d_{\mathrm{log}}$.
With these definitions,
the modular exponentiation can be written as}
\begin{align}
{g^{F(e)} \bmod N}
&{=
\left[
\prod_{i=0}^{m-1}
\left(g^{F(2^i)} \bmod N\right)^{e_i}
\right]
\bmod N}
\\
&{=
\left[
\prod_{i=0}^{m-1}
\left(g^{F(2^i)} \bmod N\right)^{e_i}
\right]
\bmod L \bmod N}
\\
&{=
\left(
\sum_{k=0}^{K-1} a_k r_k P_k
\right)
\bmod L \bmod N}
\\
&{\approx
\left[
\sum_{k=0}^{K-1}
\left(
a_k r_k P_k \bmod L \bmod N
\right)
\right]
\bmod N,
\label{eq:approx_crt_result}}
\end{align}
{where} 
\begin{equation}
{a_k =
\left[
\prod_{i=0}^{m-1}
\left(g^{F(2^i)} \bmod N\right)^{e_i}
\right]
\bmod p_k,
\label{eq: Logarithm}}
\end{equation}
{with the CRT coefficients $P_k$ and $r_k$ defined above.

The first three lines are exact because every multiplier satisfies $1\le M_i<N$ and every control bit satisfies $e_i\in\{0,1\}$. Therefore the entire unreduced product obeys}
\begin{equation}
    {1\le\prod_{i=0}^{m-1}M_i^{e_i}<N^m<L,}
\end{equation}
{so reducing this product modulo $L$ does not change it.

The approximation in Eq.~\ref{eq:approx_crt_result} arises from moving the reduction modulo \(L\) inside the CRT accumulation. 
Each omitted wraparound contributes $L$ modulo $N$. We measure its size by the normalized circular distance as defined in Ref.~\cite{Gidney2025}}
\[
    {\Delta_N(z)=\frac{\min(z\bmod N,(-z)\bmod N)}{N}.}
\]
{Thus one wraparound contributes $\Delta_N(L)$ to the modular deviation. Under \textbf{Assumption 1} of Ref.~\cite{Gidney2025}, the residue primes can be chosen such that}
\begin{equation}
{L=\prod_i p_i>N^m,
\qquad
\Delta_N(L)<2^{-f},}
\end{equation}
{in expected classical time}
\begin{equation}
{O\left(2^f\operatorname{poly}(mn)\right).}
\end{equation}
{Each reduced CRT term $a_k r_k P_k\bmod L$ lies in $[0,L)$,
so their sum lies in $[0,KL)$. Omitting its final reduction modulo $L$
therefore introduces an error of $qL$ modulo $N$, where
$0\le q\le K-1$. The normalized modular reconstruction error is bounded by}
\begin{equation}
    {\Delta_N(qL)\le q\Delta_N(L)
    \le (K-1)\Delta_N(L)<K2^{-f}.}
\end{equation}
    
{Therefore, the total error scales at most as $K\Delta_N(L)$ and
$\Delta_N(L)$ bounds the contribution of one omitted wraparound.}

\section{{Further reduction of residue arithmetic using approximate addition}}
\label{Appendix: Approximate addition}

{In this section, we introduce the additional simplifications of Shor's
algorithm required to reproduce the resource estimates of
Ref.~\cite{Gidney2025}}.

\subsection{{Replacing residue multiplications with additions}}

{In Appendix.~\ref{Appendix: CRT}, we show how an $n$-bit modular
exponentiation in $\mathbb{Z}_N$ can be reduced to multiple modular
exponentiations over the much smaller residue spaces
$\mathbb{Z}_{p_k}$, as summarized in Eq.~\ref{eq: Logarithm}.

The residue primes are chosen so that
$M_j\bmod p_k\ne0$ for every $j,k$, ensuring that the discrete
logarithms below exist. For each residue prime $p_k$, the residue}

\begin{equation}
    {a_k
    \equiv
    \prod_{j=0}^{m-1} M_j^{e_j}
    \pmod{p_k}}
\end{equation}
{would naively require $m$ controlled modular multiplications.
In Ref.~\cite{Gidney2025}, these multiplications are replaced by
controlled additions of precomputed exponents. Choose a multiplicative
generator $g_{p_k}$ of the small group $\mathbb{Z}_{p_k}^{*}$ and define}
\begin{equation}
    {\quad D_{k,j}
    =
    \operatorname{dlog}_{g_{p_k}}
    \left(M_j \bmod p_k\right)}
\end{equation}
{with $0\le D_{k,j}<p_k-1$, so that}
\begin{equation}
    {M_j
    \equiv
    g_{p_k}^{D_{k,j}}
    \pmod{p_k}.}
\end{equation}
{These are discrete logarithms in the small known modular space $\mathbb{Z}_{p_k}$.
The multipliers $M_j$ are known, so the exponents $D_{k,j}$ can be
tabulated classically before running the quantum circuit.
Consequently,}

\begin{align}
    {a_k}
    &{\equiv
    \prod_{j=0}^{m-1}
    \left(g_{p_k}^{D_{k,j}}\right)^{e_j}
    \pmod{p_k}
    \nonumber}\\
    &{\equiv
    g_{p_k}^{\sum_{j=0}^{m-1}D_{k,j}e_j}
    \pmod{p_k}.}
\end{align}
{Defining}
\begin{equation}
    {S_{p_k}
    =
    \sum_{j=0}^{m-1}D_{k,j}e_j,}
\end{equation}
{the original sequence of $m$ controlled modular multiplications is
therefore reduced to a sequence of $m$ controlled additions of
classically precomputed constants. Since the multiplicative order of
$g_{p_k}$ is $p_k-1$, only}
\begin{equation}
    {\tilde{S}_{p_k} = S_{p_k}\bmod(p_k-1)}
\end{equation}
{is required, and the residue can finally be recovered as}

\begin{equation}
    {a_k
    \equiv
    g_{p_k}^{\tilde{S}_{p_k}}
    \pmod{p_k}.
    \label{Eq. CRT exponentiation}}
\end{equation}

{This replacement reduces the number of additions needed to
compute each residue.
Before windowing, a multiplication modulo $p_k$ by a classical
constant can be implemented using $O(\lambda_k)$ controlled modular
additions. The original sequence of $m$ multiplications therefore
requires $O(m\lambda_k)$ additions. In contrast, forming $S_{p_k}$
requires only $m$ controlled direct additions and the unreduced sum $S_{p_k}$
fits in $\lambda_k+\lceil\log_2 m\rceil$ bits; reducing it modulo
$p_k-1$ requires only $O(\log m)$ comparison and subtraction steps
using binary long division~\cite{Gidney2025}. Recovering $a_k$
then requires an exponentiation with at most $\lambda_k$ exponent
bits, costing $O(\lambda_k)$ modular multiplications, or
$O(\lambda_k^2)$ additional modular additions. Thus the total
addition count per residue changes from}
\[
    {O(m\lambda_k)\quad\text{to}\quad O(m+\lambda_k^2).}
\]
{For $m=\Theta(n)$ and $\lambda_k=\Theta(\log n)$, this reduces the
count from $O(n\log n)$ to $O(n)$, giving a logarithmic
asymptotic saving.}

\subsection{{Truncating the CRT accumulation}}
\label{Appendix: CRT truncation}

{The remaining task is to accumulate the residues into an approximation
of $V(e)=g^{F(e)}\bmod N$. Although each $a_i$ is small, its CRT
contribution $a_i r_iP_i$  can have bit length comparable to that of $L$. These large
products need not be represented in quantum registers. First, consider
the reduced contribution}
\begin{equation}
    {c_i(a_i)=\bigl[(a_i r_iP_i)\bmod L\bigr]\bmod N.}
\end{equation}

{For an $n$-bit modulus $N$, choose an integer precision $1\le f\le n$ and define}
\begin{equation}
    {t=n-f,\qquad B=2^t,\qquad} {{N_{\mathrm{acc}}}}{=\left\lfloor\frac{N}{B}\right\rfloor.}
\end{equation}
{The precision parameter $f$ is the accumulator width, while
$F(e)$ denotes the exponent function. The reduced modulus
$N_{\mathrm{acc}}$ requires at most $f$ bits. Truncating a
contribution means replacing $c_i$ by
$\lfloor c_i/B\rfloor\bmod {N_{\mathrm{acc}}}$: we only keep its $f$ most significant
bits after the classical modular reductions, which leads to an error $\epsilon < B$ in each addition.

Another error caused by the truncation is the faster wraparound in the modular space $N_{\mathrm{acc}}$.
Write}
\[
    {N=BN_{\mathrm{acc}}+\delta_N,\qquad 0\le\delta_N<B.}
\]
Since {$BN_{\mathrm{acc}}\le N$, any integer $X$ satisfies}
\begin{equation}
    {X\bmod BN_{\mathrm{acc}}
    =\bigl(X\bmod BN_{\mathrm{acc}}\bigr)\bmod N.}
\end{equation}
{After rescaling by $B$, each reduction modulo $N_{\mathrm{acc}}$
contributes the normalized modular deviation}
\[
    {\Delta_N(BN_{\mathrm{acc}})
    =\Delta_N(\delta_N)
    \le\delta_N/N<B/N.}
\]
{Therefore, truncation error during the CRT reduction with $A$ additions is bounded by $O(AB)$.
Combining the errors from the truncation and CRT reduction in $\mathbb{Z}_L$, the error between the modular exponentiation $V(e)$ with modulo $N$ and the $\tilde{V}(e)$ with the truncated modulo \textcolor{revisionred}{is bounded by}}  
\begin{align}
    {\Delta_N\bigl(V(e)-B\widetilde V(e)\bigr)}
    &{\le (A-1)\Delta_N(L)+O(AB/N)\nonumber}\\
    &{=O(A2^{-f}),
    \label{eq:truncated_crt_error}}
\end{align}
{where $B/N=O(2^{-f})$ and $\Delta_N(L)<2^{-f}$ as in
Appendix.~\ref{Appendix: CRT}.}

\subsection{{Masking the approximate output}}

{Small modular deviation alone does not preserve the interference
required for period finding: distinct approximate outputs correspond to
orthogonal basis states, even for inputs with the same exact output.
Following Gidney~\cite{Gidney2025}, initialize the accumulator with}
\begin{equation}
    {\ket{\mu_W}=\frac{1}{\sqrt W}\sum_{u=0}^{W-1}\ket{u},
    \qquad W=\lceil\sigma N_{\mathrm{acc}}\rceil,}
\end{equation}
{where $0<\sigma<1$ and $1\le W<N_{\mathrm{acc}}$.
For $W=2^{b_{\mathrm{mask}}}$ and $\sigma=W/N_{\mathrm{acc}}$, this is}
\begin{equation}
    {\ket{\mu_W}
    =\ket{0}^{\otimes(f-b_{\mathrm{mask}})}
     \otimes\ket{+}^{\otimes b_{\mathrm{mask}}},}
\end{equation}
{with the most significant bits on the left. The additions shift this
state by $\widetilde V(e)$ modulo $N_{\mathrm{acc}}$, without storing $u$
in a separate register.

Choose $f$ so that $O(A2^{-f})\ll\sigma$. Masked states for inputs
with the same exact output then strongly overlap, approximately
preserving their coherence. Larger masks tolerate more arithmetic
error but reveal less information for period finding; their effect
on the success probability is analyzed in Ref.~\cite{Gidney2025}.
After clearing the residue workspace, measure the accumulator and
Fourier transform the input registers.}
\par
}

\section{Fast modular exponentiation with windowed arithmetic}
\label{Appendix: Windowed Arithmetic}

Modular exponentiations \red{in the residue number system (RNS)} are the most important subroutines of Shor's algorithm \red{as discussed in Appendix.~\ref{Appendix: CRT}. From Eq.~\ref{Eq. CRT exponentiation}, the \textcolor{revisionred}{canonical modular} exponentiation in the factoring algorithm is}

\red{\begin{equation}
{\begin{aligned}
    &U_g\sum_{e=0}^{2^m-1}
    \ket{e}\ket{\tilde{S}_{p_k}(e)}\ket0\\
    &\quad=\sum_{e=0}^{2^m-1}\ket{e}
    \ket{\tilde{S}_{p_k}(e)}
    \ket{g_{p_k}^{\tilde{S}_{p_k}(e)}\bmod p_k}.
\end{aligned}}
\end{equation}}
A naive exponentiation can use a single lookup operation to load all results to the registers~\cite{Gidney2021}. Then, factoring \textcolor{revisionred}{an n-bit number} using the residue arithmetic requires a lookup table \red{addressed by the $\lambda_k$-qubit register $\ket{\tilde{S}_{p_k}(e)}$, which needs $2^{\lambda_k} - \lambda_k - 1$ \textcolor{revisionred}{Toffoli gates.} Given $\lambda_k \approx  24$ to factor \textcolor{revisionred}{a $2048$-bit RSA integer}, the modular exponentiation  requires $\approx 1.6\times 10^{7}$ Toffoli gates and a QROM table up to $2^{24} \approx 1.6 \times 10^7$ entries. However, \textcolor{revisionred}{replacing direct QROM with quantum arithmetic can significantly improve} the circuit volume of the modular exponentiation and therefore reduce the circuit \textcolor{revisionred}{volume} of the factoring algorithm}. 

Modular exponentiations \red{in $\mathbb{Z}_{p_k}$ over a \textcolor{revisionred}{$\lambda_k$-bit quantum exponent} $\ket{\tilde{S}_{p_k}(e)} = \ket{\tilde{e}}$} can be decomposed to modular multiplications:
\red{\begin{align}
    &{g_{p_k}^{\tilde{e}}\bmod p_k = \left[\prod_{i=0}^{\lambda_k-1}(g_{p_k}^{2^i\tilde{e}_i})\bmod p_k\right]\bmod p_k.}
\end{align}}
The modular multiplication between quantum number (a) and the classical number (b) is further decomposed to modular addition:
\red{\begin{align}
    {(a \times b)\bmod p_k} & {= \left [\left(\sum_{i=0}^{\lambda_k-1} a_i2^i\right)\times  b\right ] \bmod p_k}\\
    &{= \left[\sum_{i=0}^{\lambda_k-1} \left(a_i 2^{i}b\right) \bmod p_k \right] \bmod p_k.}
\end{align}}
\red{The modular exponentiation in $\mathbb{Z}_{p_k}$ requires $\lambda_k$ modular multiplications, or equivalently $\lambda_k^2$ modular additions. Therefore, modular exponentiation requires $O(\lambda_k^3)$ Toffoli gates ($1$ modular exponentiation $\rightarrow$ $\lambda_k$ modular multiplications $\rightarrow$ $\lambda_k^2$ modular additions $\rightarrow O(\lambda_k^3)$ Toffoli gates) compared with the $O(2^{\lambda_k})$ Toffoli gates used by direct loading.}

\section{Circuit compilation of windowed arithmetic}
\label{Appendix B: Compilation of arithmetic}
\begin{figure*}[t]
    \includegraphics[width=\linewidth]{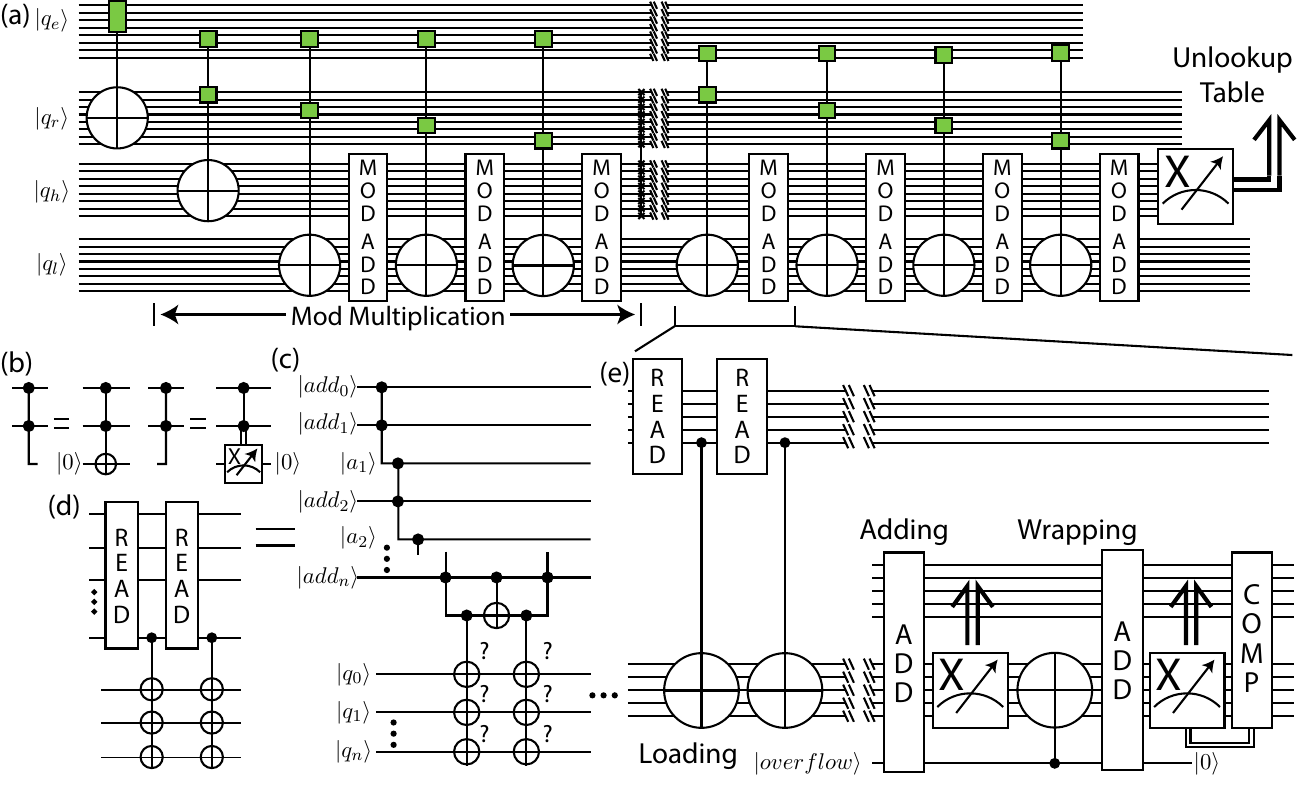}
	\caption{
        \red{\textbf{Modular exponentiations in DShor}}. (a) The modular exponentiations can be decomposed to loadings and modular additions. The addresses of loadings are represented by the light green squares. The first two modular multiplications \red{and the first addition in the modular multiplication iteration} are applied by direct loadings. (c) The circuit of unary loadings used in DShor with the logical AND and UNAND gates in (b). (d) For illustration, we simplify the cascades of AND and UNAND gates between loadings as read gates. (e) The detailed circuit of modular additions. Classical tables are loaded to $\ket{q_l}$ followed by the direct additions. $\ket{q_l}$ is cleaned by the transversal X-measurement, and the measurement result is recorded to generate the unlookup table. Another loading controlled by the overflow qubit $\ket{q_o}$ is loaded to the $\ket{q_l}$ followed by another direct addition to wrap the addition result. $\ket{q_l}$ is again cleaned by X-measurement. These two unlookup tables are merged during the un-exponentiations.
    }
    \phantomsection
    \label{fig:compilation of arithmetic}
\end{figure*}

\red{The modular exponentiation in Appendix.~\ref{Appendix: Windowed Arithmetic} can be significantly improved by windowed arithmetic~\cite{Gidney2025,Gidney2021}.  Split the $\lambda_k$-bit exponent into $w_e$-bit slices $\tilde e_{S_m}=\tilde e_{mw_e:(m+1)w_e}$, taking bits beyond $\lambda_k$ to be zero. The classically precomputed multipliers give}
\red{\begin{align}
    {L_m}&{=g_{p_k}^{2^{mw_e}\tilde e_{S_m}}\bmod p_k,\nonumber}\\
    {g_{p_k}^{\tilde e}\bmod p_k}
    &{=\left[\prod_{m=0}^{\lceil\lambda_k/w_e\rceil-1}
    L_m\right]\bmod p_k.
    \label{eq:windowed_residue_product}}
\end{align}}
\red{Each multiplication is decomposed into additions controlled by $w_m$-bit slices of the current product. A lookup therefore has at most $w_e+w_m$ address bits and costs $O(2^{w_e+w_m})$ Toffoli gates, while a modular addition costs $O(\lambda_k)$ Toffoli gates. The resulting Toffoli-count bound is}
\red{\begin{equation}
    {O\!\left(
    \left\lceil\frac{\lambda_k}{w_e}\right\rceil
    \left\lceil\frac{\lambda_k}{w_m}\right\rceil
    \left[\lambda_k+2^{w_e+w_m}\right]\right).
    \label{eq:windowed_residue_cost}}
\end{equation}}
\red{Choosing both window widths near $\tfrac12\log_2\lambda_k$ gives $O(\lambda_k^3/\log^2\lambda_k)$ Toffoli gates. Figure~\ref{fig:compilation of arithmetic} shows the lookup and modular-addition circuits}.

Modular exponentiation can be decomposed to modular additions from Appendix\red{.}~\ref{Appendix: Windowed Arithmetic}. For simplicity, we assume all arithmetic is in the modular space  \red{$\mathbb{Z}_{p_k}$}, and the quantum modular exponentiation can be written as:
\red{\begin{align}
    &{\sum_{\tilde e}\ket{\tilde e}\ket{g_{p_k}^{\tilde e}\bmod p_k}
    \nonumber}\\
    &{=\sum_{\tilde e}\ket{\tilde e}
    \ket{\left[\prod_{m=0}^{\lceil\lambda_k/w_e\rceil-1}
    g_{p_k}^{2^{mw_e}\tilde e_{S_m}}\right]\bmod p_k}}\\
    &{=\sum_{\tilde e}\ket{\tilde e}
    \ket{\left[\prod_{m=0}^{\lceil\lambda_k/w_e\rceil-1}L_m\right]\bmod p_k}.}
\end{align}}
where \red{$L_m=g_{p_k}^{2^{mw_e}\tilde e_{S_m}}\bmod p_k$ is loaded from a one-dimensional quantum lookup table addressed by $\ket{\tilde e_{S_m}}$. Write $J_e=\lceil\lambda_k/w_e\rceil$ and $J_m=\lceil\lambda_k/w_m\rceil$ for the numbers of exponent and product windows. For $J_e\ge2$, the multiplications decompose into additions as}
\red{\begin{align}
    &{\left[\prod_{m=0}^{J_e-1}L_m\right]\bmod p_k
    \nonumber}\\
    &{=\left[\mathcal P_1L_1\prod_{m=2}^{J_e-1}L_m\right]\bmod p_k}\\
    &{=\left[\sum_{a=0}^{J_m-1}2^{aw_m}(\mathcal P_1)_{S_a}L_1
    \prod_{m=2}^{J_e-1}L_m\right]\bmod p_k}\\
    &{=\left[\sum_{a=0}^{J_m-1}T_{1,a}
    \prod_{m=2}^{J_e-1}L_m\right]\bmod p_k}\\
    &{=\left[\sum_{a=0}^{J_m-1}2^{aw_m}
    (\mathcal P_{J_e-1})_{S_a}L_{J_e-1}\right]\bmod p_k}\\
    &{=\left[\sum_{a=0}^{J_m-1}T_{J_e-1,a}\right]\bmod p_k,}
\end{align}}
where \red{$\mathcal P_j=\left[\prod_{m=0}^{j-1}L_m\right]\bmod p_k$, $\mathcal P_0=1$, and $(\mathcal P_j)_{S_a}$ is its $w_m$-bit slice at positions $aw_m:(a+1)w_m$, with bits beyond $\lambda_k$ set to zero. The table entry $T_{j,a}=\left[2^{aw_m}(\mathcal P_j)_{S_a}L_j\right]\bmod p_k$ is loaded from a two-dimensional quantum lookup table addressed by the product slice $\ket{(\mathcal P_j)_{S_a}}$ and exponent slice $\ket{\tilde e_{S_j}}$.}

As the fundamental operations of modular exponentiation are modular additions, we will first provide the circuit compilation of modular addition based on Gidney's compilation~\cite{Gidney2025}. Modular additions in DShor \textcolor{revisionred}{consist of} four steps: loading, adding, wrapping and cleaning as shown in Fig. \ref{fig:compilation of arithmetic}(e).
\red{\begin{align}
    {U_{add}\sum_{i}\ket{i}\ket{a}} &{\rightarrow\sum_i \ket{i}\ket{a}\ket{f(i)}}\\ 
    &{\rightarrow\sum_i \ket{i}\ket{[a+f(i)]\bmod p_k}}
\end{align}}
The first line is to load the classical information $f(i)$ into buffer qubits,
where $f(i)$ is \textcolor{revisionred}{the table value} associated with the quantum address $\ket{i}$. In the compilation of DShor, we prefer to reduce time cost at the expense of space cost, so the values are loaded by sequential QRAM as shown in Fig.~\ref{fig:compilation of arithmetic}(c, d). We also provide the compilation of modular additions with multi-address QRAM on the dynamic array in Appendix.~\ref{Appendix: dynamic array}, which can also be applied on other architectures. The first two modular multiplications are replaced by direct loadings as shown in Fig.~\ref{fig:compilation of arithmetic}~\cite{Luongo2025}.

Since all additions are in the modular space of \red{$p_k$}, instead of decomposing modular additions to three or five direct additions, we introduce a wrapping step to decompose the modular addition with two modular additions at the price of one additional overflow qubit~\cite{Vedral1996, Beckman1996, Gidney2021, Gidney2025,VanMeter2005,Liu2023}.

\red{The residue register has $\lambda_k$ bits plus one overflow qubit (total $\lambda_k+1$ bits)}. To perform the wrapping, we replace all table values in the additions in the modular space of \red{$p_k$ to:}
\red{\begin{align}
    {\ket{i}\ket{(a+f(i)) \bmod p_k} \rightarrow \ket{i}\ket{a+f(i) - p_k + 2^{\lambda_k}}}
\end{align}}
If \red{$a+f(i) \geq p_k$, $a+f(i) - p_k + 2^{\lambda_k} \geq 2^{\lambda_k}$} and the overflow bit $\ket{q_o}$ must be $1$. Otherwise, \red{$a+f(i) - p_k + 2^{\lambda_k} < 2^{\lambda_k}$} and the overflow must be $0$. Therefore, we can determine if a ``wrapping" should happen during the addition and apply it by a single address lookup table \red{$\{0:2^{\lambda_k} + p_k, 1:2^{\lambda_k}\}$}, or equivalently a cascade of CNOTs shown in Fig. \ref{fig:compilation of arithmetic}(e):
\red{\begin{alignat}{2}
     &{\ket{i}\ket{a+f(i) - p_k + 2^{\lambda_k}} \notag}\\
     &{=\ket{i}\ket{a+f(i) - p_k + 2^{\lambda_k}, q_o = 1}}&{\,(a+f(i) \geq p_k) \notag}\\
     &{+\ket{i}\ket{a+f(i) - p_k + 2^{\lambda_k}, q_o = 0}}&{\,(a+f(i) < p_k) \notag}\\
     &{\rightarrow \ket{i}\ket{a+f(i) - p_k + 2^{\lambda_k+1}, q_o = 0} \notag}\\
     &{+ \ket{i}\ket{a+f(i) - p_k + 2^{\lambda_k+1}+p_k, q_o = 0} \notag}\\
     &{=\ket{i}\ket{a+f(i)-p_k}}&{\,(a+f(i) \geq p_k) \notag}\\
     &{+\ket{i}\ket{a+f(i)}}&{\,(a+f(i) < p_k) \notag}\\
\end{alignat}}
One trick we apply here is that the direct addition with \red{$\lambda_k+1$ bits is implicitly in the modular space of $2^{\lambda_k+1}$. Therefore, $2^{\lambda_k+1}$} is 0 in the direct addition.

The quantum information in the buffer qubits $\ket{q_l}$ is cleaned by a transversal X-basis measurement after the addition as shown in Fig. \ref{fig:compilation of arithmetic}(e):
\red{\begin{align}
    {\ket{i}\ket{a+f(i)}\ket{f(i)} \rightarrow (-1)^{\pi (f(i), M)}\ket{i}\ket{a+f(i)}}
\end{align}}
where $M$ is the measurement result. The cleaning of phase errors is combined with other memory cleaning in the Shor's algorithm shown in Appendix\red{.}~\ref{Appendix: Verification of DShor}.

\section{Distributed modular addition in DShor}
\label{Appendix: Distributed addition}
In the DShor, all qubits storing exponents ($Q_{exponent}$) and final results ($Q_{acc}$) are stored in the memory. \red{During short-logarithm loading, six exponent-memory qubits are accessed in parallel per loading layer. The shortest interval observed between consecutive layers in the sampled six-QPU circuit is $2.21$ s (Appendix.~\ref{Appendix: QEC budget}).} The other qubits actively used during modular exponentiation are distributed evenly through all QPUs as shown in Table \ref{Table: Qubit usage in Shor}.

\label{Appendix: Modular addition}

The naive compilation of Shor's algorithm on a modular quantum computer requires \red{$\lambda_k$} inter-module CNOTs controlled by the control qubit (the green register) as shown in Fig.~\ref{fig:Modular addition}. To reduce inter-module communications, we use one GHZ state to pre-connect all modules, and the table values are loaded by broadcasting the controlled qubit to all qubits in the GHZ state~\cite{Zhou2025}. This reduces the number of inter-module CNOTs per loading from \red{$\lambda_k - \lambda_k/N_{\text{QPU}}$} to $N_{\text{QPU}}$, and the burden of inter-module communications is evenly distributed to all QPUs. The modular lookup table can reduce inter-module connections if \red{$N_{\text{QPU}}<\frac{1}{2}\sqrt{\lambda_k^2 - 4\lambda_k} + \lambda_k/2 \approx \lambda_k$}. Therefore, DShor at least requires $2$ qubits of each category per QPU. In RSA $2048$, \red{$\lambda_k = 24$} and DShor can be distributed to at most $12$ QPUs.

\begin{figure}[t!]
    \includegraphics[width=\linewidth]{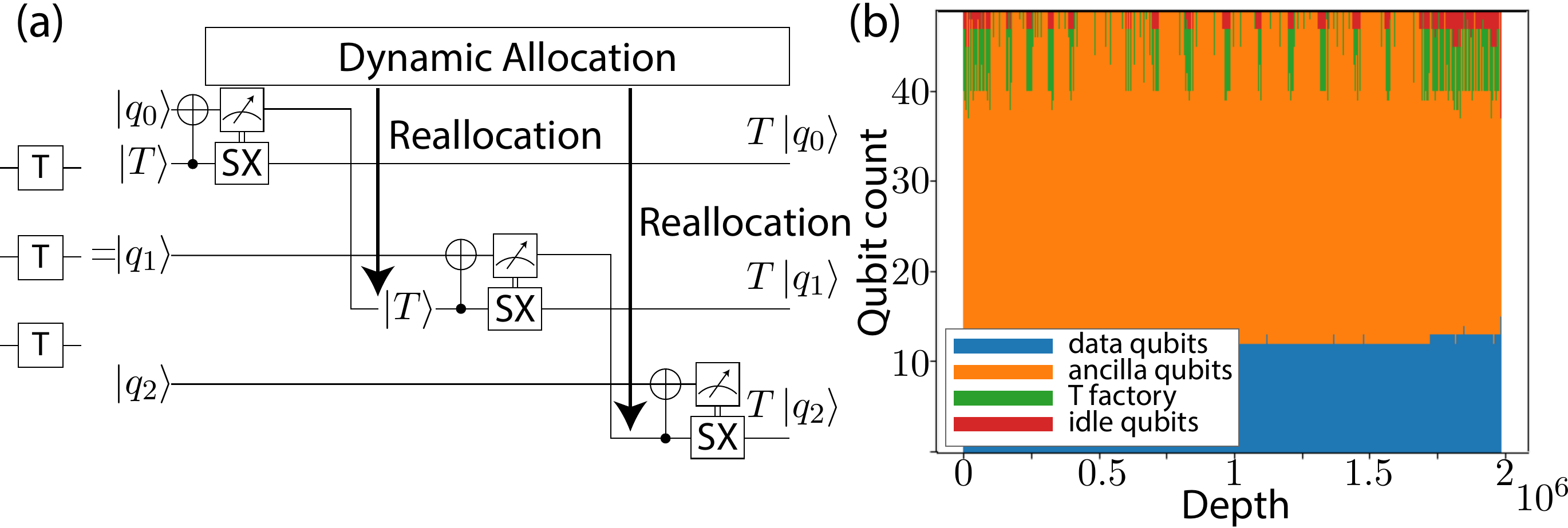}
	\caption{
        \textbf{Qubit re-allocation in DShor}. DShor marks idle qubits and reallocates them when a new qubit is requested. (a) Minimal example: three $\mathrm{T}$ gates reuse one ancilla after each magic-state injection. (b) Allocation in \red{one processing QPU during the representative $24$-bit residue-arithmetic block for RSA-2048 with six QPUs, $d=25$, $t_{\text{mea}}=1$ ms, and $\tau=25$. Short-logarithm loading and compression are excluded.} Qubits are dynamically shared between \red{cultivation, factory workspace, and arithmetic registers}.
    }
    \phantomsection
    \label{fig:qubit reallocation}
\end{figure}

The \red{$8\mathrm{T}$-to-CCX factory} also requires intensive inter-module communications if the control qubits and the target qubit are not in the same QPU. During the address reading of the quantum lookup table, the address qubits are teleported to ensure that all AND and UNAND gates are applied within one QPU as shown in Fig.~\ref{fig:Modular addition}(b). A similar technique is used in the addition ladders~\cite{Gidney2018}. 
Cleanings are implemented during the un-exponentiations, and the cleanings of phase errors can be simplified to quantum loadings with only one target~\cite{Gidney2021, Gidney2025}. To further reduce the cost of cleaning, we split \red{the local binary lookup address into two approximately equal parts and then one-hot encode each part}~\cite{Gidney2021, Gidney2025}. After decomposing cleanings to loadings, the distribution of cleanings is the same as the distribution of loadings.

\begin{figure*}[t]
    \includegraphics[width=\linewidth]{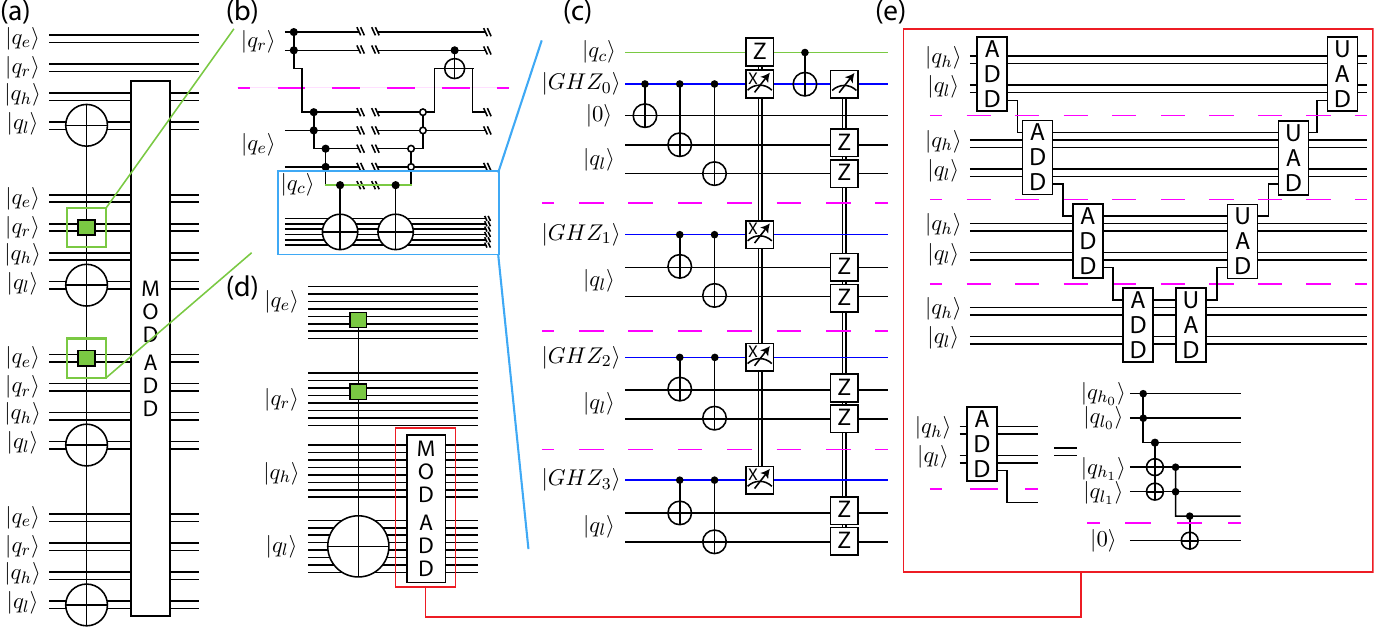}
	\caption{
        \red{\textbf{Detailed circuit of distributed modular loadings/additions}}. (a-c) The detailed circuit of modular loadings. The green squares are addresses. The naive circuit of loadings in (a) is illustrated in (b), where addresses are accessed sequentially. The qubit recording the result of the memory reading $\ket{q_c}$ is labeled by the green line. In the modular loadings, all addresses are teleported to the same QPU to avoid the intense communications in the \red{$8\mathrm{T}$-to-CCX factory}. \textcolor{revisionred}{The cascade of CNOTs in the blue box is split into} multiple pre-loadings as illustrated in (c). \textcolor{revisionred}{The reading result} $\ket{q_c}$ is teleported to \textcolor{revisionred}{all QPUs through} qubits in one GHZ state. The modular addition in (d) can be distributed with nearest-neighbor communications as illustrated in (e). The purple lines label the boundary of each QPU. The overflow qubits are teleported to the neighbor QPU during the addition ladder, and teleported back to the original QPU during the un-addition ladder.
    }
    \phantomsection
    \label{fig:Modular addition}
\end{figure*}

\section{Qubit Reallocation}
\label{Appendix: Reallocation}
\red{As mentioned in \textcolor{revisionred}{Section}~\ref{Chapter: Introduction}, QCPU and QGPU describe architectural analogies: a QCPU distributes multi-bit arithmetic across a small number of processing modules, whereas a QGPU uses many smaller modules to execute independent modular exponentiations in parallel. In both architectures, a cache is a small zone of fast local memory within each processing QPU, holding frequently accessed intermediate registers. Local access to these registers avoids transfers to the separate quantum-memory module.}

Since atoms can be flexibly rearranged on the neutral atom platform, all measured (clean) qubits are marked as idle (except some intermediate results during the modular exponentiations, such as $q_h$). Instead of requiring more qubits in total when one subroutine asks for more qubits, QPUs will first re-allocate the idle qubits. 
One simple example of qubit reallocation is illustrated in Fig.~\ref{fig:qubit reallocation}(a), where $3$ $\mathrm{T}$ gates can be injected with only $1$ ancilla qubit but with triple circuit depth. In DShor, most reallocations happen between the qubits in \red{$8\mathrm{T}$-to-CCX factory} and those in caches as shown in Fig.~\ref{fig:qubit reallocation}(b). This reallocation can increase the parallelism in each QPU and further reduce the spacetime volume of DShor.

\section{Verification of DShor}
\label{Appendix: Verification of DShor} Verifying the circuit of modular exponentiation in Appendix\red{.}~\ref{Appendix B: Compilation of arithmetic} is too computationally challenging for the quantum circuit simulation packages (e.g. Qiskit), since they track all components in the Hilbert space. For example, according to Table \ref{Table: Qubit usage in Shor},  modular exponentiation in the $2048$-bit RSA integers requires at least $99$ qubits, which requires at least $2^{99}\approx 10^{30}$ 99-bit numbers to exactly simulate the circuit, which is beyond the capability of classical computation. 

To efficiently verify our modular exponentiations and the corresponding circuits, we design an algorithm that only tracks very few samples in the Hilbert space. \red{As in Gidney's simulator~\cite{Gidney2025}, we track sampled computational-basis values and phases to check arithmetic outputs and measurement-based uncomputation. In his package, arithmetic simulation and primitive circuit checks are implemented separately, and resource estimates use operation counts and analytical models from the primitive circuits, rather than scheduling a complete generated circuit. In DShor, the same arithmetic calls update both the verification backend and emit the corresponding circuit operations. \textcolor{revisionred}{The shared calls from algorithm and circuit ensure} the \textcolor{revisionred}{consistency} between the algorithm and the simulated circuit. Moreover, we assign qubits to individual QPUs to simulate the inter-module communication latency and schedule operations using gate durations based on our hardware model.

We use the local modular exponentiation $g^{\tilde e}\bmod p_k$, with $g=g_{p_k}$ and a $\lambda_k$-bit exponent $\tilde e$, as an example:}
\red{\begin{equation}
{\sum_{\tilde e}\ket{\tilde e}\rightarrow\sum_{\tilde e}\ket{\tilde e}\ket{g^{\tilde e}\bmod p_k}
\label{eq:verification_modexp_map}}
\end{equation}}
For simplicity, we will only track one single Hilbert trajectory in the modular exponentiation: \red{$\ket{\tilde e}$. The first two exponent windows are combined by directly loading $\mathcal P_2=g^{\tilde e_{0:2w_e}}\bmod p_k$ into $q_r$}:
\red{\begin{equation}
{\ket{\tilde e}\rightarrow\ket{\tilde e}\ket{\mathcal P_2}_{q_r}\ket{0}_{q_h}
\label{eq:verification_initial_product}}
\end{equation}}
\red{As in Appendix.~\ref{Appendix B: Compilation of arithmetic}, $T_{2,a}$ denotes the table entry selected by $(\mathcal P_2)_{S_a}$ and $\tilde e_{S_2}$.} $T_{2,0}$ is loaded to $\ket{q_l}$ and added to $\ket{q_h}$:
\red{\begin{equation}
{\ket{\tilde e}\rightarrow\ket{\tilde e}\ket{\mathcal P_2}_{q_r}\ket{T_{2,0}}_{q_h}\ket{T_{2,0}}_{q_l}
\label{eq:verification_lookup_add}}
\end{equation}}
\textcolor{revisionred}{$\ket{q_l}$ is cleaned} by the transversal X measurement, reset to $\ket{0}$, and the measurement result $M_0$ is recorded:
\red{\begin{equation}
{\begin{aligned}[b]
&\ket{\tilde e}\ket{\mathcal P_2}_{q_r}\ket{T_{2,0}}_{q_h}\ket{T_{2,0}}_{q_l}\\
&\quad\longrightarrow(-1)^{\pi(M_0,T_{2,0})}\ket{\tilde e}\ket{\mathcal P_2}_{q_r}\ket{T_{2,0}}_{q_h}\ket{0}_{q_l}
\end{aligned}
\label{eq:verification_lookup_cleanup}}
\end{equation}}
The negative phase is corrected during the modular un-exponentiation. All \red{entries of $T_{2,a}$} are loaded in the same way, and all measurement results are recorded as \red{$M_a$}. The final state is:
\red{\begin{equation}
{\ket{\psi}=(-1)^{\sum_a \pi(M_a, T_{2,a})}\ket{\tilde e}\ket{\mathcal P_2}_{q_r}\ket{\mathcal P_3}_{q_h}\ket{0}_{q_l}
\label{eq:verification_product_state}}
\end{equation}}
Here we use
\red{\begin{equation}
{\begin{aligned}[b]
\left[\sum_{a=0}^{J_m-1}T_{2,a}\right]\bmod p_k
    &=[\mathcal P_2L_2]\bmod p_k\\
    &=\mathcal P_3.
\end{aligned}
\label{eq:verification_modular_product}}
\end{equation}}
$\ket{q_r}$ and $\ket{q_h}$ are swapped, and $\ket{q_h}$ is cleaned by transversal X measurement. The measurement result is recorded as \red{$\tilde M_2$.}
\red{\begin{equation}
{\ket{\psi}=(-1)^{\textcolor{revisionred}{\pi(\mathcal P_2, \tilde M_2)}+\sum_a \pi(M_a, T_{2,a})}\ket{\tilde e}\ket{\mathcal P_3}_{q_r}\ket{0}_{q_h}\ket{0}_{q_l}
\label{eq:verification_helper_cleanup}}
\end{equation}}
This loop is repeated with all $T_{j,:}$, and the final state is:
\red{\begin{equation}
{\ket{\psi}=(-1)^{\phi}\ket{\tilde e}\ket{g^{\tilde e}\bmod p_k}_{q_r}\ket{0}_{q_h}\ket{0}_{q_l}
\label{eq:verification_final_state}}
\end{equation}}
where $\phi$ depends on all measurement results of qubit cleaning. The un-exponentiation is applied by running the exponentiation inversely with phase-up cleaning at each loop~\cite{Gidney2025}.

To simulate the RSA $2048$ challenge, we start with $2^{5}$ random $24$-bit \red{$\ket{\tilde e}$} and \textcolor{revisionred}{$3$ groups of} $\ket{0}^{\otimes 25}$ for $\ket{q_r},\ket{q_h}$ and $\ket{q_l}$. As shown in Fig.~\ref{fig:Modular addition}, each addition is split to loading, adding, wrapping, and cleaning. Each Hilbert trajectory of \red{$\ket{\tilde e}$} is updated at each of these four steps \textcolor{revisionred}{similarly to} the example shown above. All measurement results are recorded and used in the modular un-exponentiation. At the same time, each time the modular addition happens, a circuit of modular addition will be applied on the corresponding quantum registers. Therefore, the parallel evolution of the quantum circuit and the back-end Hilbert trajectory guarantees the consistency between the algorithm and the circuit. We collect all qubit usage and estimate the circuit depth and qubits used from the circuit evolved during the Shor's algorithm.

More details about the estimation methods of the Shor's simulator can be found in Appendix\red{.}~\ref{Appendix: Estimation method}.



\section{Estimation methods in DShor}
\label{Appendix: Estimation method}
\begin{figure*}[t]
    \includegraphics[width=0.9\linewidth]{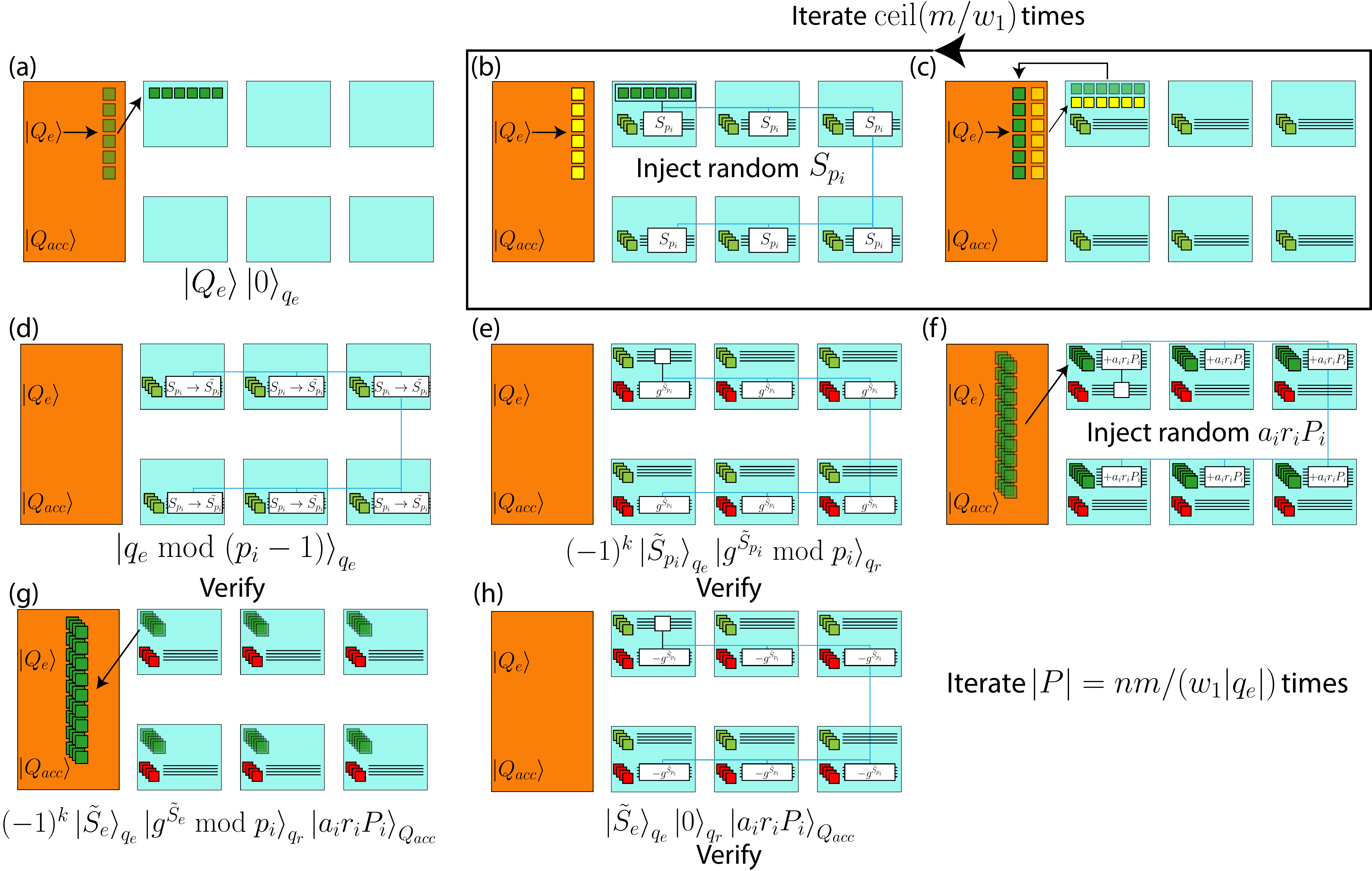}
	\caption{
        \red{\textbf{Estimation method in DShor}}.  (a-c) Iteration over all qubits of $\ket{Q_e}$ is estimated by iterating $w_1$ qubits from $\ket{Q_e}$ for $\text{ceil}(m/w_1)$ times. The Shor's simulator will simulate $40$ loadings, record the circuit time, and calculate the average circuit time for one iteration ($t_{\text{avg}}$). The circuit time of $\text{ceil}(m/w_1)$ iterations is estimated by $t_{\text{avg}}\times \text{ceil}(m/w_1)$. (a) $6$ buffer qubits are teleported from the memory to the QPUs, (b) and $S_{p_i}$ is loaded to $\ket{q_e}$. (c) Buffer qubits will be \textcolor{revisionred}{teleported back to the memory} when the loading is finished. (d) Shor's simulator will track $32$ Hilbert trajectories with random $q_e$ and verify if all the trajectories are compressed to \red{$\ket{\tilde{S}_{p_i}}_{q_e}$} at the end of the compression. (e) Shor's simulator will track $32$ Hilbert trajectories with random $q_e$ during the modular exponentiation and verify if all Hilbert trajectories are \red{$(-1)^k\ket{\tilde{S}_e}_{q_e}\ket{g^{\tilde{S}_e}\bmod p_i}_{q_r}$}. (f) The accumulator $\ket{Q_{\text{acc}}}$ is \textcolor{revisionred}{teleported} to QPUs and records the result of the modular exponentiation $\ket{q_r}$. (g) The accumulator $\ket{Q_{\text{acc}}}$ is teleported back to the memory. (h) $\ket{q_r}$ will be cleaned and the Shor's simulator will verify if all trajectories have $\ket{q_r = 0}$ without phase errors and will record the circuit time $t_{p_i}$ as the circuit time to iterate one $p_i$ in the prime system. The circuit time of one shot in the Shor's algorithm is estimated as $t_{p_i}\times nm/(|q_e|w_1)$.
    }
    \phantomsection
    \label{fig:Estimation method}
\end{figure*}
In this section, we will present the simulation procedures and the memory-QPU interactions in the DShor simulator. In the DShor, $\ket{Q_e}$ and $\ket{Q_{acc}}$ are stored in the memory, and the Shor's algorithm can be decomposed to \red{five steps} as illustrated in Fig.~\ref{fig:Shor's algorithm}. When loading the short logarithms $\ket{q_e}$ to the QPU, we simplify the iteration over all qubits in $\ket{Q_e}$ with window size $w_1$ to the iteration over the same $w_1$ qubits for $\text{ceil}(m/w_1)$ times as shown in Fig.~\ref{fig:Estimation method}(a-c) without losing generality. Buffer qubits (green squares) read $6$ qubits in the memory and are teleported from the memory to the router QPU (Fig.~\ref{fig:Estimation method}(a)), and the short logarithm is loaded from the buffer qubits (Fig.~\ref{fig:Estimation method}(b)). Buffer qubits are teleported back to the memory after the current short logarithm is loaded to $\ket{q_e}$. During the loading, the other $6$ buffer qubits in the memory will read the next $6$ qubits from $\ket{Q_e}$ as shown in Fig.~\ref{fig:Estimation method}(c). \red{We assume that the buffer qubits complete each six-qubit memory read before the next memory-loading layer. The shortest observed interval between these layers is $2.21$ s in the sampled six-QPU circuit (Appendix.~\ref{Appendix: QEC budget}).} As we estimate the iteration over $\ket{{Q_e}}$ to iterate $w_1$ qubits for $m/w_1$ times, Shor's simulator will not check the Hilbert trajectory when loading short logarithm to the QPU, and replace the table value $S_{p_i}$ to a random table as shown in Fig.~\ref{fig:Estimation method}(b). To further reduce the computation resource used by the simulator, we only iterate the loading  $40$ times and record the circuit time $t_{\text{avg}}$, and the circuit time of $m/w_1$ loadings is estimated by $t_{avg} \times \text{ceil}(m/w_1)$. When the $\ket{q_e}$ is compressed to $|p_i-1|$ bits, Shor's simulator will initialize $32$ \textcolor{revisionred}{Hilbert trajectories} with random $\ket{q_e}$, and the Shor's simulator will verify if all Hilbert trajectories are $\ket{q_e \bmod (p_i-1)}$ at the end of the circuit as shown in Fig.~\ref{fig:Estimation method}.

During the modular exponentiation, Shor's simulator will track $32$ Hilbert trajectories with random $\ket{q_e}$. $g^e$ is calculated to $\ket{q_r}$ by decomposing the modular exponentiations to the fundamental operations as discussed in Section\red{.~\ref{Chapter: residue arithmetic} and Appendix.}~\ref{Appendix B: Compilation of arithmetic}, \ref{Appendix: Modular addition}, and \ref{Appendix: Verification of DShor}, and the Shor's simulator will update all Hilbert trajectories \textcolor{revisionred}{after each fundamental operation} and record all measurement results for the delayed phase cleaning. At the end of the modular exponentiation, the Shor's simulator will verify if all Hilbert trajectories are $\ket{e}_{q_e}\ket{g^e\bmod p_i}_{q_r}$ but ignore all phase errors. $\ket{Q_{acc}}$ is teleported from the memory to the QPU, and \red{$a_ir_iP_i$} is loaded to $\ket{Q_{acc}}$ based on $\ket{q_r}$ as shown in \textcolor{revisionred}{\hypersetup{allcolors=revisionred}Fig.~\ref{fig:Estimation method}(f)}. As the prime systems are different in different RSA challenges and $N$, we replace \red{$a_ir_iP_i$} to a random table for simplicity. $\ket{Q_{acc}}$ is then teleported back to the memory in Fig.~\ref{fig:Estimation method}(g). Shor's simulator will decompose $-g^e\bmod p_i$ and memory cleaning to fundamental operations \textcolor{revisionred}{as shown in Fig.}~\ref{fig:Estimation method}(h)~\cite{Gidney2021, Gidney2025, Berry2019}, and Shor's simulator will verify if $\ket{q_r = 0}$ without phase errors in all Hilbert trajectories as shown in Fig.~\ref{fig:Estimation method}(h) and record the circuit time $t_{p_i}$ to iterate one $p_i$ in the prime system. The final circuit time of one shot in DShor is estimated by $t_{f} = t_{p_i}\times nm/(w_1|q_e|)$. The circuit time for the inverse Quantum Fourier transformation measurement is ignored as it is negligible compared with the circuit time of the modular exponentiations~\cite{Gidney2025, Copper2002,Cleve2000,Nam2020, Griffiths1996,Parker2000, Mosca1999}.

\section{Investigations of more parameters in the DShor}
\begin{figure}[t]
    \centering
    \includegraphics[width=\linewidth,height=0.64\textheight,keepaspectratio]{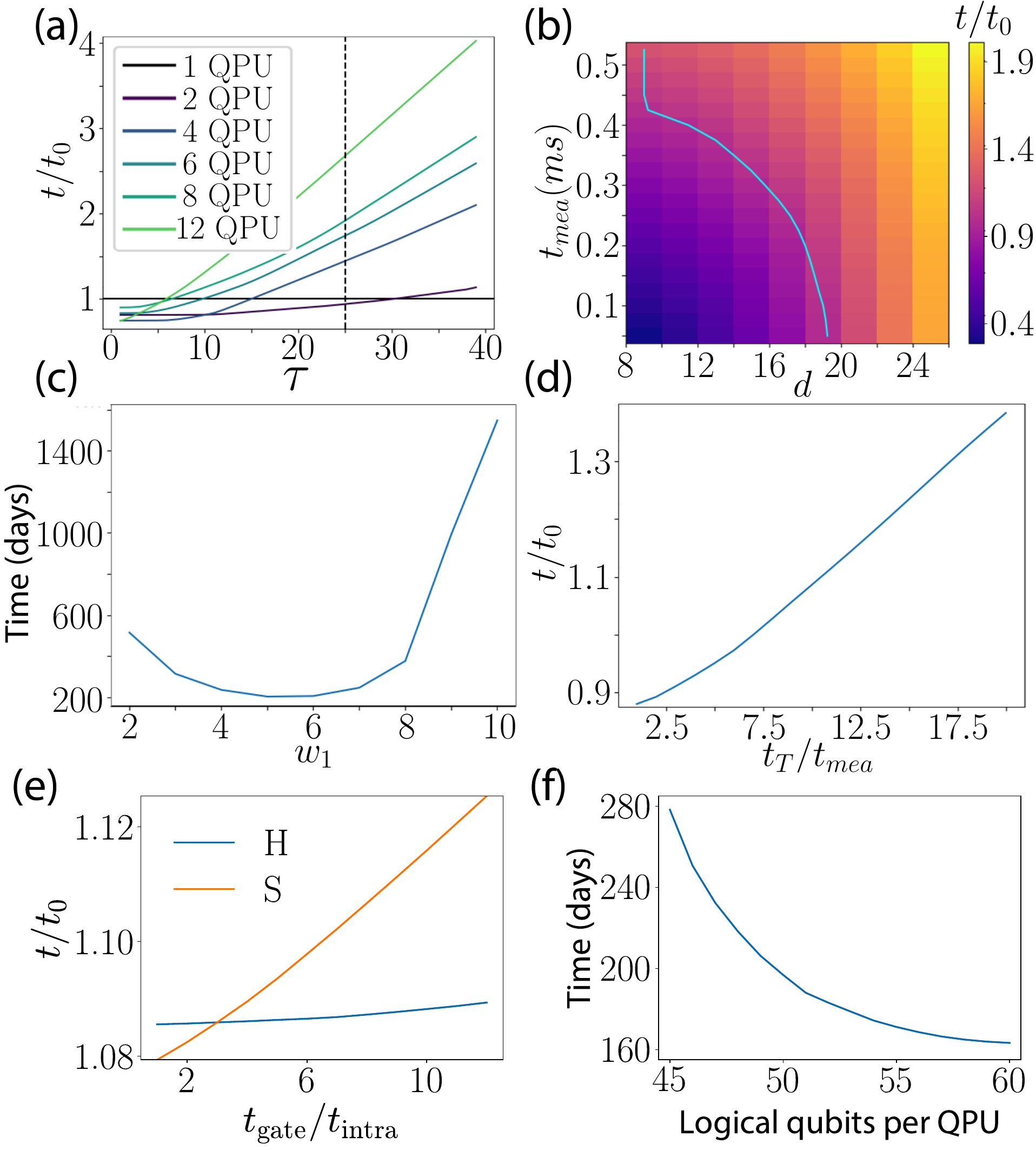}
	\caption{
        \textbf{DShor with more parameters}. (a-b) We estimate the space and time usage of the DShor with $2,4,6,8,12$ QPUs to factor RSA 2048-bit integers. The \textcolor{revisionred}{circuit} time is benchmarked by the circuit time of single-core Shor's algorithm with $180$ qubits with $t_{\text{mea}}= 250$ $\mu$s : \red{$t_0 = 74.0$} days. (a) The circuit time of DShor with $2,4,6,8,12$ QPUs at different costs of inter-module communications, where the communication cost is benchmarked by the cost of logical intra-module CNOTs (without QEC) time: $\tau = \frac{t_{\text{inter}}}{t_{\text{intra}}}$ and $t_{\text{intra}} = 250$ $\mu$s. $\tau = 25$ corresponds to $10^5$ physical inter-module Bell pairs generated per second as labeled by the dashed black line. DShor gains small advantage at low communication cost, and the delay from the inter-module communications is much more significant compared with DShor with $t_{\text{mea}} = 1$ ms as fewer operations can be hidden behind the measurements. The circuit time of 6-QPU DShor with communication rate $10^5$ Bell pairs/sec is \red{$74.1\%$} longer than the single-module Shor. (b) The circuit time of DShor with $6$ QPUs at different code distances and measurement times. The boundary of ``zero" \textcolor{revisionred}{communication} overhead is shifted to lower code distances as fewer operations can happen in parallel with the measurements. (c) The circuit time of DShor with $6$ QPUs and $t_{\text{mea}} = 1$ ms at different $w_1$. The circuit time reaches the minimum at \red{$w_1 = 5$}; \red{we retain $w_1=6$ as the canonical choice in the main text with only $1.31\%$ increase in circuit time compared with the optimal circuit time. The $w_1=10$ point uses $51$ qubits per QPU; the other window sizes use $49$.} (d) The circuit time of DShor with $6$ QPUs and $t_{\text{mea}} = 1$ ms at $t_T$ from $t_{\text{mea}}$ to $20 t_{\text{mea}}$. \red{The circuit times are normalized by the single-module reference $t_0=190$ days.} The \red{increase in circuit time from $t_T=10t_{\text{mea}}$ to $t_T=20t_{\text{mea}}$ is $27.4\%$.} (e) The circuit time of DShor with \textcolor{revisionred}{$6$ QPUs,} $t = 1\mathrm{ms}$ and $\tau = 25$ to factor 2048-bit RSA integers \textcolor{revisionred}{at different costs of} $\mathrm{H}$ and $\mathrm{S}$ gates. The circuit time of DShor depends very weakly on the cost of $\mathrm{H}$ and $\mathrm{S}$ gates. When the gate time of $\mathrm{H}$ increases from $250\mathrm{\mu s}$ to $3 \mathrm{ms}$, \textcolor{revisionred}{the circuit time increases by} $<1\%$. When the gate time of $\mathrm{S}$ gate increases from $250\mathrm{\mu s}$ to $3 \mathrm{ms}$, \textcolor{revisionred}{the circuit time increases by} $\approx 4\%$. (f) \textcolor{revisionred}{the same DShor simulation with $6$ QPUs} as $(e)$ with different logical qubits per QPU. The circuit time can further be reduced by $20.8\%$ and the circuit volume decreases by $3.05\%$ if the logical \textcolor{revisionred}{qubits per QPU increase} from $49$ to $60$\textcolor{revisionred}{.}
    }
    \phantomsection
    \label{fig:more parameters in DShor}
\end{figure}

The canonical six-QPU estimates in Fig.~\ref{fig:Result} and Table~\ref{Table: Running time in different RSA} use $t_{\text{mea}} = 1$ ms, $w_1 = 6$, $w_e = w_m = 3$, \red{$t_T = 10t_{\text{mea}}$}, $s = 8$, \red{and $t_{\text{H}} = t_{\text{S}} = 3 t_{\text{intra}}$}. In this section, we discuss the influence of other parameters in the DShor as shown in Fig.~\ref{fig:more parameters in DShor}.

In Fig.~\ref{fig:more parameters in DShor}(a), we estimate DShor with $t_{\text{mea}}=250\,\mu\mathrm{s}$, \red{using the single-module reference from Table~\ref{Table: Running time in different RSA}, $t_0=74.0$ days. For six QPUs, the "free modularization" happens at $\tau\approx9.78$. At $\tau=25$ ($10^5$ physical Bell pairs/s), the runtime is $129$ days, $74.1\%$ longer than the single-module reference.} Shorter measurements leave less time to overlap communication with local operations. Nevertheless, the six-QPU runtime is \red{$21\%$--$65\%$ lower} than for $t_{\text{mea}}=1$ ms over the sampled range $\tau=1$--$40$. Figure~\ref{fig:more parameters in DShor}(b) varies the code distance and measurement time. Shorter measurements shift the boundary $t=t_0$ toward lower code distances.

In Section.~\ref{Chapter: Result}, we fix $w_e=w_m=3$ because phase-cleanup workspace grows exponentially with these windows. Figure~\ref{fig:more parameters in DShor}(c) varies the short-logarithm loading window $w_1$ from $2$ to $10$. \red{The $w_1=10$ point uses $51$ logical qubits per QPU and takes approximately $1{,}550$ days; the other points use $49$ qubits per QPU. The minimum runtime is $204$ days at $w_1=5$ and we use $w_1=6$, which gives a very marginal increase of the circuit time: $1.31\%$ longer than the optimal setting.}

Table~\ref{Table: Running time in different RSA} uses cultivation duration \red{$t_T=10t_{\text{mea}}$}. This duration depends on the target fidelity, cultivation protocol, and code-growth procedure. Figure~\ref{fig:more parameters in DShor}(d) varies $t_T$ from $t_{\text{mea}}$ to $20t_{\text{mea}}$. \red{The six-QPU runtime increases from $167$ days at $t_T=t_{\text{mea}}$ to $206$ days at $10t_{\text{mea}}$ and $263$ days at $20t_{\text{mea}}$. The circuit time increases by $27.4\%$ when the $\mathrm{T}$-state preparation time doubles from $10$ ms to $20$ ms.}

    \red{The circuit time of DShor depends very weakly on the cost of $\mathrm{H}$ and $\mathrm{S}$ gates. In Table~\ref{Table: Running time in different RSA}, we assume the $\mathrm{H}$/$\mathrm{S}$ gate durations are $t_{\text{H}}=t_{\text{S}}=3t_{\text{intra}}=750\,\mu\mathrm{s}$. In Fig.~\ref{fig:more parameters in DShor}(e), we vary $t_{\text{H}}$ or $t_{\text{S}}$ independently from \textcolor{revisionred}{$t_{\text{intra}}=250\,\mu\mathrm{s}$ to $12t_{\text{intra}}=3\,\mathrm{ms}$}, keeping the other gate duration at $3t_{\text{intra}}$. Across this range, the circuit time increases by $0.350\%$ for $\mathrm{H}$ and $4.26\%$ for $\mathrm{S}$.

Figure~\ref{fig:more parameters in DShor}(f) varies the logical-qubit capacity of each processing QPU from $45$, the minimum that completes the current compilation, to $60$. The circuit time decreases from $278$ days at $45$ qubits to $163$ days at $60$ qubits. Although increasing the QPU size from $49$ to $60$ logical qubits can decrease the circuit time by $20.8\%$ and the processing-qubit spacetime volume ($N_{\text{QPU}}N_Lt$) by $3.05\%$, we use $49$ logical qubits per QPU in the main text to limit the size of each QPU to fit all physical qubits in the assumed $2\times 2\, \mathrm{mm^2}$ atom array.}

\label{Appendix: More parameters}
\clearpage

\section{Multi-address loading on the dynamic array architecture}
\label{Appendix: dynamic array}
Apart from the compilation of DShor on the static arrays with photonic inter-module connection, we provide the compilation of DShor on the arrays connected by the dynamic arrays as illustrated in Fig.~\ref{fig:dynamic array illustration}(a).

To provide a direct comparison with the photonic connected platforms, we focus our compilation on seven static arrays connected by four dynamic arrays. We assume both static array and dynamic arrays have dimension \red{$2\times2\,\mathrm{mm}^2$} with $7\times 7$ logical qubits as illustrated in Fig.~\ref{fig:DShor Structure}. The range of movement of dynamic array can cover all static arrays as illustrated in Fig.~\ref{fig:dynamic array illustration} but can only move collectively. We assume dynamic arrays can move with the acceleration $a_{\text{max}} = 20$ m/s$^2$, maximum speed $v_{\text{max}} = 2$ m/s and jerk $j = 600$ m/s$^3$.

As discussed in Section\red{.}~\ref{Chatper: Dynamic array}, to resolve the competition between the spatial usage of the Bell pair reservoir and the inter-module bandwidth, we provide a special compilation of DShor using multi-address loading as illustrated in Fig.~\ref{fig:dynamic loading circuit} with more details in Fig.~\ref{fig:detailed dynamic loading}. We applied a hybrid QRAM with first three qubits read linearly and the latter three qubits read in parallel by the Bucket-Brigade circuit~\cite{Hann2021, Shen2025, Giovannetti2008, Giovannetti2008Architecture, Cesa2025}. 

Fig.~\ref{fig:dynamic loading circuit} shows the circuit gadget used in the modular loading on the dynamic array. The latter three address qubits (dark green boxes) and the input qubit (light green boxes) from the linear read are teleported to the first dynamic array. The Bucket-Brigade circuit will expand the three address qubits into eight one-hot encoding qubits (yellow boxes). Four encoded qubits are teleported to the second dynamic array. The address and input qubits are ``duplicated" by maximally entangling them with four ancilla qubits, and the same procedures are repeated on the third and fourth dynamic arrays. The dynamic array will carry the "one-hot" encoding qubits and the antenna qubits (light blue boxes) for Bell pair (dark blue boxes) generations to static arrays. When the dynamic array overlaps with the static array, inter-module CNOTs will be applied by the drive by gates. An example of GHZ state preparation by drive-by gates is illustrated in Fig.~\ref{fig:dynamic loading circuit}(b). To maximally use the bandwidth of the dynamic array, half of the table values are loaded on the buffer qubits and later loaded back to the $q_l$. Therefore, one cascade of CNOTs can be split into two, and the loading can be finished within one sweep.

A more detailed protocol is shown in Fig.~\ref{fig:detailed dynamic loading}. Before the transverse move of the dynamic array, the three address qubits are expanded to eight one-hot encoding qubits on the first dynamic array, and half of the qubits are teleported to the second dynamic array (Fig.~\ref{fig:detailed dynamic loading}(a)). The same procedure happens to the third and fourth dynamic arrays. Then, the dynamic arrays sweep through the static array. We assume the static and dynamic arrays are overlapped by a $7\times 2$ logical qubit zone. Table values are loaded to the $5\times 2$ qubit zone, and the $2\times2$ qubit zone stores the Bell pairs generated by drive-by gates. During the forward sweep, four table values are loaded to the quantum registers, and each module generates two inter-module Bell pairs with \textcolor{revisionred}{each neighboring module} (Fig.~\ref{fig:detailed dynamic loading}(b)). During the backward sweep, the remaining four table values are loaded, one GHZ state across modules is generated, three quantum addresses used in Bucket-Brigade are teleported to the dynamic array (Fig.~\ref{fig:detailed dynamic loading}(c,d)), and all necessary inter-module Bell pairs between dynamic arrays are generated (Fig.~\ref{fig:detailed dynamic loading}(e)). The Bell pairs between the dynamic arrays are generated using the ancilla qubits on the static array (light blue squares), and the Bell pairs generated by the ancilla qubits share the same label in the square (e.g. the Bell pairs with label ``+" are generated by the ancilla qubits with label ``+") as shown in Fig.~\ref{fig:detailed dynamic loading}(f). After four transverse sweeps, all static arrays will use the two nearest-neighbor Bell pairs and \textcolor{revisionred}{one GHZ state} across arrays for modular addition as shown in Appendix\red{.}~\ref{Appendix: Modular addition}.

\section{\textcolor{revisionred}{Schedule of atom movements} in DShor}
\label{Appendix: Machine code}

\begin{figure}[t]
    \includegraphics[width=\linewidth]{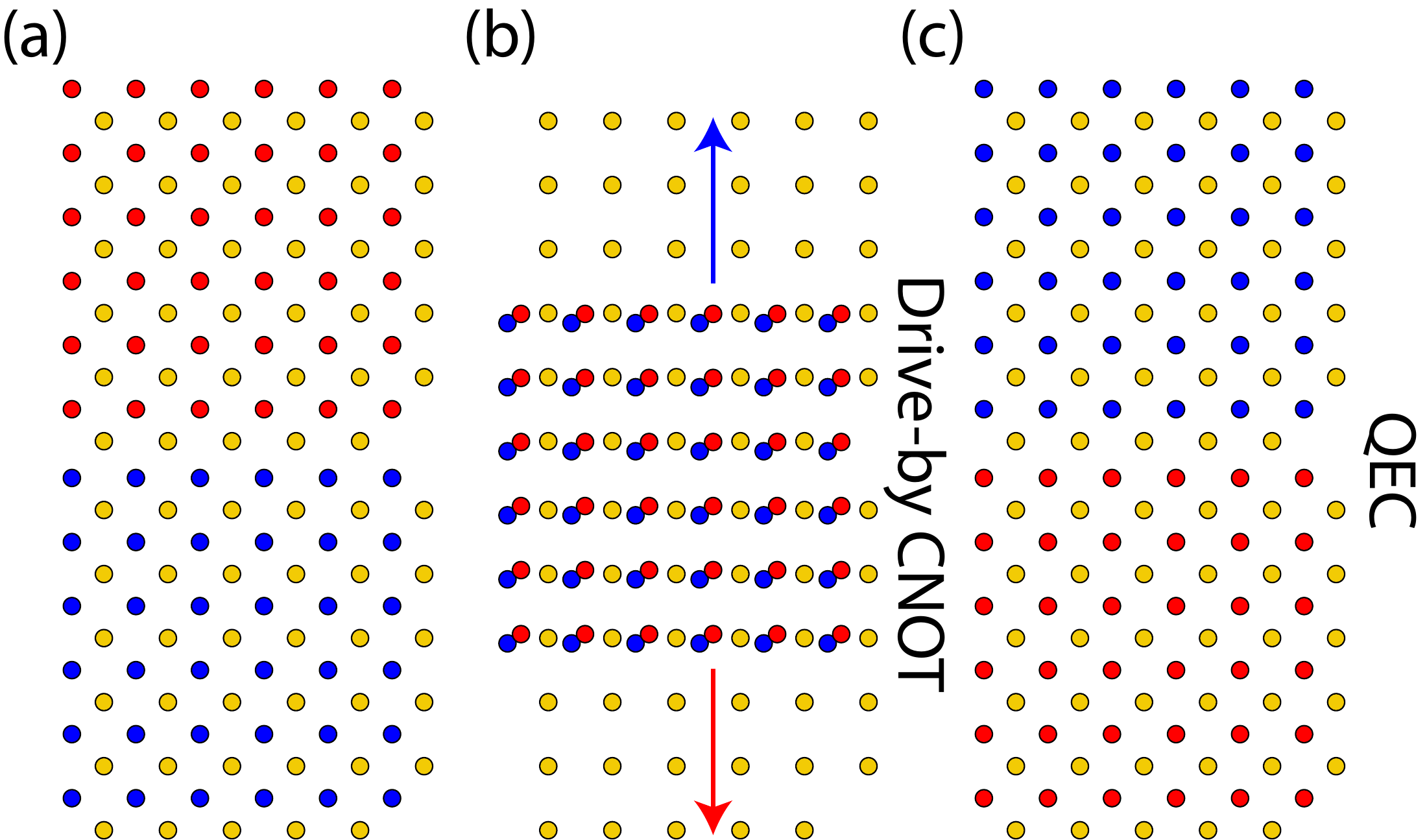}
	\caption{
        \textbf{Swap CNOT}. In the \textcolor{revisionred}{schedule of atom movements in DShor}, we assume the logical CNOT is applied by swapping the positions of two logical patches. (a-b) Two logical patches are moving toward each other, and the drive-by CNOTs are applied when two data qubit patches overlap. (c) Two patches swap their position and use the QEC ancilla qubits (yellow qubits) for the following QEC cycle.  
    }
    \phantomsection
    \label{fig:Swap CNOT}
\end{figure}
\begin{figure*}[h!]
    \includegraphics[width=\linewidth]{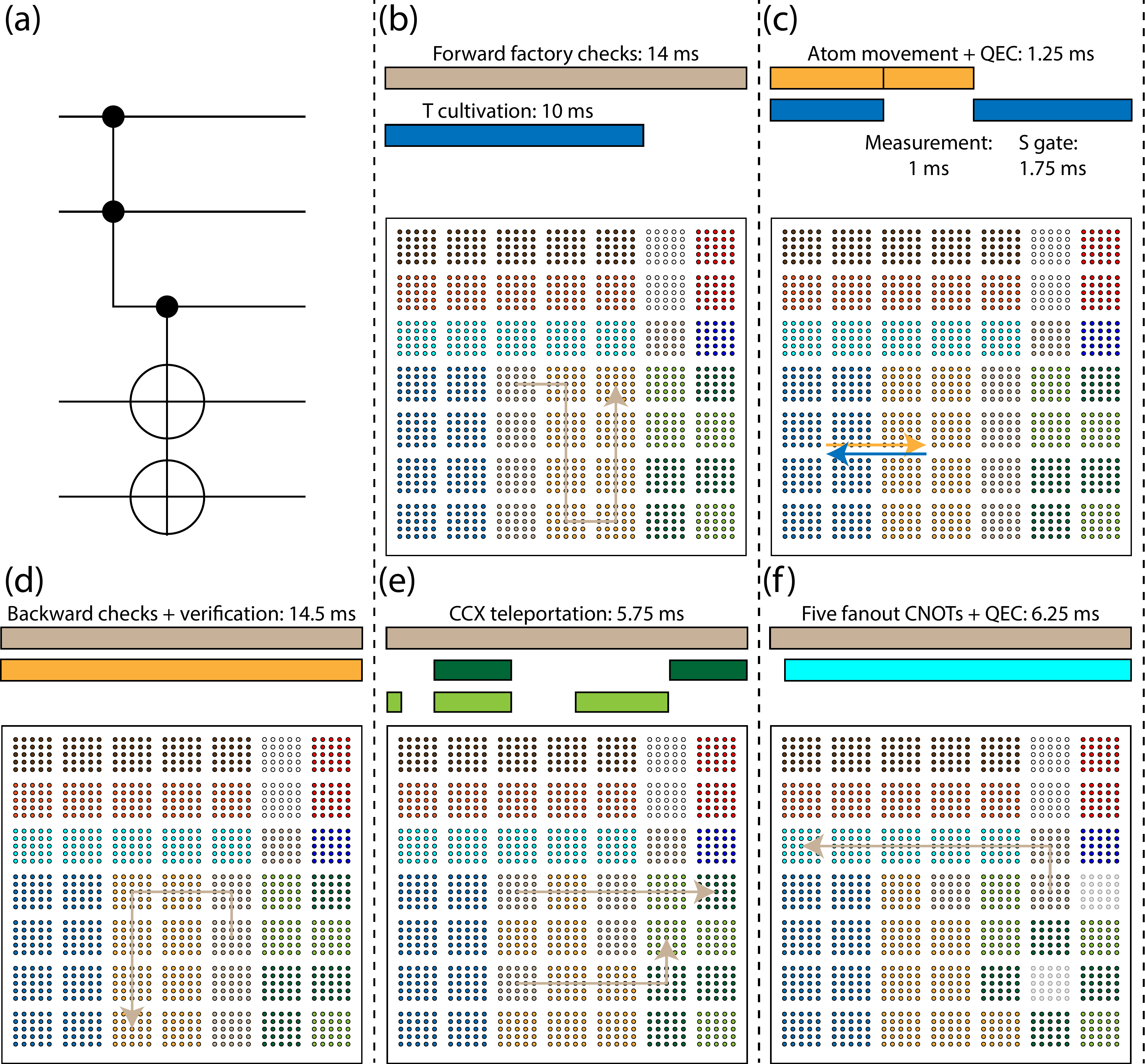}
	\caption{
        \textbf{Atom moving of the logical AND gate in DShor}. 
    The \textcolor{revisionred}{schedule of atom movements for the loading circuit is} shown in (a). \textcolor{revisionred}{The arrays in (b)--(f) show the layouts at the start of each subroutine. (b) Forward checks in the $8\mathrm{T}$-to-CCX factory.} (c) $\mathrm{T}$ gate is injected to the ancilla qubits in the factory from magic state $\ket{T}$. \textcolor{revisionred}{(d) Backward checks and verification of $\ket{CCX}$. (e) CCX teleportation from the $\ket{CCX}$ in the factory to the address qubits. Both conditional CNOTs are implemented for the worst case scenario. (f) Fanout from the output AND ancilla to the $\ket{q_l}$ register.}}
    \label{fig:machine code}
\end{figure*}
\begin{figure*}[t]
    \includegraphics[width=0.9\linewidth]{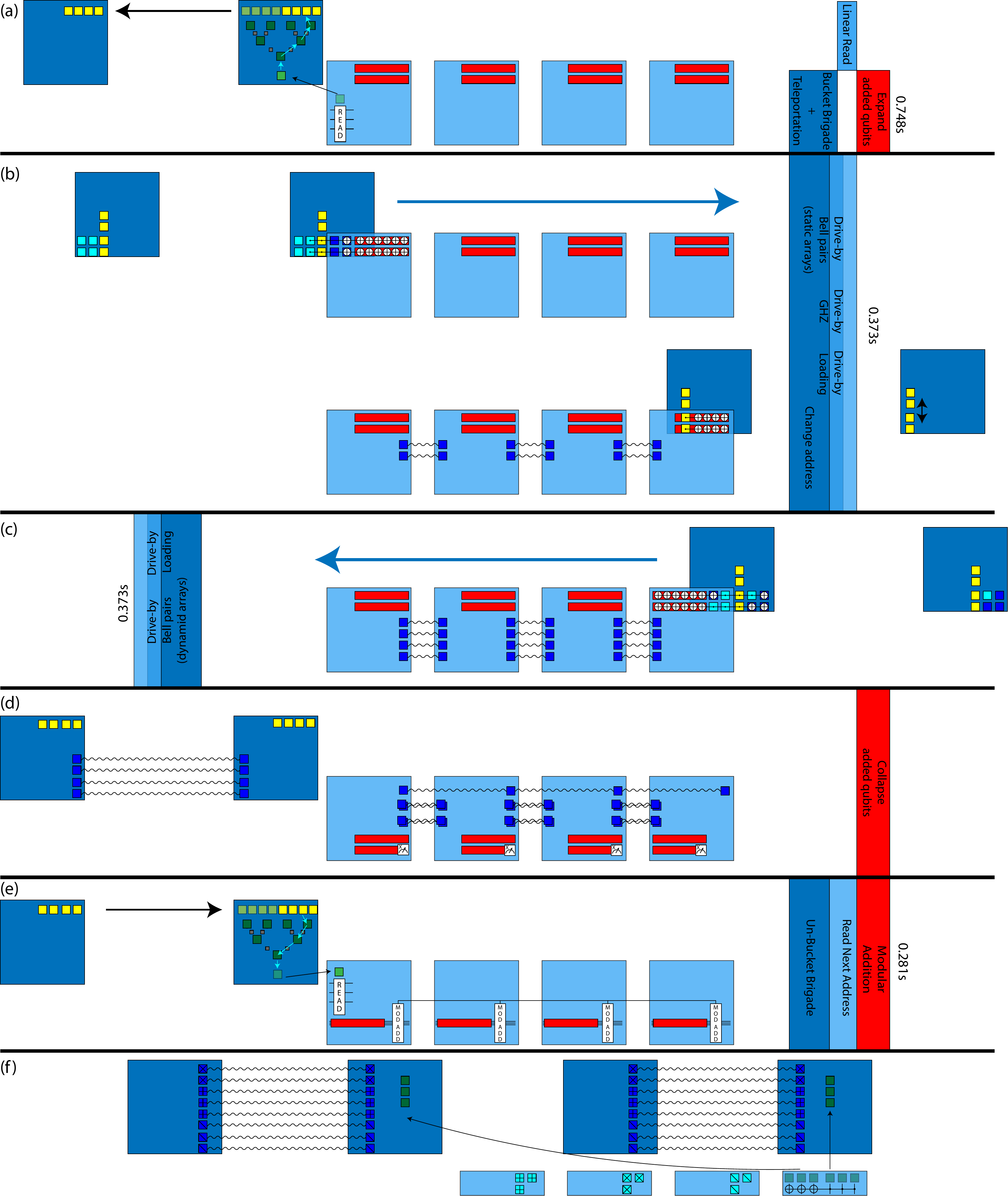}
	\caption{
        \textbf{\textcolor{revisionred}{Schedule of atom movements for} modular loadings using the dynamic array}. Colors indicate qubit activity and arrows indicate transport. (a) Expand the last three address qubits into eight one-hot selectors and distribute them between dynamic arrays. (b) Load the first four table values and generate nearest-neighbor Bell pairs during the forward sweep. (c) Load the remaining values, prepare a GHZ state, and transfer addresses during the backward sweep. (d) Entanglement after one round trip. (e) Reverse the Bucket-Brigade encoding and read the next address while performing modular addition. (f) Matching labels identify Bell pairs routed through static-array ancillas.}
    
    \phantomsection
    \label{fig:detailed dynamic loading}
\end{figure*}

Figure~\ref{fig:machine code} and the supplementary video illustrate an AOD-based atom-motion schedule for an \red{$8\mathrm{T}$-to-CCX factory and loading subroutine} \textcolor{revisionred}{including the exact schedule of atom transport, Rydberg gates, and measurements.} The ``swap'' CNOT in Fig.~\ref{fig:Swap CNOT} moves two patches in opposite directions; a transversal drive-by CNOT is applied as they overlap, followed by QEC after they exchange positions. \textcolor{revisionred}{Under the gate timings shown in Fig.~\ref{fig:machine code}, the schedule takes $44.75$ ms, compared with $35.00$ ms for the corresponding compiler schedule, an increase of $9.75$ ms ($27.9\%$). } The overhead depends on the transport hardware, including fast SLMs\red{~\cite{Wei2026,Deters2026} or independently controlled AODs.} We leave optimization of the full transport schedule to future work.
\textcolor{revisionred}{Table.~\ref{tab:atom-movement-timings} compares the five subroutines in the DShor compiler and the exact atom schedule as shown in Fig.~\ref{fig:machine code}. The main latency from the exact atom schedule is the forward and backward cascades of CNOTs, which lead to $8.75\,\mathrm{ms}$ additional latency among the total $9.75ms$ latency. The $\mathrm{T}$ injection is faster in the exact atom schedule because the forward checks in the exact schedule are slower than the $t_\mathrm{T} = 10 \mathrm{ms}$, so the $\ket{T}$ can be injected without additional waiting. However, in DShor's compiler where the moving latency is ignored, the forward cascades of CNOTs are faster than the cultivation time, which leads to additional $0.25\mathrm{ms}$ waiting time during the T injection.}
\begin{table}[H]
    \centering
    \color{revisionred}
    \small
    \setlength{\tabcolsep}{4pt}
    \renewcommand{\arraystretch}{1.18}
    \begin{tabular}{@{}lcc@{}}
        \toprule
        \multicolumn{1}{@{}c}{Subroutine} & Compiler (ms) & \makecell[c]{Exact atom\\schedule (ms)} \\
        \midrule
        (b) Forward checks & 9.75 & 14.00 \\
        \makecell[l]{(c) $\mathrm{T}$ injection} & 4.25 & 4.00 \\
        \makecell[l]{(d) Backward checks\\and verification} & 10.00 & 14.50 \\
        \makecell[l]{(e) CCX teleportation\\after factory readiness} & 4.75 & 5.75 \\
        (f) Fanout and routing & 6.25 & 6.50 \\
        \midrule
        \textbf{Total runtime} & \textbf{35.00} & \textbf{44.75} \\
        \bottomrule
    \end{tabular}
    \caption{Runtime of the five subroutines during Loadings as shown in Fig.~\ref{fig:machine code}. $8.75\mathrm{ms}$ of the $9.75 \mathrm{ms}$ moving latency is from the forward and backward checks during the $8\mathrm{T}$-to-CCX factory. The T injection in \textcolor{revisionred}{DShor's compiler} is $0.25\mathrm{ms}$ slower than that from the exact schedule because the \textcolor{revisionred}{forward-check time} in \textcolor{revisionred}{DShor's compiler} is $0.25\mathrm{ms}$ faster than the code cultivation time $t_{\mathrm{T}} = 10\mathrm{ms}$, so there is additional $0.25\mathrm{ms}$ waiting time during the $\mathrm{T}$ gate injection in \textcolor{revisionred}{DShor's compiler}.}
    \label{tab:atom-movement-timings}
\end{table}

{\color{revisionred}
Our compiled runtimes also ignore the atom-refilling latency. We approximate the latency from atom loss by the SWAP-based loss-detecting SECs in Ref.~\cite{Baranes2026Loss}. According to Fig.~15(a) in Ref.~\cite{Baranes2026Loss}, the logical error changes weakly for a period-five sequence of four conventional SECs followed by one loss-detecting SEC. We adopt this sequence and refill lost atoms after each loss-detecting SEC. Assuming $\textcolor{revisionred}{10^{-3}}$ loss per atom per $1.25$ ms cycle~\cite{Sharma2026Loss}, the expected number of lost atoms among the $61{,}250$ atoms in one QPU is
\[
61{,}250\bigl[1-(1-10^{-3})^5\bigr]\simeq306.
\]

At the assumed supply rate of $10^6$ atoms/s per QPU~\cite{LiThompson2025}, refilling these atoms takes $0.306$ ms, which is shorter than the duration of five SECs. This refilling can happen in parallel with the SECs. Therefore, the main latency from the atom refill is to move the atoms from the refilling reservoir to the data zone. For the \textcolor{revisionred}{$2\times2\,\mathrm{mm}^2$} array and $1$ m/s atom moving speed we assume in this paper, the average transport time of the replacement atoms is $1\,\mathrm{ms}$. Therefore, atom-refilling latency is $1\,\mathrm{ms}$ every five SECs, or $0.2\,\mathrm{ms}$ per logical operation, corresponding to $16\%$ overhead on the logical clock rate. Combining the time overheads from long-range atom transport and atom refilling, the logical clock rate can increase by $16\% + 27.9\% = 43.9\%$. We round this allowance up to $50\%$, as mentioned in Section~\ref{Chapter: Hardware and Architecture}.
}

\red{
\section{Comparison with other works}
\label{Appendix: Comparison with other works}

In this section, we will \textcolor{revisionred}{compare the circuit volume of DShor with} Gidney's compilation~\cite{Gidney2025} (surface code based computation) and the compilation from Oratomic~\cite{Cain2026} (neutral atom platform with qLDPC code).

\subsection{Comparison with Gidney}
\label{Appendix: Comparison with Gidney}

{\color{revisionred}\hypersetup{allcolors=revisionred}
In this section, we compare the circuit volume of DShor with that of the compilation in Ref.~\cite{Gidney2025}. We define the circuit volume as
\begin{equation}
V=N_{\mathrm{peak}}D,
\end{equation}
where $D$ is the circuit depth per shot and $N_{\mathrm{peak}}$ is the peak allocated number of physical qubits used in the compilation, including memory. However, Ref.~\cite{Gidney2025} does not provide the explicit depth of the full circuit. We therefore replace $D$ with an effective circuit depth to remove the overall gate-time scale assumed for the different platforms:
\begin{equation}
D_{\mathrm{eff}}=\frac{t_{\mathrm{shot}}}{t_{\mathrm{layer}}},
\qquad
V_{\mathrm{depth}}=N_{\mathrm{peak}}D_{\mathrm{eff}}.
\end{equation}
Here, $t_{\mathrm{shot}}$ is the runtime per shot and $t_{\mathrm{layer}}$ is the reference duration of a logical lattice surgery.

Using the reported runtime of $12.07$ hours per shot and the $25\,\mu\mathrm{s}$ duration of an ordinary lattice-surgery layer at $d=25$, \textcolor{revisionred}{Gidney's estimated effective depth} is
\begin{equation}
D_{\mathrm{eff}}^{\mathrm{Gidney}}
\simeq\frac{12.07\times3600\,\mathrm{s}}{25\times10^{-6}\,\mathrm{s}}
\simeq1.74\times10^9.
\end{equation}
Gidney's compilation uses $897{,}864$ physical qubits~\cite{Gidney2025}, giving $V_{\mathrm{depth}}^{\mathrm{Gidney}}\simeq1.56\times10^{15}$ physical-qubit layers.

For DShor, we use the single-module compilation with $300$ $d = 25$ surface code patches plus memory for comparison to remove the time and space overhead from the distributed implementation from DShor. The total number of physical qubits is $485{,}000$. The compiler gives an estimated per-shot depth of $5.55\times10^{10}$ clock slots, where one slot is $250\,\mu\mathrm{s}$. One logical CNOT ($250\,\mu\mathrm{s}$) followed by QEC ($1\,\textcolor{revisionred}{\mathrm{ms}}$) occupies five slots, so $D_{\mathrm{eff}}^{\mathrm{DShor}}\simeq1.11\times10^{10}$ logical layers. Therefore, the circuit volume per shot is $V_{\mathrm{depth}}^{\mathrm{DShor}}\simeq5.38\times10^{15}$ physical-qubit layers. The circuit volume ratio is therefore
\begin{equation}
\frac{V_{\mathrm{depth}}^{\mathrm{DShor}}}{V_{\mathrm{depth}}^{\mathrm{Gidney}}}
\simeq\frac{485{,}000\times(1.11\times10^{10})}{897{,}864\times(1.74\times10^9)}
\simeq3.45.
\label{eq:gidney-volume-comparison}
\end{equation}
}
Although \textcolor{revisionred}{our circuit volume is $3.45$ times that of Gidney's compilation}, the two compilations use similar numbers of Toffoli gates. Gidney's public resource estimator with the RSA-2048 parameters in Table~5 of Ref.~\cite{Gidney2025} gives approximately $7.13\times10^8$ Toffoli gates per shot, compared with $7.63\times10^8$ in this single-module compilation. The \textcolor{revisionred}{$7.01\%$} increase is mainly from the different arithmetic parameters retained from DShor's distributed compilation.

{\color{revisionred}\hypersetup{allcolors=revisionred}
The circuit volume inflation is largely due to the different implementations of qubit fanout in the two compilations. On the superconducting platform, Ref.~\cite{Gidney2025} assumes a $1\,\mu\mathrm{s}$ QEC cycle and nearest-neighbor interactions, and uses lattice surgery. On the neutral-atom platform, where measurement is much slower ($1$ ms assumed here) and non-local interactions are available, we instead use transversal gates. These choices lead to different implementations of qubit fanout. Lattice surgery can implement fanout in one logical parity-measurement layer containing $d$ syndrome-extraction rounds, taking $25\,\mu\mathrm{s}$ under the assumptions of Ref.~\cite{Gidney2025}. DShor instead uses sequential transversal CNOTs, so a fanout to $k$ targets costs $k$ CNOT-plus-QEC layers. 

One may wonder if replacing the sequential transversal \textcolor{revisionred}{CNOTs used in DShor with the lattice surgery} from Ref.~\cite{Gidney2025} can remove the $k\times$ overhead from the qubit fanout. The answer is no. Under our neutral-atom timing assumptions, a $d=25$ surgery layer would take $25.0$ ms, or $20.0$ times the duration of one logical CNOT plus QEC ($1.25$ ms). In the six-QPU compilation, the largest local fanout has six targets and takes $7.50$ ms; a $25.0$ ms surgery layer would therefore take $3.33$ times as long. Although a GHZ resource can provide the efficient qubit fanout with transversal gates~\cite{Zhou2025}, this method leads to additional ancillas so the assumed $2\times 2 mm^2$ atom array can no longer hold all required qubits.

To isolate the fanout cost, we replace each CNOT cascade in the single-module compilation by one logical fanout layer costed at the same $1.25$ ms CNOT-plus-QEC duration (same as the assumption from Ref.~\cite{Gidney2025}). The compiler gives an estimated per-shot depth of $1.98\times10^{10}$ clock slots, corresponding to $D_{\mathrm{eff}}\simeq3.95\times10^9$ and $V_{\mathrm{depth}}\simeq1.92\times10^{15}$ physical-qubit layers. This reduces the effective depth and volume by a factor of $2.81$, leaving a volume ratio of approximately $1.23$ relative to Gidney. Uniformly reducing all DShor operation durations by $50$ leaves these normalized depths and volumes unchanged. The remaining $1.23\times$ overhead may come from the longer assumed CCX preparation and teleportation time in DShor ($452\,\mu\mathrm{s}$ compared with $150\,\mu\mathrm{s}$ per CCX state per factory in Ref.~\cite{Gidney2025}) and the more detailed gate accounting in DShor's compiler.
}

\subsection{Comparison with Oratomic}

We also compare the single-core DShor compilation with $300$ logical qubits {\color{revisionred}\hypersetup{allcolors=revisionred}(using the configuration in Appendix~\ref{Appendix: Comparison with Gidney})} with the space-efficient and balanced Oratomic architectures~\cite{Cain2026}. Since Ref.~\cite{Cain2026} uses the same compilation as Ref.~\cite{Gidney2025}, the \textcolor{revisionred}{numbers of Toffoli gates used by the two methods are} still very similar using the analysis from Appendix.~\ref{Appendix: Comparison with Gidney}.

Since the circuit time from Ref.~\cite{Cain2026} is derived from the high-level circuit time scaling with the number of Toffoli gates used in the algorithm, \textcolor{revisionred}{directly comparing the logical-operation depths of Oratomic's compilation and DShor's compilation is challenging}. We instead use the space-time circuit volume definition where \textcolor{revisionred}{$V_{\mathrm{time}} = t_{\mathrm{shot}}N_{\mathrm{peak}}$}. {\color{revisionred}\hypersetup{allcolors=revisionred}The space-efficient and balanced points in Fig.~3(b) of Ref.~\cite{Cain2026} use $11{,}033$ and $13{,}255$ physical qubits. We convert their reported expected total runtimes of approximately $43{,}000$ and $10{,}000$ days to per-shot runtimes by dividing by the $9.2$ expected shots in Table~5 of Ref.~\cite{Gidney2025}, giving approximately $4.67\times10^3$ and $1.09\times10^3$ days per shot, respectively. Their per-shot volumes are therefore approximately $5.16\times10^7$ and $1.44\times10^7$ physical-qubit days. For comparison, the single-core DShor compilation used in Appendix~\ref{Appendix: Comparison with Gidney} uses $485{,}000$ physical qubits and has a runtime of approximately $161$ days per shot, giving a per-shot volume of $7.79\times10^7$ physical-qubit days.} 

DShor's volume is therefore approximately $1.51$ times the space-efficient estimate and $5.41$ times the balanced estimate. Since the circuit time from Ref.~\cite{Cain2026} is from the high-level approximation based on the number of Toffoli gates used in the circuit, the circuit time from DShor based on the exact circuit can lead to a more inflated circuit. Besides, the \textcolor{revisionred}{hardware assumptions from Ref.~\cite{Cain2026} are more optimistic} than those from DShor. Oratomic uses high-rate lifted-product codes and assumes a $1$ ms stabilizer-measurement cycle, with $2d/3$ such cycles per surgery~\cite{Cain2026}, which requires more advanced atom transport technology and \textcolor{revisionred}{more efficient decoding assumptions} compared with the assumptions in DShor.
}

\clearpage
\bibliography{library}

\end{document}